%Paper: hep-ph/9409393
%From: LANCE@SLACVM.SLAC.Stanford.EDU
%Date: Thu, 22 Sep 1994 16:22 -0800 (PST)
%Date (revised): Sun, 09 Oct 1994 09:43 -0800 (PST)
%Date (revised): Tue, 04 Apr 1995 14:01 -0800 (PST)

%%%%%%%%%%%%%%%%%%%%%%%%%%%%%%%%%%%%%%%%%%%%%%%%%%%%%%%%%%%%%%%%
% One-Loop Corrections to Two-Quark Three-Gluon Amplitudes,
%  by Z. Bern, L. Dixon and D.A. Kosower,
%  46 pages, SLAC-PUB-6663, UCLA/94/TEP/33, Saclay/SPhT-T94/108.
%
% The file should be TeXed TWICE for proper cross-referencing.
% Thirteen postscript files containing 13 figures accompany this
% file, and are to be encapsulated using epsf (included in this file):
%
%  Rules.psd
%  ColorOrder.psd
%  LeftRightA.psd
%  LeftRightB.psd
%  NoTree.psd
%  ColorFlow.psd
%  TraceReverse.psd
%  Exclude.psd
%  SubleadCancelA.psd
%  SubleadCancelB.psd
%  ColorStrip.psd
%  Gluino.psd
%  Photon.psd
%
%%%%%%%%%%%%%%%%%%%%%%%%%%%%%%%%%%%%%%%%%%%%%%%%%%%%%%%%%%%%%%%%%

\magnification=\magstephalf

%%%%%%%%%%%%%%%%%%%%%%%%%%%%%%%%%%%%%%%%%%%%%%%%%%%%%%%%%%%%%%%%%
%\input header
\newbox\SlashedBox
\def\slashed#1{\setbox\SlashedBox=\hbox{#1}
\hbox to 0pt{\hbox to 1\wd\SlashedBox{\hfil/\hfil}\hss}#1}
\def\hboxtosizeof#1#2{\setbox\SlashedBox=\hbox{#1}
\hbox to 1\wd\SlashedBox{#2}}

% The following is necessary so that we can get a partial slash
% inside a math display... sigh.
\def\mathslashed#1{\setbox\SlashedBox=\hbox{$#1$}
\hbox to 0pt{\hbox to 1\wd\SlashedBox{\hfil/\hfil}\hss}#1}

\def\ifsmall{\iffalse}  % default is unreduced.
\def\titlepagefont{}  % default is ordinary font.

% the ps: landscape must be the first special command in order
% to get the first page in landscape mode -- so we go through some
% contortions to define TeXgraphics in the default case.
\def\DefineTeXgraphics{%
\special{ps::[global] /TeXgraphics { } def}}  % No need to do anything

\def\today{\ifcase\month\or January\or February\or March\or April\or May
\or June\or July\or August\or September\or October\or November\or
December\fi\space\number\day, \number\year}
\def\eatPrefix19{}
\def\Year{\expandafter\eatPrefix\the\year}
\newcount\hours \newcount\minutes
\def\monthname{\ifcase\month\or
January\or February\or March\or April\or May\or June\or July\or
August\or September\or October\or November\or December\fi}
\def\shortmonthname{\ifcase\month\or
Jan\or Feb\or Mar\or Apr\or May\or Jun\or Jul\or
Aug\or Sep\or Oct\or Nov\or Dec\fi}

\def\TimeStamp{\hours\the\time\divide\hours by60%
\minutes -\the\time\divide\minutes by60\multiply\minutes by60%
\advance\minutes by\the\time%
${\rm \shortmonthname}\cdot\if\day<10{}0\fi\the\day\cdot\the\year%
\qquad\the\hours:\if\minutes<10{}0\fi\the\minutes$}

%\DefineTeXgraphics}

%\DefineTeXgraphics}

%\DefineTeXgraphics}

\def\Title#1{%
\vskip 1in{\titlefont\centerline{#1}}\vskip .5in}
%\DefineTeXgraphics}

\def\Date#1{\leftline{#1}\tenrm\supereject%
\global\hsize=\hsbody\global\hoffset=\hbodyoffset%
\footline={\hss\tenrm\folio\hss}}% restores pagenumbers

\newif\ifdraftmode
\newif\ifleftlabels  % Labels in left margins as well, for European-size paper

% Stolen from harvmac.tex 04/08/92
%       use \nolabels to get rid of eqn, ref, and fig labels in draft mode
\def\nolabels{\def\wrlabeL##1{}\def\eqlabeL##1{}\def\reflabeL##1{}}
\def\writelabels{\def\wrlabeL##1{\leavevmode\vadjust{\rlap{\smash%
{\line{{\escapechar=` \hfill\rlap{\sevenrm\hskip.03in\string##1}}}}}}}%
\def\eqlabeL##1{{\escapechar-1\rlap{\sevenrm\hskip.05in\string##1}}}%
\def\reflabeL##1{\noexpand\rlap{\noexpand\sevenrm[\string##1]}}}
\def\writeleftlabels{\def\wrlabeL##1{\leavevmode\vadjust{\rlap{\smash%
{\line{{\escapechar=` \hfill\rlap{\sevenrm\hskip.03in\string##1}}}}}}}%
\def\eqlabeL##1{{\escapechar-1%
\rlap{\sixrm\hskip.05in\string##1}%
\llap{\sevenrm\string##1\hskip.03in\hbox to \hsize{}}}}%
\def\reflabeL##1{\noexpand\rlap{\noexpand\sevenrm[\string##1]}}}
\nolabels

\newdimen\fullhsize
\newdimen\hstitle
\hstitle=\hsize % default
\newdimen\hsbody
\hsbody=\hsize % default
\newdimen\hbodyoffset
\hbodyoffset=\hoffset % default
\newbox\leftpage
\def\abstract#1{#1}
\def\rotated{\special{ps: landscape}
\magnification=1000  % This line must come before we change vsize,
                     % since \magnification sets it to a fixed value.
\baselineskip=14pt
\global\hstitle=9truein\global\hsbody=4.75truein
\global\vsize=7truein\global\voffset=-.31truein
\global\hoffset=-0.54in\global\hbodyoffset=-.54truein
\global\fullhsize=10truein
\def\DefineTeXgraphics{%
\special{ps::[global]
/TeXgraphics {currentpoint translate 0.7 0.7 scale
              -80 0.72 mul -1000 0.72 mul translate} def}}
 % 0.7 is slightly less than the ratio of horizontal sizes: 4.75 to 6.5
\let\lr=L
\def\ifsmall{\iftrue}
\def\titlepagefont{\twelvepoint}
\trueseventeenpoint
\def\almostshipout##1{\if L\lr \count1=1
      \global\setbox\leftpage=##1 \global\let\lr=R
   \else \count1=2
      \shipout\vbox{\hbox to\fullhsize{\box\leftpage\hfil##1}}
      \global\let\lr=L\fi}

\output={\ifnum\count0=1 %%% This is the HUTP version
 \shipout\vbox{\hbox to \fullhsize{\hfill\pagebody\hfill}}\advancepageno
 \else
 \almostshipout{\leftline{\vbox{\pagebody\makefootline}}}\advancepageno
 \fi}

\def\abstract##1{{\leftskip=1.5in\rightskip=1.5in ##1\par}} }

% Messages on lines by themselves
\def\linemessage#1{\immediate\write16{#1}}

% tagged sec numbers
\global\newcount\secno \global\secno=0
\global\newcount\appno \global\appno=0
\global\newcount\meqno \global\meqno=1
\global\newcount\subsecno \global\subsecno=0
% and figure numbers
\global\newcount\figno \global\figno=0

\newif\ifAnyCounterChanged
% If we are comparing numbers, there's no special problem.
% But if we are comparing roman numerals, we must be careful, because
% stuff read in from the aux file would be made up of ordinary
% characters (category code = 11), whereas \romannumeral generates
% characters with category code = 12..., so the stuff from the
% current run won't appear equal to the previous definition, as far
% as \warnIfChanged is concerned.
% To get around this, we have a macro \makeNormal, which converts
% letters `ivxlcdmIVXLCDM' to normal letters, no matter what their category
% code.  The macro has the convoluted form it does, with aftergroup's & all,
% to avoid blowing up TeX...
% The macro is used below in makeNormalizedRomappno, by which means we
% define the appendix counters to be strings containing vanilla versions
% of the letters... Sigh
\let\terminator=\relax
% The string to be normalized must not contain { and } tokens...
\def\normalize#1{\ifx#1\terminator\let\next=\relax\else%
\if#1i\aftergroup i\else\if#1v\aftergroup v\else\if#1x\aftergroup x%
\else\if#1l\aftergroup l\else\if#1c\aftergroup c\else%
\if#1m\aftergroup m\else%
\if#1I\aftergroup I\else\if#1V\aftergroup V\else\if#1X\aftergroup X%
\else\if#1L\aftergroup L\else\if#1C\aftergroup C\else%
\if#1M\aftergroup M\else\aftergroup#1\fi\fi\fi\fi\fi\fi\fi\fi\fi\fi\fi\fi%
\let\next=\normalize\fi%
\next}
% makes #1 a normalized version of #2...
\def\makeNormal#1#2{\def\doNormalDef{\edef#1}\begingroup%
\aftergroup\doNormalDef\aftergroup{\normalize#2\terminator\aftergroup}%
\endgroup}
% makes a normalized version of its argument:

\def\warnIfChanged#1#2{%
\ifundef#1% skip it
\else\begingroup%
\edef\oldDefinitionOfCounter{#1}\edef\newDefinitionOfCounter{#2}%
%\message{old: \oldDefinitionOfCounter}%
%\message{new: \newDefinitionOfCounter}%
\ifx\oldDefinitionOfCounter\newDefinitionOfCounter%
\else%
\linemessage{Warning: definition of \noexpand#1 has changed.}%
\global\AnyCounterChangedtrue\fi\endgroup\fi}

\def\Section#1{\global\advance\secno by1\relax\global\meqno=1%
\global\subsecno=0%
\bigbreak\bigskip% (combination \goodbreak\bigskip\bigskip)
\centerline{\twelvepoint \bf %
\the\secno. #1}%
\par\nobreak\medskip\nobreak}
\def\tagsection#1{%
\warnIfChanged#1{\the\secno}%
\xdef#1{\the\secno}%
\ifWritingAuxFile\immediate\write\auxfile{\noexpand\xdef\noexpand#1{#1}}\fi%
}
\def\section{\Section}
\def\Subsection#1{\global\advance\subsecno by1\relax\medskip %
\leftline{\bf\the\secno.\the\subsecno\ #1}%
\par\nobreak\smallskip\nobreak}
\def\tagsubsection#1{%
\warnIfChanged#1{\the\secno.\the\subsecno}%
\xdef#1{\the\secno.\the\subsecno}%
\ifWritingAuxFile\immediate\write\auxfile{\noexpand\xdef\noexpand#1{#1}}\fi%
}

\def\subsection{\Subsection}

\def\romappno{\uppercase\expandafter{\romannumeral\appno}}
\def\makeNormalizedRomappno{%
\expandafter\makeNormal\expandafter\normalizedromappno%
\expandafter{\romannumeral\appno}%
\edef\normalizedromappno{\uppercase{\normalizedromappno}}}
\def\Appendix#1{\global\advance\appno by1\relax\global\meqno=1\global\secno=0%
\global\subsecno=0%
\bigbreak\bigskip% (combination \goodbreak\bigskip\bigskip)
\centerline{\twelvepoint \bf Appendix %
\romappno. #1}%
\par\nobreak\medskip\nobreak}
\def\tagappendix#1{\makeNormalizedRomappno%
\warnIfChanged#1{\normalizedromappno}%
\xdef#1{\normalizedromappno}%
\ifWritingAuxFile\immediate\write\auxfile{\noexpand\xdef\noexpand#1{#1}}\fi%
}
\def\appendix{\Appendix}
\def\Subappendix#1{\global\advance\subsecno by1\relax\medskip %
\leftline{\bf\romappno.\the\subsecno\ #1}%
\par\nobreak\smallskip\nobreak}
\def\tagsubappendix#1{\makeNormalizedRomappno%
\warnIfChanged#1{\normalizedromappno.\the\subsecno}%
\xdef#1{\normalizedromappno.\the\subsecno}%
\ifWritingAuxFile\immediate\write\auxfile{\noexpand\xdef\noexpand#1{#1}}\fi%
}

\def\eqn#1{\makeNormalizedRomappno%
\ifnum\secno>0%
  \warnIfChanged#1{\the\secno.\the\meqno}%
  \eqno(\the\secno.\the\meqno)\xdef#1{\the\secno.\the\meqno}%
     \global\advance\meqno by1
\else\ifnum\appno>0%
  \warnIfChanged#1{\normalizedromappno.\the\meqno}%
  \eqno({\rm\romappno}.\the\meqno)%
      \xdef#1{\normalizedromappno.\the\meqno}%
     \global\advance\meqno by1
\else%
  \warnIfChanged#1{\the\meqno}%
  \eqno(\the\meqno)\xdef#1{\the\meqno}%
     \global\advance\meqno by1
\fi\fi%
\eqlabeL#1%
\ifWritingAuxFile\immediate\write\auxfile{\noexpand\xdef\noexpand#1{#1}}\fi%
}
\def\defeqn#1{\makeNormalizedRomappno%
\ifnum\secno>0%
  \warnIfChanged#1{\the\secno.\the\meqno}%
  \xdef#1{\the\secno.\the\meqno}%
     \global\advance\meqno by1
\else\ifnum\appno>0%
  \warnIfChanged#1{\normalizedromappno.\the\meqno}%
  \xdef#1{\normalizedromappno.\the\meqno}%
     \global\advance\meqno by1
\else%
  \warnIfChanged#1{\the\meqno}%
  \xdef#1{\the\meqno}%
     \global\advance\meqno by1
\fi\fi%
\eqlabeL#1%
\ifWritingAuxFile\immediate\write\auxfile{\noexpand\xdef\noexpand#1{#1}}\fi%
}
\def\anoneqn{\makeNormalizedRomappno%
\ifnum\secno>0
  \eqno(\the\secno.\the\meqno)%
     \global\advance\meqno by1
\else\ifnum\appno>0
  \eqno({\rm\normalizedromappno}.\the\meqno)%
     \global\advance\meqno by1
\else
  \eqno(\the\meqno)%
     \global\advance\meqno by1
\fi\fi%
}
\def\mfig#1#2{\global\advance\figno by1%
\relax#1\the\figno%
\warnIfChanged#2{\the\figno}%
\edef#2{\the\figno}%
\reflabeL#2%
\ifWritingAuxFile\immediate\write\auxfile{\noexpand\xdef\noexpand#2{#2}}\fi%
}

\def\fig#1{\mfig{fig.\ ~}#1}

\catcode`@=11 % borrow the private macros of PLAIN (with care)

% \LoadFigure is used to put a figure into the text.  Its first argument
% is the symbolic name for the figure (if it isn't defined, a new number
% will be assigned);  the second argument is a caption;
% the third argument size information in the form
% \epsfxsize=3.0in\epsfysize=3.5in (this argument may be blank and
% may contain any valid preparatory argument used by the epsf package);
% the fourth and last argument is the name of the file which contains the
% figure.
% The macro is basically just a front-end for \epsfbox; its purpose is
% to allow figures to be switched from placement in the running text
% to placement on a separate page at the end of the text.  This choice
% is made using the flag \FiguresInText{true,false}; in the latter case,
% figures are placed at the end, size information is ignored (figures
% will be full-size), and the captions are listed separately on a page
% when the \listfigs command is invoked, followed by the figures, each
% on a separate page.
%  The epsf package must be loaded by the user.
\newif\ifFiguresInText\FiguresInTexttrue
\newif\if@FigureFileCreated
\newwrite\capfile
\newwrite\figfile

\def\PlaceTextFigure#1#2#3#4{%
\vskip 0.5truein%
#3\hfil\epsfbox{#4}\hfil\break%
\hfil\vbox{Figure #1. #2}\hfil%
\vskip10pt}
\def\PlaceEndFigure#1#2{%
\epsfxsize=\hsize\epsfbox{#2}\vfill\centerline{Figure #1.}\eject}

\def\LoadFigure#1#2#3#4{%
\ifundef#1{\phantom{\mfig{}#1}}%  Write out definition only if it's new.
\warnIfChanged#1{\the\figno}%
\ifWritingAuxFile\immediate\write\auxfile{\noexpand\xdef\noexpand#1{#1}}\fi\fi%
\ifFiguresInText% Figure is immediate
\PlaceTextFigure{#1}{#2}{#3}{#4}%
\else% Figure is at the end
\if@FigureFileCreated\else%
\immediate\openout\capfile=\jobname.caps%
\immediate\openout\figfile=\jobname.figs%
\@FigureFileCreatedtrue
\fi%
\immediate\write\capfile{\noexpand\item{Figure \noexpand#1.\ }{#2}
\vskip10pt
}%
\immediate\write\figfile{\noexpand\PlaceEndFigure\noexpand#1{\noexpand#4}}%
\fi}

\def\listfigs{\ifFiguresInText\else%
\vfill\eject\immediate\closeout\capfile%\parindent=20pt
\immediate\closeout\figfile%
\centerline{{\bf Figures}}\bigskip\frenchspacing%
\catcode`@=11 % borrow the private macros of PLAIN (with care)
\def\ninerm{\tenrm}
\input \jobname.caps\vfill\eject\nonfrenchspacing%
\catcode`\@=\active
\catcode`@=12  % No longer.
\input\jobname.figs\fi}

%\font\titlefont=cmr10 at 16pt
\font\ninerm=cmr9
\font\eightrm=cmr8
\font\sixrm=cmr6

\def\loadtrueseventeenpoint{
 \font\seventeenrm=cmr10 at 17.28truept
 \font\seventeeni=cmmi10 at 17.28truept
 \font\seventeenbf=cmbx10 at 17.28truept
 \font\seventeenit=cmti10 at 17.28truept
 \font\seventeensl=cmsl10 at 17.28truept
 \font\seventeensy=cmsy10 at 17.28truept
}
\def\loadfourteenpoint{
\font\fourteenrm=cmr10 at 14.4pt
\font\fourteeni=cmmi10 at 14.4pt
\font\fourteenit=cmti10 at 14.4pt
\font\fourteensl=cmsl10 at 14.4pt
\font\fourteensy=cmsy10 at 14.4pt
\font\fourteenbf=cmbx10 at 14.4pt
}
\def\loadtruetwelvepoint{
\font\twelverm=cmr10 at 12truept
\font\twelvei=cmmi10 at 12truept
\font\twelveit=cmti10 at 12truept
\font\twelvesl=cmsl10 at 12truept
\font\twelvesy=cmsy10 at 12truept
\font\twelvebf=cmbx10 at 12truept
}

\font\ninei=cmmi9
\font\eighti=cmmi8
\font\sixi=cmmi6
\skewchar\ninei='177 \skewchar\eighti='177 \skewchar\sixi='177

\font\ninesy=cmsy9
\font\eightsy=cmsy8
\font\sixsy=cmsy6
\skewchar\ninesy='60 \skewchar\eightsy='60 \skewchar\sixsy='60

\font\ninebf=cmbx9
\font\eightbf=cmbx8
\font\sixbf=cmbx6

\font\ninett=cmtt9
\font\eighttt=cmtt8

\hyphenchar\tentt=-1 % inhibit hyphenation in typewriter type
\hyphenchar\ninett=-1
\hyphenchar\eighttt=-1

\font\ninesl=cmsl9
\font\eightsl=cmsl8

\font\nineit=cmti9
\font\eightit=cmti8

 % unslanted text italic

\newskip\ttglue
\def\tenpoint{\def\rm{\fam0\tenrm}%
  \textfont0=\tenrm \scriptfont0=\sevenrm \scriptscriptfont0=\fiverm
  \textfont1=\teni \scriptfont1=\seveni \scriptscriptfont1=\fivei
  \textfont2=\tensy \scriptfont2=\sevensy \scriptscriptfont2=\fivesy
  \textfont3=\tenex \scriptfont3=\tenex \scriptscriptfont3=\tenex
  \def\it{\fam\itfam\tenit}\textfont\itfam=\tenit
  \def\sl{\fam\slfam\tensl}\textfont\slfam=\tensl
  \def\bf{\fam\bffam\tenbf}\textfont\bffam=\tenbf \scriptfont\bffam=\sevenbf
  \scriptscriptfont\bffam=\fivebf
  \normalbaselineskip=12pt
  \let\sc=\eightrm
  \let\big=\tenbig
  \setbox\strutbox=\hbox{\vrule height8.5pt depth3.5pt width\z@}%
  \normalbaselines\rm}

\def\twelvepoint{\def\rm{\fam0\twelverm}%
  \textfont0=\twelverm \scriptfont0=\ninerm \scriptscriptfont0=\sevenrm
  \textfont1=\twelvei \scriptfont1=\ninei \scriptscriptfont1=\seveni
  \textfont2=\twelvesy \scriptfont2=\ninesy \scriptscriptfont2=\sevensy
  \textfont3=\tenex \scriptfont3=\tenex \scriptscriptfont3=\tenex
  \def\it{\fam\itfam\twelveit}\textfont\itfam=\twelveit
  \def\sl{\fam\slfam\twelvesl}\textfont\slfam=\twelvesl
  \def\bf{\fam\bffam\twelvebf}\textfont\bffam=\twelvebf%
  \scriptfont\bffam=\ninebf
  \scriptscriptfont\bffam=\sevenbf
  \normalbaselineskip=12pt
  \let\sc=\eightrm
  \let\big=\tenbig
  \setbox\strutbox=\hbox{\vrule height8.5pt depth3.5pt width\z@}%
  \normalbaselines\rm}

\def\fourteenpoint{\def\rm{\fam0\fourteenrm}%
  \textfont0=\fourteenrm \scriptfont0=\tenrm \scriptscriptfont0=\sevenrm
  \textfont1=\fourteeni \scriptfont1=\teni \scriptscriptfont1=\seveni
  \textfont2=\fourteensy \scriptfont2=\tensy \scriptscriptfont2=\sevensy
  \textfont3=\tenex \scriptfont3=\tenex \scriptscriptfont3=\tenex
  \def\it{\fam\itfam\fourteenit}\textfont\itfam=\fourteenit
  \def\sl{\fam\slfam\fourteensl}\textfont\slfam=\fourteensl
  \def\bf{\fam\bffam\fourteenbf}\textfont\bffam=\fourteenbf%
  \scriptfont\bffam=\tenbf
  \scriptscriptfont\bffam=\sevenbf
  \normalbaselineskip=17pt
  \let\sc=\elevenrm
  \let\big=\tenbig
  \setbox\strutbox=\hbox{\vrule height8.5pt depth3.5pt width\z@}%
  \normalbaselines\rm}

\def\seventeenpoint{\def\rm{\fam0\seventeenrm}%
  \textfont0=\seventeenrm \scriptfont0=\fourteenrm \scriptscriptfont0=\tenrm
  \textfont1=\seventeeni \scriptfont1=\fourteeni \scriptscriptfont1=\teni
  \textfont2=\seventeensy \scriptfont2=\fourteensy \scriptscriptfont2=\tensy
  \textfont3=\tenex \scriptfont3=\tenex \scriptscriptfont3=\tenex
  \def\it{\fam\itfam\seventeenit}\textfont\itfam=\seventeenit
  \def\sl{\fam\slfam\seventeensl}\textfont\slfam=\seventeensl
  \def\bf{\fam\bffam\seventeenbf}\textfont\bffam=\seventeenbf%
  \scriptfont\bffam=\fourteenbf
  \scriptscriptfont\bffam=\twelvebf
  \normalbaselineskip=21pt
  \let\sc=\fourteenrm
  \let\big=\tenbig
  \setbox\strutbox=\hbox{\vrule height 12pt depth 6pt width\z@}%
  \normalbaselines\rm}

\def\ninepoint{\def\rm{\fam0\ninerm}%
  \textfont0=\ninerm \scriptfont0=\sixrm \scriptscriptfont0=\fiverm
  \textfont1=\ninei \scriptfont1=\sixi \scriptscriptfont1=\fivei
  \textfont2=\ninesy \scriptfont2=\sixsy \scriptscriptfont2=\fivesy
  \textfont3=\tenex \scriptfont3=\tenex \scriptscriptfont3=\tenex
  \def\it{\fam\itfam\nineit}\textfont\itfam=\nineit
  \def\sl{\fam\slfam\ninesl}\textfont\slfam=\ninesl
  \def\bf{\fam\bffam\ninebf}\textfont\bffam=\ninebf \scriptfont\bffam=\sixbf
  \scriptscriptfont\bffam=\fivebf
  \normalbaselineskip=11pt
  \let\sc=\sevenrm
  \let\big=\ninebig
  \setbox\strutbox=\hbox{\vrule height8pt depth3pt width\z@}%
  \normalbaselines\rm}

\def\eightpoint{\def\rm{\fam0\eightrm}%
  \textfont0=\eightrm \scriptfont0=\sixrm \scriptscriptfont0=\fiverm%
  \textfont1=\eighti \scriptfont1=\sixi \scriptscriptfont1=\fivei%
  \textfont2=\eightsy \scriptfont2=\sixsy \scriptscriptfont2=\fivesy%
  \textfont3=\tenex \scriptfont3=\tenex \scriptscriptfont3=\tenex%
  \def\it{\fam\itfam\eightit}\textfont\itfam=\eightit%
  \def\sl{\fam\slfam\eightsl}\textfont\slfam=\eightsl%
  \def\bf{\fam\bffam\eightbf}\textfont\bffam=\eightbf \scriptfont\bffam=\sixbf%
  \scriptscriptfont\bffam=\fivebf%
  \normalbaselineskip=9pt%
  \let\sc=\sixrm%
  \let\big=\eightbig%
  \setbox\strutbox=\hbox{\vrule height7pt depth2pt width\z@}%
  \normalbaselines\rm}

 % use after $ in ninepoint sections
\def\tenbig#1{{\hbox{$\left#1\vbox to8.5pt{}\right.\n@space$}}}
\def\ninebig#1{{\hbox{$\textfont0=\tenrm\textfont2=\tensy
  \left#1\vbox to7.25pt{}\right.\n@space$}}}
\def\eightbig#1{{\hbox{$\textfont0=\ninerm\textfont2=\ninesy
  \left#1\vbox to6.5pt{}\right.\n@space$}}}

% Page layout
%\newinsert\footins
\def\footnote#1{\edef\@sf{\spacefactor\the\spacefactor}#1\@sf
      \insert\footins\bgroup\eightpoint
      \interlinepenalty100 \let\par=\endgraf
        \leftskip=\z@skip \rightskip=\z@skip
        \splittopskip=10pt plus 1pt minus 1pt \floatingpenalty=20000
        \smallskip\item{#1}\bgroup\strut\aftergroup\@foot\let\next}
\skip\footins=12pt plus 2pt minus 4pt % space added when footnote is present
%\count\footins=1000 % footnote magnification factor (1 to 1)
\dimen\footins=30pc % maximum footnotes per page

\newinsert\margin
\dimen\margin=\maxdimen
%\count\margin=0 \skip\margin=0pt % marginal inserts take up no space
\def\titlefont{\seventeenpoint}
\loadtruetwelvepoint % At FNAL...
\loadtrueseventeenpoint

% \use\cs
% puts in the expansion of `\cs' if it's defined, the literal "\cs" otherwise.
\def\eatOne#1{}
\def\ifundef#1{\expandafter\ifx%
\csname\expandafter\eatOne\string#1\endcsname\relax}
\def\notTrue{\iffalse}\def\isTrue{\iftrue}
\def\ifdef#1{{\ifundef#1%
\aftergroup\notTrue\else\aftergroup\isTrue\fi}}
\def\use#1{\ifundef#1\linemessage{Warning: \string#1 is undefined.}%
{\tt \string#1}\else#1\fi}

%     \ref\label{text}
% generates a number, assigns it to \label, generates an entry.
% To list the refs on a separate page,  \listrefs
% \nref does the same without generating any text at the reference
% point
% June 26 1994: \preref postpones the generation of an entry, along with
% the text, until the first use of the reference

\global\newcount\refno \global\refno=1
\newwrite\rfile
\newlinechar=`\^^J
\def\@ref#1#2{\the\refno\n@ref#1{#2}}
\def\n@ref#1#2{\xdef#1{\the\refno}%
\ifnum\refno=1\immediate\openout\rfile=\jobname.refs\fi%
\immediate\write\rfile{\noexpand\item{[\noexpand#1]\ }#2.}%
\global\advance\refno by1}
\def\nref{\n@ref} % Hide to allow redefinitions of \ref,\nref to \preref
\def\ref{\@ref}   % without breaking the latter...
% To start a long reference...
\def\lref#1#2{\the\refno\xdef#1{\the\refno}%
\ifnum\refno=1\immediate\openout\rfile=\jobname.refs\fi%
\immediate\write\rfile{\noexpand\item{[\noexpand#1]\ }#2\semi}%
\global\advance\refno by1}
% To continue a long reference...
\def\cref#1{\immediate\write\rfile{#1\semi}}
% To end a long reference...

\def\preref#1#2{\gdef#1{\@ref#1{#2}}}

\def\semi{;\hfil\noexpand\break}

\def\listrefs{\vfill\eject\immediate\closeout\rfile%\parindent=20pt
\centerline{{\bf References}}\bigskip\frenchspacing%
\input \jobname.refs\vfill\eject\nonfrenchspacing}

\def\inputAuxIfPresent#1{\immediate\openin1=#1
\ifeof1\message{No file \auxfileName; I'll create one.
}\else\closein1\relax\input\auxfileName\fi%
}
% For references, some journal names
\def\NPB{Nucl.\ Phys.\ B}
\def\PRL{Phys.\ Rev.\ Lett.\ }

% An .aux file --- for forward references...
\newif\ifWritingAuxFile
\newwrite\auxfile
\def\SetUpAuxFile{%
\xdef\auxfileName{\jobname.aux}%
% Read it in if it exists
\inputAuxIfPresent{\auxfileName}%
% Now write a new one.
\WritingAuxFiletrue%
\immediate\openout\auxfile=\auxfileName}

% Some generally useful notation
\def\L{\left(}\def\R{\right)}
\def\LP{\left.}\def\RP{\right.}
\def\LB{\left[}\def\RB{\right]}

\def\RV{\right|}

% Warn about changed counters...

\catcode`\@=\active
\catcode`@=12  % No longer.
\catcode`\"=\active

%%%%%%%%%%%%%%%%%%%%%%%%%%%%%%%%%%%%%%%%%%%%%%%%%%%%%%%%%%%%%%%%%%%%%%%%%%%%%%

%\input gaugedefs

\def\L{\left(}
\def\R{\right)}

\def\Tr{\mathop{\rm Tr}\nolimits}
\def\Gr{\mathop{\rm Gr}\nolimits}

\def\si{\sigma}

\def\eps{\epsilon}

\def\LP{\left.}\def\RP{\right.}

\def\dl^#1_#2{\delta^{#1}{}_{#2}}

\def\Li{\mathop{\rm Li}\nolimits}

\def\Ord{{\cal O}}

\def\A#1{{\cal A}_{#1}}

\catcode`@=11  % Make @ letter.
\def\meqalign#1{\,\vcenter{\openup1\jot\m@th
   \ialign{\strut\hfil$\displaystyle{##}$ && $\displaystyle{{}##}$\hfil
             \crcr#1\crcr}}\,}
\catcode`@=12  % No longer.

%%%%%%%%%%%%%%%%%%%%%%%%%%%%%%%%%%%%%%%%%%%%%%%%%%%%%%%%%%%%%%%%%%%%%%%%%%%%%%

%\input spinordef

\def\si{\sigma}

\def\Tr{\mathop{\rm Tr}\nolimits}

\def\A#1{{\cal A}_{#1}}

\def\L{\left(}\def\R{\right)}
\def\LP{\left.}\def\RP{\right.}
\def\spa#1.#2{\left\langle#1\,#2\right\rangle}
\def\spb#1.#2{\left[#1\,#2\right]}
\def\lor#1.#2{\left(#1\,#2\right)}
\def\sand#1.#2.#3{%
\left\langle\smash{#1}{\vphantom1}^{-}\right|{#2}%
\left|\smash{#3}{\vphantom1}^{-}\right\rangle}
\def\sandp#1.#2.#3{%
\left\langle\smash{#1}{\vphantom1}^{-}\right|{#2}%
\left|\smash{#3}{\vphantom1}^{+}\right\rangle}
\def\sandpp#1.#2.#3{%
\left\langle\smash{#1}{\vphantom1}^{+}\right|{#2}%
\left|\smash{#3}{\vphantom1}^{+}\right\rangle}
\catcode`@=11  % Make @ letter.
\def\meqalign#1{\,\vcenter{\openup1\jot\m@th
   \ialign{\strut\hfil$\displaystyle{##}$ && $\displaystyle{{}##}$\hfil
             \crcr#1\crcr}}\,}
\catcode`@=12  % No longer.

\newread\epsffilein    % file to \read
\newif\ifepsffileok    % continue looking for the bounding box?
\newif\ifepsfbbfound   % success?
\newif\ifepsfverbose   % report what you're making?
\newdimen\epsfxsize    % horizontal size after scaling
\newdimen\epsfysize    % vertical size after scaling
\newdimen\epsftsize    % horizontal size before scaling
\newdimen\epsfrsize    % vertical size before scaling
\newdimen\epsftmp      % register for arithmetic manipulation
\newdimen\pspoints     % conversion factor
\pspoints=1bp          % Adobe points are `big'
\epsfxsize=0pt         % Default value, means `use natural size'
\epsfysize=0pt         % ditto
\def\epsfbox#1{\global\def\epsfllx{72}\global\def\epsflly{72}%
   \global\def\epsfurx{540}\global\def\epsfury{720}%
   \def\lbracket{[}\def\testit{#1}\ifx\testit\lbracket
   \let\next=\epsfgetlitbb\else\let\next=\epsfnormal\fi\next{#1}}%
\def\epsfgetlitbb#1#2 #3 #4 #5]#6{\epsfgrab #2 #3 #4 #5 .\\%
   \epsfsetgraph{#6}}%
\def\epsfnormal#1{\epsfgetbb{#1}\epsfsetgraph{#1}}%
\def\epsfgetbb#1{%
%
%   The first thing we need to do is to open the
%   PostScript file, if possible.
%
\openin\epsffilein=#1
\ifeof\epsffilein\errmessage{I couldn't open #1, will ignore it}\else
%
%   Okay, we got it. Now we'll scan lines until we find one that doesn't
%   start with %. We're looking for the bounding box comment.
%
   {\epsffileoktrue \chardef\other=12
    \def\do##1{\catcode`##1=\other}\dospecials \catcode`\ =10
    \loop
       \read\epsffilein to \epsffileline
       \ifeof\epsffilein\epsffileokfalse\else
%
%   We check to see if the first character is a % sign;
%   if not, we stop reading (unless the line was entirely blank);
%   if so, we look further and stop only if the line begins with
%   `%%BoundingBox:'.
%
          \expandafter\epsfaux\epsffileline:. \\%
       \fi
   \ifepsffileok\repeat
   \ifepsfbbfound\else
    \ifepsfverbose\message{No bounding box comment in #1; using defaults}\fi\fi
   }\closein\epsffilein\fi}%
%
%   Now we have to calculate the scale and offset values to use.
%   First we compute the natural sizes.
%
\def\epsfclipstring{}% do we clip or not?  If so,
\def\epsfsetgraph#1{%
   \epsfrsize=\epsfury\pspoints
   \advance\epsfrsize by-\epsflly\pspoints
   \epsftsize=\epsfurx\pspoints
   \advance\epsftsize by-\epsfllx\pspoints
%
%   If `epsfxsize' is 0, we default to the natural size of the picture.
%   Otherwise we scale the graph to be \epsfxsize wide.
%
   \epsfxsize\epsfsize\epsftsize\epsfrsize
   \ifnum\epsfxsize=0 \ifnum\epsfysize=0
      \epsfxsize=\epsftsize \epsfysize=\epsfrsize
      \epsfrsize=0pt
%
%   We have a sticky problem here:  TeX doesn't do floating point arithmetic!
%   Our goal is to compute y = rx/t. The following loop does this reasonably
%   fast, with an error of at most about 16 sp (about 1/4000 pt).
%
     \else\epsftmp=\epsftsize \divide\epsftmp\epsfrsize
       \epsfxsize=\epsfysize \multiply\epsfxsize\epsftmp
       \multiply\epsftmp\epsfrsize \advance\epsftsize-\epsftmp
       \epsftmp=\epsfysize
       \loop \advance\epsftsize\epsftsize \divide\epsftmp 2
       \ifnum\epsftmp>0
          \ifnum\epsftsize<\epsfrsize\else
             \advance\epsftsize-\epsfrsize \advance\epsfxsize\epsftmp \fi
       \repeat
       \epsfrsize=0pt
     \fi
   \else \ifnum\epsfysize=0
     \epsftmp=\epsfrsize \divide\epsftmp\epsftsize
     \epsfysize=\epsfxsize \multiply\epsfysize\epsftmp
     \multiply\epsftmp\epsftsize \advance\epsfrsize-\epsftmp
     \epsftmp=\epsfxsize
     \loop \advance\epsfrsize\epsfrsize \divide\epsftmp 2
     \ifnum\epsftmp>0
        \ifnum\epsfrsize<\epsftsize\else
           \advance\epsfrsize-\epsftsize \advance\epsfysize\epsftmp \fi
     \repeat
     \epsfrsize=0pt
    \else
     \epsfrsize=\epsfysize
    \fi
   \fi
%
%  Finally, we make the vbox and stick in a \special that dvips can parse.
%
   \ifepsfverbose\message{#1: width=\the\epsfxsize, height=\the\epsfysize}\fi
   \epsftmp=10\epsfxsize \divide\epsftmp\pspoints
   \vbox to\epsfysize{\vfil\hbox to\epsfxsize{%
      \ifnum\epsfrsize=0\relax
        \includegraphics{#1}%
      \else
        \epsfrsize=10\epsfysize \divide\epsfrsize\pspoints
        \includegraphics{#1}%
      \fi
      \hfil}}%
\global\epsfxsize=0pt\global\epsfysize=0pt}%
%
%   We still need to define the tricky \epsfaux macro. This requires
%   a couple of magic constants for comparison purposes.
%
{\catcode`\%=12 \global\let\epsfpercent=%\global\def\epsfbblit{%BoundingBox}}%
%
%   So we're ready to check for `%BoundingBox:' and to grab the
%   values if they are found.
%
\long\def\epsfaux#1#2:#3\\{\ifx#1\epsfpercent
   \def\testit{#2}\ifx\testit\epsfbblit
      \epsfgrab #3 . . . \\%
      \epsffileokfalse
      \global\epsfbbfoundtrue
   \fi\else\ifx#1\par\else\epsffileokfalse\fi\fi}%
%
%   Here we grab the values and stuff them in the appropriate definitions.
%
\def\epsfempty{}%
\def\epsfgrab #1 #2 #3 #4 #5\\{%
\global\def\epsfllx{#1}\ifx\epsfllx\epsfempty
      \epsfgrab #2 #3 #4 #5 .\\\else
   \global\def\epsflly{#2}%
   \global\def\epsfurx{#3}\global\def\epsfury{#4}\fi}%
%
%   We default the epsfsize macro.
%
\def\epsfsize#1#2{\epsfxsize}
%
%   Finally, another definition for compatibility with older macros.
%

%%%%%%%%%%%%%%%%%%%%%%%%%%%%%%%%%%%%%%%%%%%%%%%%%%%%%%%%%%%%%%%%%%%%%%%%%%%%%%
\SetUpAuxFile
\loadfourteenpoint
\FiguresInTexttrue
\nopagenumbers\hsize=\hstitle\vskip1in
\overfullrule 0pt
\hfuzz 35 pt
% Suppress vbox warnings
\vbadness=10001
%
%  Some definitions for this paper:
%

\def\hf{{\textstyle{1\over2}}}
\def\Tr{\mathop{\rm Tr}\nolimits}

\def\Li{\mathop{\rm Li}\nolimits}

\def\e{\epsilon}

\def\coeff#1#2{{\textstyle{#1\over#2}}}
\def\hf{{\textstyle{1\over2}}}

% Avoid problems with refs.tmp
\def\"#1{{\accent127 #1}}

\def\lr{\leftrightarrow}
%Let's be sneaky
\def\spa#1.#2{\left\langle\mskip-1mu#1\,#2\mskip-1mu\right\rangle}
\def\spb#1.#2{\left[\mskip-1mu#1\,#2\mskip-1mu\right]}

\def\cg{c_\Gamma}

\def\Ls{\mathop{\rm Ls}\nolimits}
\def\Ll{\mathop{\rm L}\nolimits}

\def\e{\epsilon}
\def\gluon{{\rm gluon}}
\def\gluino{{\tilde g}}
\def\qb{{\bar q}}
\def\susy{{\rm SUSY}}

\def\tree{{\rm tree}}
\def\si{\sigma}
\def\ns{n_{\mskip-2mu s}}\def\nf{n_{\mskip-2mu f}}
\def\ib{{\bar\imath}}
\def\Split{\mathop{\rm Split}\nolimits}
%
%\draft

%%%%%%%%%%%%%%%%%%%%%%%%%%%%%%%%%%%%%%%%%%%%%%%%%%%

\noindent
hep-ph/9409393 \hfill {SLAC--PUB--6663}\break
\rightline{Saclay/SPhT-T94/108}
\rightline{UCLA/94/TEP/33}
\rightline{September, 1994}
\rightline{(T)}
\rightline{   }

\leftlabelstrue
\vskip -1.0 in
\Title{One-Loop Corrections to Two-Quark Three-Gluon Amplitudes}

\centerline{Zvi Bern${}^{\sharp}$}
\baselineskip12truept
\centerline{\it Department of Physics}
\centerline{\it University of California, Los Angeles}
\centerline{\it Los Angeles, CA 90024}
\centerline{\tt bern@physics.ucla.edu}

\smallskip\smallskip

\baselineskip17truept
\centerline{Lance Dixon${}^{\star}$}
\baselineskip12truept
\centerline{\it Stanford Linear Accelerator Center}
\centerline{\it Stanford, CA 94309}
\centerline{\tt lance@slac.stanford.edu}

\smallskip \centerline{and} \smallskip

\baselineskip17truept
\centerline{David A. Kosower}
\baselineskip12truept
\centerline{\it Service de Physique Th\'eorique${}^{\dagger}$}
\centerline{\it Centre d'Etudes de Saclay}
\centerline{\it F-91191 Gif-sur-Yvette cedex, France}
\centerline{\tt kosower@amoco.saclay.cea.fr}

\vskip 0.2in\baselineskip13truept

\vskip 0.5truein
\centerline{\bf Abstract}

{\narrower

We present the one-loop QCD amplitudes for two external massless
quarks and three external gluons ($\bar{q}qggg$).
This completes the set of
one-loop amplitudes needed for the next-to-leading-order corrections
to three-jet production at hadron colliders.  We also discuss how to
use group theory and supersymmetry to minimize the amount of
calculation required for the more general case of one-loop two-quark
$n$-gluon amplitudes.  We use collinear limits to provide a stringent
check on the amplitudes.  }

\vskip 0.3truein

\centerline{\sl Submitted to Nuclear Physics B}

\vfill
\vskip 0.1in
\noindent\hrule width 3.6in\hfil\break
\noindent
${}^{\sharp}$Research supported in part by the US Department of Energy
under grant DE-FG03-91ER40662 and in part by the
Alfred P. Sloan Foundation under grant BR-3222. \hfil\break
${}^{\star}$Research supported by the US Department of
Energy under grant DE-AC03-76SF00515.\hfil\break
${}^{\dagger}$Laboratory of the
{\it Direction des Sciences de la Mati\`ere\/}
of the {\it Commissariat \`a l'Energie Atomique\/} of France.\hfil\break

\Date{}

\line{}

\baselineskip17pt
%

%%%%%%%%%%%%%%%%%%%%%%%%%%%%%%%%%%%%%%%%%%%%%%%%%%%%%%%%%%%%%%%%
%  Refs.

\preref\NLOTwoJets{%
S.D.\ Ellis, Z. Kunszt and D.E.\ Soper,
Phys.\ Rev.\ D40:2188 (1989);
Phys.\ Rev.\ Lett.\ 64:2121 (1990);
Phys.\ Rev.\ Lett.\ 69:1496 (1992)\semi
F. Aversa, M. Greco, P. Chiappetta and J.P.\
Guillet, Phys.\ Rev.\ Lett.\ 65:401 (1990)\semi
F. Aversa, L. Gonzales, M. Greco, P. Chiappetta
and J.P.\ Guillet, Z. Phys.\ C49:459 (1991)}

\preref\Ellis{R.K. Ellis and J.C. Sexton, Nucl.\ Phys.\ B269:445 (1986)}

\preref\FiveGluon{%
Z. Bern, L. Dixon and D.A. Kosower, Phys.\ Rev. Lett.\
70:2677 (1993)}

\preref\Kunsztqqqqg{%
Z. Kunszt, A. Signer and Z. Tr\'ocs\'anyi, preprint ETH-TH/94-14,
hep-ph/9405386}

\preref\KunsztPrivate{%
Z. Kunszt, private communication}

\preref\TreeColor{%
F.A. Berends and W.T. Giele,
Nucl.\ Phys.\ B294:700 (1987)\semi
M.\ Mangano, S. Parke and Z.\ Xu, Nucl.\ Phys.\ B298:653 (1988)}

\preref\ManganoReview{%
M. Mangano and S.J. Parke, Phys.\ Rep.\ 200:301 (1991)}

\preref\GG{W.T.\ Giele and E.W.N.\ Glover,
Phys.\ Rev.\ D46:1980 (1992)\semi
W.T.\ Giele, E.W.N.\ Glover and D. A. Kosower,
Nucl.\ Phys.\ B403:633 (1993)}

\preref\KunsztSoper{Z. Kunszt and D. Soper, Phys.\ Rev.\ D46:192 (1992)}

\preref\SpinorHelicity{%
F.A.\ Berends, R.\ Kleiss, P.\ De Causmaecker, R.\ Gastmans and T.\ T.\ Wu,
        Phys.\ Lett.\ 103B:124 (1981)\semi
P.\ De Causmaeker, R.\ Gastmans,  W.\ Troost and  T.T.\ Wu,
Nucl. Phys. B206:53 (1982)\semi
R.\ Kleiss and W.J.\ Stirling,
   Nucl.\ Phys.\ B262:235 (1985)\semi
   J.F.\ Gunion and Z.\ Kunszt, Phys.\ Lett.\ 161B:333 (1985)\semi
 R.\ Gastmans and T.T.\ Wu,
{\it The Ubiquitous Photon: Helicity Method for QED and QCD}
(Clarendon Press, 1990)\semi
Z. Xu, D.-H.\ Zhang and L. Chang, Nucl.\ Phys.\ B291:392 (1987)}

\preref\TreeColorB{%
M.\ Mangano, Nucl.\ Phys.\ B309:461 (1988)}

\preref\Color{%
Z. Bern and D.A.\ Kosower, Nucl.\ Phys.\ B362:389 (1991)}

\preref\Long{Z. Bern and D.A.\ Kosower \NPB 379:451 (1992)}

\preref\StringBased{
Z. Bern and D.A.\ Kosower, \PRL 66:1669 (1991)\semi
Z. Bern and D.A.\ Kosower, in {\it Proceedings of the PASCOS-91
Symposium}, eds.\ P. Nath and S. Reucroft (World Scientific, 1992)\semi
Z. Bern, Phys.\ Lett.\  296B:85 (1992)}

\preref\Mapping{Z. Bern and D.C.\ Dunbar,  Nucl.\ Phys.\ B379:562 (1992)}

\preref\Subsequent{%
K. Roland, Phys.\ Lett.\ 289B:148 (1992)\semi
M.J.\ Strassler,  Nucl.\ Phys.\ B385:145 (1992)\semi
C.S. Lam, Nucl.\ Phys.\ B397:143 (1993); Phys.\ Rev.\ D48:873 (1993)\semi
Z. Bern, D.C. Dunbar and T. Shimada, Phys.\ Lett.\ 312B:277 (1993)\semi
G. Cristofano, R. Marotta and K. Roland, Nucl.\ Phys.\ B392:345 (1993)\semi
M.G.\ Schmidt and C. Schubert, Phys.\ Lett.\ 318B:438 (1993);
Phys.\ Lett.\ B331:69 (1994)\semi
D. Fliegner, M.G.\ Schmidt and C. Schubert, preprint HD-THEP-93-44,
hep-ph/9401221}

\preref\Cutting{L.D.\ Landau, Nucl.\ Phys.\ 13:181 (1959)\semi
 S. Mandelstam, Phys.\ Rev.\ 112:1344 (1958), 115:1741 (1959)\semi
 R.E.\ Cutkosky, J.\ Math.\ Phys.\ 1:429 (1960)}

\preref\Susy{%
M.T.\ Grisaru, H.N.\ Pendleton and P.\ van Nieuwenhuizen,
Phys. Rev. {D15}:996 (1977)\semi
M.T. Grisaru and H.N. Pendleton, Nucl.\ Phys.\ B124:81 (1977)\semi
S.J. Parke and T. Taylor, Phys.\ Lett.\ 157B:81 (1985)\semi
Z. Kunszt, Nucl.\ Phys.\ B271:333 (1986)}

\preref\KunsztFourPoint{%
Z. Kunszt, A. Signer and Z. Tr\'ocs\'anyi, Nucl.\ Phys.\ B411:397 (1994)}

\preref\Tasi{%
Z. Bern, hep-ph/9304249, in {\it Proceedings of Theoretical
Advanced Study Institute in High Energy Physics (TASI 92)},
eds.\ J. Harvey and J. Polchinski (World Scientific, 1993)}

\preref\WeakInt{Z.\ Bern and A.G.\ Morgan, Phys.\ Rev.\ D49:6155 (1994)}

\preref\SusyFour{Z. Bern, L. Dixon, D.C. Dunbar and D.A. Kosower,
Nucl.\ Phys.\ B425:217 (1994)}

\preref\SusyOne{Z. Bern, L. Dixon, D.C. Dunbar and D.A. Kosower,
preprint SLAC--PUB--6563, hep-ph/9409265}

\preref\BDKconf{Z. Bern, L. Dixon and D.A. Kosower, hep-th/9311026,
in {\it Proceedings of Strings 1993}, eds. M.B. Halpern, A. Sevrin
and G. Rivlis (World Scientific, 1994)}

\preref\AllPlus{Z. Bern, G. Chalmers, L. Dixon and D.A. Kosower,
Phys.\ Rev.\ Lett.\ 72:2134 (1994)}

\preref\Mahlon{G.D.\ Mahlon, Phys.\ Rev.\ D49:2197 (1994);
               Phys.\ Rev.\ D49:4438 (1994)}

\preref\Siegel{W. Siegel, Phys.\ Lett.\ 84B:193 (1979)\semi
D.M.\ Capper, D.R.T.\ Jones and P. van Nieuwenhuizen, Nucl.\ Phys.\
B167:479 (1980)\semi
L.V.\ Avdeev and A.A.\ Vladimirov, Nucl.\ Phys.\ B219:262 (1983)}

\preref\CollinsBook{J.C.\ Collins, {\it Renormalization}
(Cambridge University Press, 1984)}

\preref\DoubleLine{%
 G. 't Hooft, Nucl. Phys. B72:461 (1974); Nucl. Phys. B75:461 (1974)\semi
 see also P. Cvitanovi\'c, {\it Group Theory} (Nordita, 1984)}

\preref\EpsHel{D.A. Kosower, Phys.\ Lett.\ B254:439 (1991)}

\preref\Background{G. 't Hooft,
in Acta Universitatis Wratislavensis no.\
38, 12th Winter School of Theoretical Physics in Karpacz, {\it
Functional and Probabilistic Methods in Quantum Field Theory},
Vol. 1 (1975)\semi
B.S.\ DeWitt, in {\it Quantum gravity II}, eds. C. Isham, R.\ Penrose and
D.\ Sciama (Oxford, 1981)\semi
L.F.\ Abbott, Nucl.\ Phys.\ B185:189 (1981)\semi
L.F. Abbott, M.T. Grisaru and R.K. Schaefer,
Nucl.\ Phys.\ B229:372 (1983)}

\preref\GN{J.L.\ Gervais and A. Neveu, Nucl.\ Phys.\ B46:381 (1972)}

\preref\IntegralsShort{Z. Bern, L. Dixon and D.A. Kosower,
Phys.\ Lett.\ 302B:299 (1993); erratum {\it ibid.} 318:649 (1993)}

\preref\IntegralsLong{Z. Bern, L. Dixon and D.A. Kosower,
\NPB 412:751 (1994)}

\preref\Lewin{L.\ Lewin, {\it Dilogarithms and Associated Functions\/}
(Macdonald, 1958)}

\preref\TreeCollinear{F.A. Berends and W.T. Giele, Nucl.\ Phys.\
B313:595 (1989)}

\preref\GordonConf{G. Chalmers, preprint UCLA-94-TEP-24, hep-ph/9405393,
to appear in proceedings of the XXII ITEP International Winter School
of Physics (Gordon and Breach, 1995)}

\preref\Morgan{E.W.N.\ Glover and A. Morgan, Z. Phys.\ C60:175 (1993)}

\preref\KunsztSingular{%
Z. Kunszt, A. Signer and Z. Tr\'ocs\'anyi, Nucl.\ Phys.\
B420:550 (1994)}

%%%%%%%%%%%%%%%%%%%%%%%%%%%%%%%%%%%%%%%%%%%%%%%%%%%%%%%%%%%%%%%%%

$\null$
\vskip -3. cm

\section{Introduction}
\tagsection\IntroSection

Jet physics at hadron colliders allows one to confront the theoretical
predictions of QCD with experimental results and thereby probe for new
physics at the highest possible energies.  Yet precise comparisons
between theory and experiment are hampered by the lack of calculations
beyond the leading order of perturbation theory, for all but the
simplest processes.  In pure QCD, the next-to-leading-order
corrections computed to date~[\use\NLOTwoJets] have relied on the
one-loop amplitudes for four external partons, first calculated by
Ellis and Sexton~[\use\Ellis].  More recently, we have calculated the
one-loop amplitudes for five external gluons
($ggggg$)~[\use\FiveGluon], and Kunszt, Signer, and
Tr\'ocs\'anyi~(KST) have calculated the amplitudes for four quarks and
a gluon ($\bar{q}q\bar{q}qg$)~[\use\Kunsztqqqqg].  In this paper we
present the remaining one-loop five-parton amplitudes, for
two (massless) quarks and three gluons ($\bar{q}qggg$).
Combining these analytic results
with the known six-parton tree
amplitudes~[\use\TreeColor,\use\ManganoReview], one can now construct
numerical programs for next-to-leading-order corrections to three-jet
production at hadron colliders, and examine the structure of jets, for
example dependence of cross-sections on the cone size, beyond the
leading non-trivial order probed in next-to-leading order two-jet
programs~[\use\NLOTwoJets,\use\GG].
Computation of the ratio of three-jet to two-jet events at
hadron colliders at next-to-leading order in $\alpha_s$ would also make
possible the measurement of $\alpha_s$ in purely hadronic
processes and at the largest energy scales available.

Many methods developed in recent years can be used to simplify
the computation of one-loop multi-parton amplitudes,
including spinor helicity methods~[\use\SpinorHelicity],
color decomposition of
amplitudes~[\use\TreeColor,\use\TreeColorB,\use\Color],
string-based techniques
[\use\Long,\use\StringBased,\use\Mapping,\use\Subsequent,\use\FiveGluon],
supersymmetry Ward identities~[\use\Susy,\use\KunsztFourPoint],
supersymmetry-based decompositions
[\use\FiveGluon,\use\Tasi,\use\WeakInt], and perturbative
unitarity~[\use\Cutting,\use\SusyFour,\use\SusyOne];
all of these techniques have
been used to obtain the amplitudes presented in this paper.

We have found it useful to organize the calculation in terms of
gauge-invariant, color-ordered building blocks, dubbed {\it primitive
amplitudes}.
We show in the
next section that all of the kinematic coefficients (partial
amplitudes) appearing in the color decomposition of amplitudes with
two quarks and $(n-2)$ gluons can be expressed as sums over
permutations of gauge invariant primitive amplitudes.  The analytic
structure of a primitive amplitude is generally simpler than that of a
partial amplitude; a primitive amplitude receives
contributions only from diagrams with a fixed ordering of external legs,
while the generic partial amplitude receives contributions
from multiple orderings. Thus, fewer kinematic invariants appear in each
primitive amplitude.
Although this organization was motivated in part by string
theory, our discussion is entirely field-theoretic.

We use supersymmetry to reduce the number of quantities
to be calculated. QCD amplitudes may be decomposed in terms of
supersymmetric and non-supersymmetric parts.  Through use of
supersymmetry Ward identities, the supersymmetric parts of amplitudes
with two external quarks and three external gluons may be
obtained directly from the previously calculated five-gluon amplitudes
[\use\FiveGluon].

We have also made use of the cut-reconstruction method described in
refs.~[\use\SusyFour,\use\SusyOne].  If certain power-counting criteria
are satisfied, amplitudes are entirely constructible from their cuts.
Although QCD amplitudes are generally not cut-constructible, by taking
linear combinations of QCD amplitudes with ones involving scalars
and/or gluinos, the QCD amplitudes may be separated into
cut-constructible and non-cut-constructible parts.
We have used this unitarity-based technique to
obtain the cut-constructible components of some of the primitive
amplitudes for $\bar{q}qggg$
(those that enter into the subleading-in-color contributions to
the virtual part of the cross-section).
Here the cut-constructible components are formed by adding
to the desired diagrams a new set of diagrams, which differs only in
the replacement of virtual gluons in the loop by scalars.
For a specific choice of the Yukawa coupling between the scalars and
the quark line, the sum of gluon and scalar diagrams satisfies the
power-counting criteria (see ref.~[\use\SusyOne]).
We then calculate the scalar contributions directly;
they are not cut-constructible, but they are easier to calculate
directly than the full gluon contributions.
Finally we reassemble the desired gluon contributions.

In order to ensure the correctness of the amplitudes, we have
performed a number of checks.  As the momenta of two external legs
become collinear the amplitudes must factorize properly.  We have
verified this factorization for all amplitudes in all channels.  This
provides an extremely stringent constraint on the amplitudes.  In
fact, this constraint is sufficiently powerful that it has been used to
construct ans\"atze for a number of amplitudes with fixed helicities
but an arbitrary number of external legs
[\use\BDKconf,\use\AllPlus,\use\SusyFour], which were then proven
correct by either recursive [\use\Mahlon] or unitarity
[\use\SusyFour,\use\SusyOne] techniques.  (The recursive and unitarity
techniques have also been used to construct a variety of other
one-loop amplitudes with an arbitrary number of external
legs~[\use\Mahlon,\use\SusyFour,\use\SusyOne].)

We have performed additional checks on certain helicity amplitudes
by computing all diagrams that enter into a
supersymmetry Ward identity~[\use\Susy], and explicitly verifying the
identity.
Not only does this provide a check on amplitudes presented in this
paper, but also on the supersymmetric combinations of the five-gluon
amplitudes presented in ref.~[\use\FiveGluon].
(A similar supersymmetry check using the five-gluon amplitudes has
been carried out~[\use\KunsztPrivate] for the $\bar{q}q\bar{q}qg$
amplitudes reported in ref.~[\use\Kunsztqqqqg].)
As a final check, we have verified that the cuts in some
amplitudes obtained by more direct diagrammatic means are
consistent with unitarity.

In section~\use\ColorSection, we give the $SU(N_c)$ color
decomposition for amplitudes involving two external quarks and $n-2$
external gluons, as a sum of color factors multiplied by
partial amplitudes.
We also give a formula for the sum over colors of the interference
between tree and one-loop $\bar{q}qggg$ amplitudes, in terms of
partial amplitudes; this formula is required for
the virtual part of the color-summed parton-level cross-section.
The primitive amplitudes, which form the
gauge-invariant building blocks for the amplitudes, are
described in section~\use\PrimitiveSection.  The precise relation of
the primitive amplitudes to the partial amplitudes is given in
section~\use\PartialAmplSection.
In section~\use\qqgggSection\ we give the main results of the paper,
the primitive amplitudes for $\bar{q}qggg$.
Section~\use\ConclusionSection\ contains our conclusions.
Four appendices contain technical details related
to color algebra and collinear checks.
Appendix~\use\SubleadAppendix\ provides a
derivation of the relation between primitive and partial amplitudes.
Appendix~\use\FourPointAppendix\ collects the one-loop four-point
amplitudes~[\use\KunsztFourPoint,\use\Long] that appear in collinear
limits of $\bar{q}qggg$ amplitudes, namely $gggg$ and $\bar{q}qgg$.
Appendix~\use\CollinearAppendix\ then illustrates the procedure for
carrying out collinear checks, using these amplitudes and ``splitting
amplitudes'' from ref.~[\use\SusyFour].
Finally, appendix~\use\MixedAppendix\ shows how to use the
two-quark $(n-2)$-gluon primitive amplitudes to construct
amplitudes where some of the gluons are replaced by photons.

%%%%%%%%%%%%%%%%%%%%%%%%%%%%%%%%%%%

\section{Color Decomposition for Two-Quark $(n-2)$-Gluon Amplitudes}
\tagsection\ColorSection

In this section we describe a color decomposition of the one-loop
two-quark $(n-2)$-gluon amplitude $\bar{q}qg\ldots g$, in terms of
group-theoretic factors (color structures) multiplied by kinematic
functions called {\it partial amplitudes}.  In the following sections,
we shall give formulae for all of the partial amplitudes in terms of
color-ordered, gauge-invariant building blocks called {\it primitive
amplitudes}.  A primitive amplitude is defined as the sum of all
one-loop diagrams in which the $n$ external legs have a fixed order
around the loop (the color order), with some additional restrictions
to be described in the following section.

For the $\bar{q}qg\ldots g$ amplitudes, let particle 1 be an
antiquark, transforming in the $\overline{N}_c$ representation of
$SU(N_c)$, with color index $\bar i_1$, and let particle 2 be a quark,
transforming in the $N_c$ with index $i_2$.  Denote these particles by
$1_\qb$ and $2_q$ in order to distinguish them from the remaining
gluons, particles 3 to $n$, transforming in the adjoint representation
with indices $a_3,\ldots, a_n$.  We also allow for $\nf$ Weyl fermions
and $\ns$ complex scalars circulating in the loop, both in the
$(N_c+\overline{N}_c)$ representation ($\nf$ flavors of massless
quarks and $\ns$ massless scalars).

The general strategy for obtaining multi-parton color decompositions
is to rewrite the $SU(N_c)$ structure constants $f^{abc}$ in terms of
the group generators in the fundamental representation $T^a$,
normalized so that $\Tr(T^aT^b) = \delta^{ab}$,
$$
  f^{abc}\ =\ -{i\over\sqrt2} \Tr\L \LB T^a, T^b\RB T^c\R.
\eqn\structure
$$
Then one applies the $SU(N_c)$ Fierz identity
$$
   (X_1 \, T^a \, X_2)\ (Y_1 \, T^a \, Y_2)
   \ =\  (X_1 \, Y_2)\  (Y_1 \, X_2)
    - {1\over N_c} (X_1 \, X_2)\  (Y_1 \, Y_2)\ ,
\eqn\SUNFierz
$$
where $X_i,Y_i$ are strings of generator matrices $T^{a_i}$,
in order to remove contracted color indices.
In discussing amplitudes where all external particles are in the
adjoint representation (such as amplitudes in supersymmetric QCD with
no matter content), or trees consisting of only particles in the adjoint
representation, one may (and should) replace the $SU(N_c)$ Fierz identity
with the corresponding $U(N_c)$ identity, since it is simpler: the
`photonic' term decouples [\use\TreeColor,\use\TreeColorB,\use\Color].
For the two-quark amplitudes under consideration, at tree level
only one string survives
and the color decomposition is~[\use\TreeColor]
$$
 {\cal A}_n^\tree(1_{\bar{q}},2_q,3,\ldots,n)
 \ =\  g^{n-2} \sum_{\sigma\in S_{n-2}}
   (T^{a_{\sigma(3)}}\ldots T^{a_{\sigma(n)}})_{i_2}^{~\ib_1}\
    A_n^\tree(1_{\bar{q}},2_q;\sigma(3),\ldots,\sigma(n))\ ,
\eqn\gentreeqqdecomp
$$
where $S_{n-2}$ is the permutation group on $n-2$ elements,
and $A_n^\tree$ are the tree-level partial amplitudes.  They are
identical to the tree-level partial amplitudes for the process
with gluinos replacing quarks, and are thereby related to the
tree-level all-gluon amplitudes by supersymmetry Ward
identities~[\use\Susy].
We adopt throughout the convention that all momenta are taken
to be outgoing.
Because the color indices have been stripped off from the
partial amplitudes,
there is no need to distinguish a quark leg from an anti-quark leg;
charge conjugation relates the two choices.

At one loop an additional trace may survive, and the color
decomposition is
$$
 \A{n}(1_{\bar{q}},2_q,3,\ldots,n)
 \ =\ g^n \sum_{j=1}^{n-1} \sum_{\sigma\in S_{n-2}/S_{n;j}}
    \Gr_{n;j}^{(\bar{q}q)}(\sigma(3,\ldots,n))\
  A_{n;j}(1_{\bar{q}},2_q;\sigma(3,\ldots,n))\ ,
\eqn\genqqdecomp
$$
where the color structures $\Gr_{n;j}^{(\bar{q}q)}$ are defined by
$$
\eqalign{
 \Gr_{n;1}^{(\bar{q}q)}(3,\ldots,n)
  \ &=\ N_c\ (T^{a_3}\ldots T^{a_n})_{i_2}^{~\ib_1}\ ,\cr
 \Gr_{n;2}^{(\bar{q}q)}(3;4,\ldots,n)
  \ &=\ 0\ ,\cr
 \Gr_{n;j}^{(\bar{q}q)}(3,\ldots,j+1;j+2,\ldots,n)
 \ &=\ \Tr(T^{a_3}\ldots T^{a_{j+1}})\ \
   (T^{a_{j+2}}\ldots T^{a_n})_{i_2}^{~\ib_1}\ ,
  \quad j=3,\ldots,n-2, \cr
 \Gr_{n;n-1}^{(\bar{q}q)}(3,\ldots,n)\ &=\ \Tr(T^{a_3}\ldots T^{a_n})\ \
     \delta_{i_2}^{~\ib_1} \ ,\cr}
\eqn\grqq
$$
and $S_{n;j} = Z_{j-1}$ is the subgroup of $S_{n-2}$ that
leaves $\Gr_{n;j}^{(\bar{q}q)}$ invariant.
When the permutation $\si$ acts on a list of indices, it is to be
applied to each index separately:
$\si(3,\ldots,n) \equiv \si(3),\ldots,\si(n)$, etc.
We refer to $A_{n;1}$ as the leading-color partial amplitude,
and to the $A_{n;j>1}$ as subleading,
because for large $N_c$, $A_{n;1}$ alone gives the
leading contribution to the color-summed correction to the cross-section,
obtained by interfering $\A{n}^\tree$ with $\A{n}$.
The explicit $N_c$ in the definition of the leading color
structure $\Gr_{n;1}^{(\bar{q}q)}$ --- which is otherwise identical to
the tree color structure --- ensures that $A_{n;1}$ is $\Ord(1)$ for
large $N_c$.

\def\MSbar{$\overline{\rm MS}$}
It is useful to recall the analogous color decomposition
for $n$ external particles in the adjoint representation,
in particular for the pure super-Yang-Mills amplitude
for two external gluinos and $n-2$ gluons (similar to the
decomposition for $n$-gluon amplitudes~[\use\Color]),
$$
 {\cal A}_n^\susy(1_\gluino,2_\gluino,3,\ldots,n)
 \ =\ g^n \sum_{j=1}^{\lfloor n/2 \rfloor + 1}
    \sum_{\sigma\in S_n/\tilde S_{n;j}}
    \Gr_{n;j}(\sigma(1_\gluino,2_\gluino,3,\ldots,n))\
  A_{n;j}^\susy(\sigma(1_\gluino,2_\gluino,3,\ldots,n))\ ,
\eqn\gengluinodecomp
$$
where
$$
\eqalign{
 \Gr_{n;1}(1,2,\ldots,n)\ &=\ N_c\ \Tr(T^{a_1}T^{a_2}\ldots T^{a_n})\ ,\cr
 \Gr_{n;j}(1,\ldots,j-1;j,\ldots,n)\ &=\ \Tr(T^{a_1}\ldots T^{a_{j-1}})\ \
   \Tr(T^{a_{j}}\ldots T^{a_n})\ ,
  \quad j=2,\ldots,\lfloor n/2 \rfloor + 1, \cr}
\eqn\grgluino
$$
$\lfloor x \rfloor$ is the largest integer less than or equal to
$x$, $\gluino$ stands for the gluino legs, and $\tilde S_{n;j}$ is the
subset of $S_n$ that leaves $\Gr_{n;j}$ invariant.  The similar
structure of the gluino amplitudes and quark amplitudes will help in
understanding how to simplify the organization of the latter,
particularly the subleading-color contributions, $A_{n;j>1}$.

The partial amplitudes $A_{n;j}^\susy$ for $\gluino\gluino ggg$ can be
obtained from five-gluon results~[\use\FiveGluon] and supersymmetry
Ward identities~[\use\Susy]; these can in turn be used to reduce the
work required in the quark case.
Throughout this paper, we consider only supersymmetric amplitudes
with no matter content.
The use of supersymmetry in loop amplitudes implies a calculation
using a supersymmetry-preserving regulator, such as dimensional
reduction~[\use\Siegel] or the four-dimensional helicity
scheme~[\use\Long] (which are very closely related at one
loop).  To obtain results for the quark amplitudes in other schemes,
such as the conventional dimensional regularization method
[\use\CollinsBook] (often called the `\MSbar' scheme in the literature),
one must shift the partial amplitudes presented here by a quantity
proportional to the tree amplitude; the constant of proportionality
has been determined by Kunszt, Signer, and
Tr\'ocs\'anyi~[\use\KunsztFourPoint] (see eqn.~(\use\schemeshift)).

Unlike the all-external-gluon case, we can subdivide the
two-gluino $(n-2)$-gluon color structures further,
depending on the trace to which the fermion charge matrices belong.
Define $\Gr_{n;j}^{s}$ and $\Gr_{n;j}^{d}$ to be respectively
$\Gr_{n;j}$ with fermion charge matrices in the same and different traces;
furthermore, require for $\Gr_{n;j}^s$ that the fermion charge matrices
lie in the second trace, which increases the
number of such structures from $\lfloor n/2 \rfloor + 1$ to $n-1$.
We may then write
$$\eqalign{
 {\cal A}_n^\susy(1_\gluino,2_\gluino,3,\ldots,n)
  &= g^n \sum_{j=1}^{n-1}
    \sum_{\sigma\in \Sigma^s_{n;j}}
    \Gr_{n;j}^s(\sigma(1_\gluino,2_\gluino,3,\ldots,n))\;
  A_{n;j}^\susy(\sigma(1_\gluino,2_\gluino,3,\ldots,n))\cr
 &\hskip 2mm +g^n \sum_{j=2}^{\lfloor n/2\rfloor + 1}
    \sum_{\sigma\in \Sigma^d_{n;j}}
    \Gr_{n;j}^d(\sigma(1_\gluino,2_\gluino,3,\ldots,n))\;
  A_{n;j}^\susy(\sigma(1_\gluino,2_\gluino,3,\ldots,n))\;,\cr
}\eqn\gengluinodecompII
$$
where $\Sigma_{n;j}^{s}$ and $\Sigma_{n;j}^{d}$ are the sets of
permutations of $n$ elements that act nontrivially on $\Gr_{n;j}^{s}$
and $\Gr_{n;j}^{d}$ respectively, and preserve the assignment of
fermion charge matrices to the two traces.
In appendix~\use\SubleadAppendix,
we will find this separation useful when relating adjoint
representation partial amplitudes to fundamental representation
partial amplitudes.

Returning to the quark case, the explicit decomposition of the
$\bar{q}qgg$ one-loop amplitude is
$$
\eqalign{
 \A{4}(1_{\bar{q}},2_q,3,4)\ &=\ g^4 \biggl[
      N_c\,(T^{a_3}T^{a_4})_{i_2}^{~\ib_1}
              \ A_{4;1}(1_{\bar{q}},2_q;3,4)
 \ +\ N_c\,(T^{a_4}T^{a_3})_{i_2}^{~\ib_1}
              \ A_{4;1}(1_{\bar{q}},2_q;4,3) \cr
&\qquad
   +\ \Tr(T^{a_3}T^{a_4}) \, \delta_{i_2}^{~\ib_1}
              \ A_{4;3}(1_{\bar{q}},2_q;3,4)\biggr] \ ,\cr}
\eqn\qqggdecomp
$$
in agreement with the decomposition used by KST~[\use\KunsztFourPoint]
(apart from notational differences such as the explicit factor of $N_c$ in
$\Gr_{4;1}^{(\bar{q}q)}$ and the ordering of external legs).

The decomposition of the $\bar{q}qggg$ amplitude is~[\use\KunsztSingular]
$$
\eqalign{
 \A{5}(1_{\bar{q}},2_q,&3,4,5)\ =\ g^5 \biggl[
      N_c\, \sum_{\sigma\in S_3}
 (T^{a_{\sigma(3)}}T^{a_{\sigma(4)}}T^{a_{\sigma(5)}})_{i_2}^{~\ib_1}
    \ A_{5;1}(1_{\bar{q}},2_q;\sigma(3),\sigma(4),\sigma(5)) \cr
  &+ \sum_{\sigma\in Z_3} \Tr\L T^{a_{\sigma(3)}} T^{a_{\sigma(4)}}\R\;
              \L T^{a_{\sigma(5)}}\R_{i_2}^{~\ib_1}
              \  A_{5;3}(1_{\bar{q}},2_q;\sigma(3),\sigma(4);\sigma(5))\cr
  & +\ \Tr(T^{a_3}T^{a_4}T^{a_5}) \, \delta_{i_2}^{~\ib_1}
              \  A_{5;4}(1_{\bar{q}},2_q;3,4,5)
   \ +\ \Tr(T^{a_5}T^{a_4}T^{a_3}) \, \delta_{i_2}^{~\ib_1}
              \  A_{5;4}(1_{\bar{q}},2_q;5,4,3)
  \biggr].\cr}
\eqn\fiveqqdecomp$$
In the partial amplitude $A_{5;3}$ an additional semicolon separates
the gluon sandwiched between the quark indices
(the last gluon in $A_{5;3}$) from the other two gluons.

The virtual part of the next-to-leading order correction to the
parton-level cross-section is given by the sum over colors of the
interference between the tree amplitude $\A{n}^{{\rm tree}}$
and the one-loop amplitude $\A{n}$.
Using the color decompositions~(\use\gentreeqqdecomp) and
(\use\genqqdecomp), and the Fierz rules~(\use\SUNFierz) it is
straightforward to evaluate this color-sum in terms of partial
amplitudes.
Here we give the formula for the five-point case, $\bar{q}qggg$,
$$
\eqalign{
2\,{\rm Re} & \sum_{{\rm colors}} [\A{5}^{{\rm tree}\,*} \A{5}]
  \ =\ 2g^8\, {N_c^2-1 \over N_c} \, {\rm Re} \sum_{\si\in S_3}
    A_5^{{\rm tree}\,*}(1_{\bar{q}},2_q;\si(3),\si(4),\si(5)) \cr
  \times \biggl[
   & \ (N_c^2-1)^2\ A_{5;1}(1_{\bar{q}},2_q;\si(3),\si(4),\si(5)) \cr
   & -\ (N_c^2-1)\ \Bigl(
         A_{5;1}(1_{\bar{q}},2_q;\si(4),\si(3),\si(5))
       + A_{5;1}(1_{\bar{q}},2_q;\si(3),\si(5),\si(4)) \cr
   &\qquad\qquad
       - A_{5;3}(1_{\bar{q}},2_q;\si(4),\si(5);\si(3))
       - A_{5;3}(1_{\bar{q}},2_q;\si(3),\si(4);\si(5)) \Bigr) \cr
   & +\ (N_c^2+1)\ A_{5;1}(1_{\bar{q}},2_q;\si(5),\si(4),\si(3))
   \ +\ (N_c^2-2)\ A_{5;4}(1_{\bar{q}},2_q;\si(3),\si(4),\si(5)) \cr
   & +\ A_{5;1}(1_{\bar{q}},2_q;\si(4),\si(5),\si(3))
      + A_{5;1}(1_{\bar{q}},2_q;\si(5),\si(3),\si(4)) \cr
   & -\ A_{5;3}(1_{\bar{q}},2_q;\si(5),\si(3);\si(4))
   - 2\ A_{5;4}(1_{\bar{q}},2_q;\si(5),\si(4),\si(3))
         \biggr]\ , \cr}
\eqn\NLOCrossSection
$$
where the sum is over all permutations of the three gluons.

%%%%%%%%%%%%%%%%%%%%%%%%%%%%%%%%%%%%%%%%%%%%%%%

\section{Primitive Amplitudes}
\tagsection\PrimitiveSection

In this section we introduce a set of gauge-invariant, color-ordered
building blocks, which we call {\it primitive amplitudes\/}, that
suffice to determine all the two-quark $(n-2)$-gluon partial
amplitudes $A_{n;j}$. Explicit expressions for primitive amplitudes
tend to be much more compact than those for the generic partial
amplitude, because the legs are ordered.  Only a subset of the
kinematic invariants $(\sum k_i)^2$ appear as arguments of logarithms
or dilogarithms in a color-ordered set of diagrams, namely those where
all the momenta are adjacent with respect to the ordering, and this
leads to simpler analytic structure for such objects.

It is not obvious {\it a priori} that every partial amplitude can be
expressed in terms of primitive amplitudes, because the generic
partial amplitude receives contributions from diagrams with several
{\it different} cyclic orderings of the external legs, and it does
{\it not} receive contributions from certain classes of diagrams
present in the primitive amplitudes.  Nevertheless, it is possible to
write every two-quark $(n-2)$-gluon partial amplitude as a sum over
permutations of primitive amplitudes; one can show (using
string-inspired groupings of diagrams) that the unwanted diagrams
cancel out in the sum.  We perform the necessary $SU(N_c)$ group
theory manipulations in the double-line formalism~[\use\DoubleLine].
The color decomposition for one-loop $\bar{q}qg\ldots g$ amplitudes is
analogous to that presented in ref.~[\use\Color] for one-loop
amplitudes with $n$ external gluons, and the manipulations used to
simplify subleading color contributions are similar to ones used in
ref.~[\use\SusyFour].  We also employ supersymmetry
[\use\Susy,\use\Tasi,\use\KunsztFourPoint], in order to
reduce the number of primitive amplitudes that have to be calculated
directly for $\bar{q}qggg$.

To present the primitive amplitudes, we will find it convenient to use
the language of color-ordered diagrams.  These are basically Feynman
diagrams from which color indices have been stripped, the relevant
signs encoded in the topology of the graph.  The vertices are no
longer symmetric under exchanges of legs, and thus the external
ordering of the legs becomes important: two diagrams which would be
equivalent as Feynman diagrams are generally no longer so as
color-ordered diagrams.  The reader may find a description of
tree-level color-ordered Feynman rules in
refs.~[\use\ManganoReview,\use\Tasi].  One obtains them by using the
trace representation~(\use\structure) of the structure constants
$f^{abc}$, as well as eqn.~(\use\SUNFierz) or its $U(N_c)$ counterpart
for adjoint states.  One then dresses the Feynman diagrams using the
doubled color line notation for the adjoint-representation
propagators, and single color lines for fundamental-representation
ones~[\use\DoubleLine].  The two terms in equation~(\use\structure),
and in equation~(\use\SUNFierz), mean that one Feynman diagram can
generate many different color-flow diagrams.  By convention, we draw
diagrams so that the ordering of legs is clockwise around the loop.

In computing the coefficient of a particular color structure, or
configuration of generator matrices $T^{a_i}$,
one can then remove group theory factors from the vertices,
arriving at the color-ordered rules shown in \fig\RulesFigure,
where the gluon Lorentz indices are $\mu,\nu,\rho,\lambda$ and
outgoing momenta are $k,p,q$, and the fermion $\gluino$ is taken to be
in the adjoint representation.
The cyclic ordering of the legs in fig.~\use\RulesFigure\ is
important since there is a relative
sign between the two orderings of the three point vertices;
if one interchanges any two of the legs the vertex changes by a sign.
This sign follows from the antisymmetry of the structure constants
$f^{abc}$.

\LoadFigure\RulesFigure{\baselineskip 13 pt
\noindent\narrower\ninerm The color-ordered Feynman rules
have antisymmetric three-vertices.
Straight lines represent fermions, and wavy lines gluons.}
{\epsfysize 3.0truein}{Rules.psd}{}

In various steps of our explicit calculations it is
advantageous~[\use\Mapping,\use\Tasi,\use\WeakInt] to use a different
gauge than the standard Feynman gauge used in fig.~\use\RulesFigure,
such as background-field gauge~[\use\Background] and Gervais-Neveu
gauge~[\use\GN].  In the various gauges the color-ordered three- and
four-gluon vertices look different.  However, partial amplitudes and
primitive amplitudes are gauge invariant (we prove the invariance of
the latter at the end of this section); therefore the formulae below,
expressing partial amplitudes as sums of primitive amplitudes, which
are derived using the rules in fig.~\use\RulesFigure, will hold in any
gauge.

If the fermion is in the fundamental representation, the same rules
hold, but one of the three color lines at the vertex, the one flowing
along side the fermion line, should be removed or `stripped off'.
This procedure of `color-stripping' equates a contribution to an
amplitude with external fundamental-representation fermions to a
contribution to an amplitude with external adjoint-representation fermions.
The latter differ from the former in a way we shall make precise in
appendix~\use\SubleadAppendix .

In order to motivate the precise definition of primitive amplitudes for
two quarks and $(n-2)$ gluons, we first discuss amplitudes with
external gluons only.  Consider the set of diagrams contributing to
the leading all-gluon partial amplitude $A_{n;1}(1,2,\ldots,n)$.  The
only diagrams that contribute are those with a single ordering
$1\ldots n$ of legs around the loop, the ordering matching the
associated color structure $\Tr(T^{a_1}T^{a_2}\cdots T^{a_n})$,
as depicted in \fig\ColorOrderFigure.
The color-ordered diagrams for this partial amplitude may be further
distinguished by whether a gluon, a fermion, or a scalar circulates
around the loop; each contribution is separately gauge-invariant
because the coefficients of $\nf$ and $\ns$ in the full amplitude must
be gauge-invariant.  We denote them by $A_{n;1}^{[J]}$, with $J$ the
spin of the circulating particle, $J=1,{1\over2},0$.

\LoadFigure\ColorOrderFigure{\baselineskip 13 pt
\noindent\narrower\ninerm
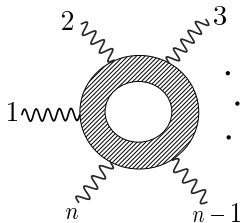
The external legs of color-ordered diagrams
have a fixed clockwise ordering.}{\epsfysize 1.2truein}
{ColorOrder.psd}{}

The subleading $n$-gluon partial amplitudes $A_{n;j>1}$ (which are
only present when an adjoint particle, not a fundamental, circulates
in the loop) generically receive contributions from diagrams with many
different cyclic orderings.  However, it is possible~[\use\SusyFour]
to express the $A_{n;j}^{[J]}$ as a sum over permutations of the
$A_{n;1}^{[J]}$, as we review in appendix~\use\SubleadAppendix.
Therefore the partial amplitudes $A_{n;1}^{[J]}$ suffice to construct
the full $n$-gluon amplitude.  On the other hand, other than
separating the contributions of internal particles of different spin,
it is not possible to find gauge-invariant subsets of the diagrams
that contribute to $A_{n;1}^{[J]}$.  When all external legs are
gluons, then, the $A_{n;1}^{[J]}$ are the irreducible gauge-invariant
pieces of the amplitude, and serve as primitive amplitudes.

If we were interested only in the two-gluino $(n-2)$-gluon (supersymmetric)
amplitude we could take the primitive amplitudes simply to be the
leading-color partial amplitudes $A_{n;1}$, just as in the $n$-gluon case,
because the same subleading-from-leading permutation formula that holds in
the $n$-gluon case applies to mixed gluino-gluon amplitudes as well.
However, for the $\bar{q}qg\ldots g$ amplitudes, one has to divide
the sets of color-ordered diagrams into finer pieces before such an
approach can succeed.  The need for a finer division of the diagrams
arises from the color flow which goes solely in one direction along
a quark line, but in both directions along a gluino line.
The gluino amplitudes $A_{n;1}$ {\it can\/} be decomposed further
in a gauge-invariant way;
the pieces of this decomposition are the primitive amplitudes,
which are also the pieces out of which we may form the amplitude
for external quarks.

To ferret out these gauge-invariant subparts of the supersymmetric
amplitude, we must trace the external fermion line through the diagram.
Since the fermion line has an arrow, we can distinguish one-loop
diagrams according to which side of the loop the fermion line is on.
We define a `left' class of diagrams to be those where,
following the arrow, the fermion line passes to the left of the loop;
the remaining diagrams are in the `right' class.
For example, in \fig\LeftRightFigureA{a}\ the fermion line
enters the loop and turns left, passing to the left of the loop,
so this is a `left' diagram; fig.~\use\LeftRightFigureA{b}\
is a `right' diagram.
In \fig\LeftRightFigureB{a}\ the external fermion line does not
actually enter the loop, but still it passes to the left,
so this is a `left' diagram; fig.~\use\LeftRightFigureB{b}\
is a `right' diagram.
In diagrams of the type shown in fig.~\use\LeftRightFigureB\
the particle circulating in the closed loop may be either a gluon,
a fermion or a scalar; the left/right designation is applied
in the same way.  (The external fermion line is always used to
make the distinction, not the fermion that might be in the closed
loop.)

\LoadFigure\LeftRightFigureA{\baselineskip 13 pt
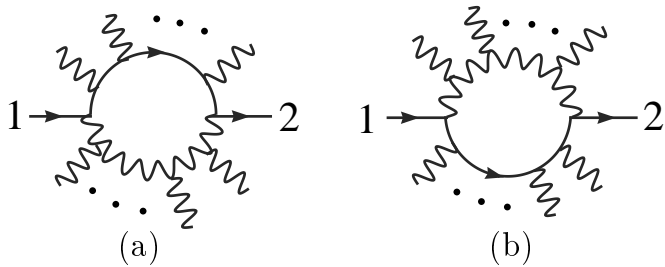
\noindent\narrower\ninerm In diagram (a) the fermion line (following
the arrow) turns `left' on entering the loop, and in diagram
(b) it turns `right'.}
{\epsfysize 1.5truein}
{LeftRightA.psd}{}

\LoadFigure\LeftRightFigureB{\baselineskip 13 pt
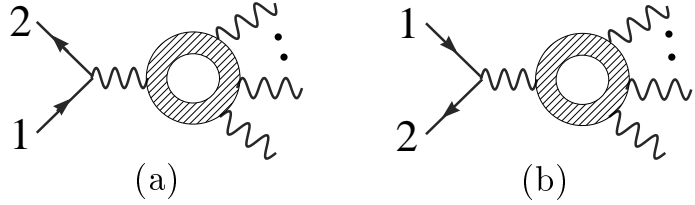
\noindent\narrower\ninerm In diagram (a) the external fermion line
passes to the `left' of the loop, and in diagram (b) it passes to the
`right'.  A gluon, fermion or scalar may circulate in the closed loop
(hatched region).}
{\epsfysize 1.2truein}
{LeftRightB.psd}{}

The `left' and `right' diagrams have to be separated from each other
when the external fermions are in the fundamental representation
because their color factors are different.
(For gluinos in the adjoint representation the color factors
are identical, and in fact the `left' and `right' diagrams must be
added together to get supersymmetric partial amplitudes.)
We shall show below that this division into `left' diagrams and
`right' diagrams is gauge invariant; those `left' and `right' diagrams
with a closed fermion or scalar loop form further gauge-invariant sets.
We thus take the primitive amplitudes for $\bar{q}qg\ldots g$ to be
$$
\eqalign{
   & A_n^{L,[J]}(1_{\bar{q}},3,4,\ldots,2_q,\ldots,n),  \cr
   & A_n^{R,[J]}(1_{\bar{q}},3,4,\ldots,2_q,\ldots,n),  \cr}
   \hskip 1 cm  J=1,\coeff{1}{2},0,
\eqn\qqgprimitive
$$
corresponding to the sum of all diagrams with the indicated cyclic ordering
of external legs, where the fermion line from $1_{\bar{q}}$ to $2_q$
passes to the left ($L$) or to the right ($R$)
of the loop, and where $J={1\over2}$ ($J=0$) represents
the subset of diagrams with a closed fermion loop (closed scalar loop).
The normalization is such that two helicity states (Weyl fermions or
complex scalars) circulate in the loop.
Diagrams without closed fermion or scalar loops are assigned to $J=1$;
they may or may not contain a closed gluon loop, as the two types of
diagrams mix under gauge transformations.
We shall often suppress the superscript ``$[1]$''; this
creates no ambiguity.
In the next section and in appendix~\use\SubleadAppendix\
we show that all of the quark-gluon
partial amplitudes $A_{n;j}$ can be obtained from these
primitive amplitudes.

Not all of the primitive amplitudes~(\use\qqgprimitive) are independent.
By flipping over the set of diagrams where $1_{\bar{q}}$ turns right,
one obtains (up to a sign) the set of diagrams where $1_{\bar{q}}$ turns
left, with the cyclic ordering also reversed,
$$
  A_n^{R,[J]}(1_{\bar{q}},3,4,\ldots,2_q,\ldots,n-1,n)
  \ =\ (-1)^n  A_n^{L,[J]}(1_{\bar{q}},n,n-1,\ldots,2_q,\ldots,4,3).
\eqn\flipsym
$$
Also, the super-Yang-Mills partial amplitudes for two gluinos and $n-2$
gluons $A_n^\susy \equiv A_{n;1}^\susy$ are given
by the sum (with all cyclic orderings identical)
$$
 A_n^\susy\ \equiv\
 A_{n;1}^\susy\ =\ A_n^{L}\ +\ A_n^{R}\ +\ A_n^{L,[1/2]}\ +\ A_n^{R,[1/2]},
\eqn\susysum
$$
because the `left' and `right'  diagrams have the same group-theory
weight for an adjoint-representation fermion.
In this equation supersymmetric cancellations occur between
the `left' and `right' primitive amplitudes; in general, $A_n^\susy$
is a much simpler quantity than either $ A_n^{L}$ or $A_n^{R}$.
Equation~(\use\susysum) allows one to obtain
one of the four terms on the right for free (say $A_n^{R}$), given
$A_{n;1}^\susy$.

Finally, the following fermion-loop contributions vanish,
$$
\eqalign{
 A_n^{R,[1/2]}(1_\qb,2_q,3,4,\ldots,n)\ &=\
 A_n^{L,[1/2]}(1_\qb,n,\ldots,4,3,2_q)\ =\ 0, \cr
 A_n^{R,[1/2]}(1_\qb,3,2_q,4,\ldots,n)\ &=\
 A_n^{L,[1/2]}(1_\qb,n,\ldots,4,2_q,3)\ =\ 0,  \cr}
\eqn\floopvanish
$$
and similarly for the scalar-loop contributions.
The restriction to `left' or `right' diagrams
combines with the ordering of the
external legs to leave only tadpole and massless external
bubble diagrams behind; but these are zero in dimensional
regularization.

We conclude this section by proving that the primitive
amplitudes~(\qqgprimitive) are gauge-invariant.%
\footnote{${}^\dagger$}{
One can motivate the argument using string theory, because each
primitive amplitude arises from a distinct string sector.
Assuming that the complete set of string states is inserted
between gluon $n$ and fermion $1$, then $A_n^L$ arises from the
Neveu-Schwarz sector and $A_n^R$ from the Ramond sector.}
Suppose that the two external fermions ($f_1$ and $f_2$) are cyclicly
separated from each other by $n_a$ gluons going one way around the loop,
and by $n_b$ gluons going the other way around ($n_a+n_b=n-2$),
and set aside for a moment the closed fermion- or scalar-loop
contributions.
The sum $A_n^{L}+A_n^{R}$ is obviously gauge invariant
because it equals the gauge-invariant, color-ordered partial amplitude
$A_{n;1}^\susy - A_{n;1}^{[1/2]}$,
i.e. the coefficient of the leading color structure
$$
  N_c\ {\rm Tr}(T^{f_1}T^{a_1}\cdots T^{a_{n_a}}
             T^{f_2}T^{b_1}\cdots T^{b_{n_b}})
\anoneqn
$$
in a theory where the fermion line is in the adjoint representation,
and omitting the fermion loop contribution.
(Different color structures are orthogonal in the large $N_c$ limit,
so their coefficients, when independent of $N_c$, must be individually
gauge invariant~[\use\ManganoReview].)

To prove that $A_n^{L}$ and $A_n^{R}$ are each invariant independently,
we consider a different gauge theory,
with gauge group $SU(N_a)\times SU(N_b)$,
where the $n_a$ gluons belong to $SU(N_a)$,
the $n_b$ gluons belong to $SU(N_b)$, fermion 2 belongs to the
representation $(N_a,N_b)$, and anti-fermion 1 belongs to the
representation $(\overline{N}_a,\overline{N}_b)$.
The coefficients of the color structures
$$
  N_a\ (T^{a_{n_a}}\cdots T^{a_1})_{i_2}^{~\bar i_1}
     \ (T^{b_1}\cdots T^{b_{n_b}})_{i_2^\prime}^{~\bar i_1^\prime}
\anoneqn
$$
and
$$
  N_b\ (T^{a_{n_a}}\cdots T^{a_1})_{i_2}^{~\bar i_1}
     \ (T^{b_1}\cdots T^{b_{n_b}})_{i_2^\prime}^{~\bar i_1^\prime}
\anoneqn
$$
must be separately gauge invariant, since $N_a$ and $N_b$ can be
chosen independently.  But the coefficient of the first of these color
structures is given precisely by the set of `right' diagrams,
 so it is just $A_n^{R}$.
This result holds because when the internal gluon line runs past the $n_a$
gluons from $SU(N_a)$, and the fermion line runs past the $n_b$ gluons from
$SU(N_b)$, then the internal gluon must belong to $SU(N_a)$, and
an extra factor of $N_a$ is generated in each such graph.
Similarly, in the case that the internal gluon line runs past the
$n_b$ gluons from $SU(N_b)$, an extra factor of $N_b$ is generated in
each graph, and so the second coefficient is $A_n^{L}$.
Thus $A_n^{L}$ and $A_n^{R}$ are separately gauge invariant.
This argument works even if there are no external gluons between
fermions $f_1$ and $f_2$, that is, $n_a=0$.
A similar argument can be carried out for the fermion (scalar) loop
contributions $A_n^{L,[1/2]}$ (respectively $A_n^{L,[0]})$
 simply by adding some extra fermion
(scalar) flavors transforming under $SU(N_a)$ but not under $SU(N_b)$,
in order to separate the $L$ and $R$ diagrams.

%%%%%%%%%%%%%%%%%%%%%%%%%%%%%%%%%%%%%%%%%%%%%%%

\section{From Primitive Amplitudes to Partial Amplitudes}
\tagsection\PartialAmplSection

In this section, we present the relation of the $\bar{q}qg\ldots g$
partial amplitudes $A_{n;j}$ to
the primitive amplitudes $A_n^{L,[J]}$, $A_n^{R,[J]}$.
The easiest case is the leading partial amplitude
$A_{n;1}(1_{\bar{q}},2_q;3,\ldots,n)$, the coefficient of
the leading color coefficient
$N_c\ (T^{a_{3}}\ldots T^{a_{n}})_{i_2}^{~\ib_1}$.
Inspecting color flows in the double-line formalism, it is easy to see that
this amplitude receives contributions only from diagrams where
the two fermions are adjacent.
The $A_n^L(1_{\bar{q}},2_q,3,\ldots,n)$ diagrams contribute
with weight $1$, after noting that the factor of $N_c$ in the diagrams
supplies the explicit $N_c$ in the definition~(\use\grqq) of
$\Gr_{n;1}^{(\bar{q}q)}$.
The $A_n^R(1_{\bar{q}},2_q,3,\ldots,n)$ diagrams would not
contribute at all to $A_{n;1}$, because of their wrong color flow,
were it not for the $-1/N_c$ term in the
$SU(N_c)$ Fierz identity~(\use\SUNFierz).
Because of this term, the $A_n^R(1_{\bar{q}},2_q,3,\ldots,n)$ diagrams
contribute with weight $-1/N_c^2$.
The fermion-loop piece $A_n^{L,[1/2]}(1_{\bar{q}},2_q,3,\ldots,n)$
contributes with weight $\nf/N_c$, and similarly for scalar loops.
As discussed near eqn.~(\use\floopvanish) the primitive
amplitude $A_n^{R,[1/2]}(1_{\bar{q}},2_q,3,\ldots,n)$
vanishes because all its diagrams are tadpoles.
Putting the contributions together, $A_{n;1}$ is given by
$$
\eqalign{
  A_{n;1}(1_{\bar{q}},2_q;3,\ldots,n)\ &=\
   A_n^L(1_{\bar{q}},2_q,3,\ldots,n)
  \ -\ {1\over N_c^2} A_n^R(1_{\bar{q}},2_q,3,\ldots,n) \cr
 &\quad +\ {\nf\over N_c} A_n^{L,[1/2]}(1_{\bar{q}},2_q,3,\ldots,n)
 \ +\ {\ns\over N_c} A_n^{L,[0]}(1_{\bar{q}},2_q,3,\ldots,n). \cr}
\eqn\Anoneformula
$$

One can make use of supersymmetry~(\use\susysum) to
rewrite~(\use\Anoneformula) as
$$
\eqalign{
  A_{n;1}(1_{\bar{q}},2_q;3,\ldots,n)\ &=
   A_n^\susy(1_{\bar{q}},2_q,3,\ldots,n)
   - \L 1+{1\over N_c^2}\R A_n^R(1_{\bar{q}},2_q,3,\ldots,n) \cr
 &\quad + \L 1-{\nf\over N_c} \R A_n^{f}(1_{\bar{q}},2_q,3,\ldots,n)
   + \L 1+{\ns\over N_c}-{\nf\over N_c}\R
      A_n^{s}(1_{\bar{q}},2_q,3,\ldots,n), \cr
&= \L 1+{1\over N_c^2}\R\,A_n^L\L 1_{\bar{q}},2_q,3,\ldots,n\R
  -{1\over N_c^2}A_n^\susy(1_{\gluino},2_\gluino,3,\ldots,n) \cr
 &\quad
  -\L {\nf\over N_c}+{1\over N_c^2}\R\,
   A_n^{f}(1_{\bar{q}},2_q,3,\ldots,n)
  + \L{\ns-\nf\over N_c}-{1\over N_c^2}\R\,
     A_n^s(1_{\bar{q}},2_q,3,\ldots,n), \cr}
\eqn\susyAnoneformula
$$
where
$$
\eqalign{
  A_n^f\ &=\ -A_n^{L,[0]} - A_n^{L,[1/2]}, \cr
  A_n^s\ &=\  A_n^{L,[0]}. \cr}
\eqn\sfdefn
$$
By the expression $A_n^\susy(1_{\bar{q}},2_q,3,\ldots,n)$
we simply mean the supersymmetric $\gluino\gluino g\ldots g$ amplitude
$A_{n;1}^\susy(1_\gluino,2_\gluino,3,\ldots,n)$ with
gluino labels replaced by antiquark/quark labels.
The combination $A_n^f$ is simpler than $A_n^{L,[1/2]}$ or $A_n^{L,[0]}$,
because it can be viewed as a chiral matter supermultiplet contribution
to a supersymmetric amplitude.

In appendix~\use\SubleadAppendix\ we prove that the subleading-color
partial amplitudes $A_{n;j>1}$ may be expressed as a permutation sum over
primitive amplitudes,
$$
\eqalign{
 A_{n;j}(1_\qb,2_q; 3,\ldots,j+1;j+2,j+3,\ldots,n)
 \ =\
 (-1)^{j-1} \sum_{\sigma\in COP\{\alpha\}\{\beta\}}
    &\Biggl[ A_n^{L,[1]} \L \sigma(1_{\bar{q}},2_q,3,\ldots,n) \R  \cr
   - {\nf\over N_c} A_n^{R,[1/2]} \L \sigma(1_{\bar{q}},2_q,3,\ldots,n) \R
   - {\ns\over N_c} &A_n^{R,[0]} \L \sigma(1_{\bar{q}},2_q,3,\ldots,n) \R
     \Biggr]\ , \cr}
\eqn\subltotal
$$
where $\alpha_i \in \{\alpha\} \equiv \{j+1,j,\ldots,4,3\}$,
$\beta_i \in \{\beta\} \equiv \{1,2,j+2,j+3,\ldots,n-1,n\}$,
and  $COP\{\alpha\}\{\beta\}$ is the set of all
permutations of $\{1,2,\ldots,n\}$ with leg $1$ held fixed
that preserve the cyclic
ordering of the $\alpha_i$ within $\{\alpha\}$ and of the $\beta_i$
within $\{\beta\}$, while allowing for all possible relative orderings
of the $\alpha_i$ with respect to the $\beta_i$.
For example if $\{\alpha\} = \{4,3\}$ and
$\{\beta\} = \{1,2,5\}$, then $COP\{\alpha\}\{\beta\}$
contains the twelve elements
$$
\eqalign{
 &(1,2,5,4,3),\quad (1,2,4,5,3),\quad (1,4,2,5,3),\quad
  (1,2,4,3,5),\quad (1,4,3,2,5),\quad (1,4,2,3,5),\quad \cr
 &(1,2,5,3,4),\quad (1,2,3,5,4),\quad (1,3,2,5,4),\quad
  (1,2,3,4,5),\quad (1,3,4,2,5),\quad (1,3,2,4,5) \quad \cr}
\eqn\OrderingExample
$$
(cyclic ordering for a two-element set is meaningless).
Note that the ordering of the first set of indices is reversed
with respect to the second.  Formula (\use\subltotal) is analogous to the
one proven for adjoint representation external states in
ref.~[\use\SusyFour].  (In ref.~[\use\SusyFour] leg $n$
was held fixed; the choice of fixed leg is completely arbitrary, and
here we find it convenient to hold fixed a fermion leg, labeled by 1.)

The terms independent of $\nf$ and $\ns$ in formula~(\use\subltotal)
(or equivalently, formula~(\use\sublanswerfund))
may be heuristically understood in terms of `parent' diagrams,
which have no trees attached to the loop,
as depicted in \fig\NoTreeFigure.
Performing a color decomposition on ordinary Feynman diagrams, using
eqns.~(\use\structure) and (\use\SUNFierz), it is easy to see that the
set of all parent diagrams feed into both $A_{n;j}$ and
$A_n^{L,[1]}$ in the correct way so that eqn.~(\use\sublanswerfund)
is satisfied for this class of diagrams.
Roughly speaking, gauge invariance requires all other diagrams to tag
along properly with the parent diagrams.  The right-hand sides of
(\use\subltotal) and (\use\sublanswerfund) do, however, contain diagrams
not appearing on the left-hand side;
in appendix~\use\SubleadAppendix\ we prove that all such unwanted
diagrams cancel in the permutation sum, and also treat the $\nf$- and
$\ns$-dependent terms.

\LoadFigure\NoTreeFigure{\baselineskip 13 pt
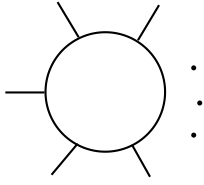
\noindent\narrower\ninerm Parent diagrams have no trees attached to the
loop.  The diagram lines represent any particles in the theory.}
{\epsfysize 1.0truein}{NoTree.psd}{}

Equation~(\use\subltotal) contains in the permutation sum primitive
amplitudes $A_n^L \equiv A_n^{L,[1]}$ with any number of gluons
sandwiched between the fermions $1_\qb$ and $2_q$.
If we know the supersymmetric $n$-gluon partial amplitudes,
then we can use the reflection symmetry~(\use\flipsym) plus
supersymmetry~(\use\susysum) to eliminate those $A_n^L$ with
more than $(n-2)/2$ gluons sandwiched between $1_\qb$ and $2_q$.
Below we carry out this procedure for the four- and five-point cases.

As an explicit example, we present the subleading-color
four-point $\qb qgg$ amplitude
relations, which are particularly simple,
$$
\eqalign{
A_{4;3}(1_\qb, 2_q ; 3, 4)\ &=\
  A_4^L(1_\qb, 2_q, 3, 4)
+ A_4^L(1_\qb, 2_q, 4, 3)
+ A_4^L(1_\qb, 3, 2_q, 4)
+ A_4^L(1_\qb, 4, 2_q, 3)  \cr
&\qquad\qquad\qquad
+ A_4^L(1_\qb, 3, 4, 2_q)
+ A_4^L(1_\qb, 4, 3, 2_q)  \cr
&=\
  A_{4;1}^{\susy} (1_\qb, 2_q, 3, 4)\
+ A_{4;1}^{\susy} (1_\qb, 2_q, 4, 3)\
+ A_{4;1}^{\susy} (1_\qb, 3, 2_q, 4)\ ,   \cr}
\eqn\Afourthree
$$
where we used eqns.~(\use\flipsym) and (\use\susysum).
In $A_{4;3}$ the fermion and scalar loop contributions~(\use\sublnfns)
vanish.  The only orderings that do not vanish by
equation~(\use\floopvanish) are
$A_4^{R,[1/2]}(1_\qb, 3,4, 2_q) + A_4^{R,[1/2]}(1_\qb, 4,3, 2_q)$,
and the same combination with $[1/2]$ replaced by $[0]$;
these combinations cancel due to Furry's theorem.
In fact the expression for $A_{4;3}$ reduces to a supersymmetric quantity.
One can verify this relation using the explicit amplitudes given by
KST~[\use\KunsztFourPoint].

The five-point relations are of course a bit more complicated,
$$
\eqalign{
A_{5;3}&(1_\qb, 2_q ; 4,5; 3) =
      \sum_{\si\in S_3}  A_5^L(1_\qb, 2_q, \si(3), \si(4), \si(5))
     +A_5^L(1_\qb, 4, 5, 2_q, 3)
     +A_5^L(1_\qb, 5, 4, 2_q, 3) \cr
 &    +A_5^L(1_\qb, 5, 2_q, 3, 4)
      +A_5^L(1_\qb, 4, 2_q, 5, 3)
      +A_5^L(1_\qb, 4, 2_q, 3, 5)
      +A_5^L(1_\qb, 5, 2_q, 4, 3)\; , \cr
}\eqn\AfivethreeA
$$
$$
\eqalign{
A_{5;4}(1_\qb, 2_q ; 3, 4,5)\
=\ \sum_{\sigma \in Z_3}
\Biggl[
&     - A_5^L(1_\qb, 2_q , \sigma(5), \sigma(4), \sigma(3))
      - A_5^L(1_\qb,  \sigma(5), 2_q, \sigma(4), \sigma(3)) \cr
&     - A_5^L(1_\qb, \sigma(5), \sigma(4), 2_q, \sigma(3))
      - A_5^L(1_\qb,  \sigma(5), \sigma(4), \sigma(3), 2_q)  \cr
  - {\ns - \nf \over N_c} \Bigl(
&       A_5^{s}(1_\qb, 2_q , \sigma(3), \sigma(4), \sigma(5))
      + A_5^{s}(1_\qb, \sigma(3), 2_q, \sigma(4), \sigma(5)) \Bigr) \cr
  + {\nf\over N_c}          \Bigl(
&       A_5^{\! f}(1_\qb, 2_q , \sigma(3), \sigma(4), \sigma(5))
      + A_5^{\! f}(1_\qb, \sigma(3), 2_q, \sigma(4), \sigma(5)) \Bigr)
\Biggr]\ , \cr}
\eqn\AfivefourA
$$
where $A_5^s$ and $A_5^{\! f}$ are defined by equation~(\use\sfdefn).
For $A_{5;3}$ the $n_{\! f,s}$ terms cancel
because of equation~(\floopvanish) plus Furry's theorem; the
contribution with ordering $(1_\qb,4,5,2_q,3)$ cancels the
contribution with ordering $(1_\qb,5,4,2_q,3)$.
We have also used the reflection identity~(\use\flipsym) to
convert $R$-type fermion/scalar loop contributions into $L$-type
ones.  We can further use the reflection identity on the $J=1$
primitive amplitudes, followed by the supersymmetry
identity~(\use\susysum) combined with Furry's theorem
to write these two partial amplitudes as follows,
$$\eqalign{
A_{5;3}(1_\qb, 2_q ; 4,5; 3) &=
      \sum_{\si\in S_3}  \LB A_5^L(1_\qb, 2_q, \si(3), \si(4), \si(5))
                        + A_5^L(1_\qb, \si(3), 2_q, \si(4), \si(5)) \RB\cr
 &\hskip 5mm -A^\susy_5(1_\qb,3,2_q,4,5)-A^\susy_5(1_\qb,3,2_q,5,4)\; ,\cr
A_{5;4}(1_\qb, 2_q ; 3, 4,5) &=
     -\sum_{\si\in S_3}  \LB A_5^L(1_\qb, 2_q, \si(3), \si(4), \si(5))
                        + A_5^L(1_\qb, \si(3), 2_q, \si(4), \si(5)) \RB\cr
 &\hskip 5mm
     + \sum_{\si\in Z_3} \LB\vphantom{\sum_X}
           A_5^\susy(1_\qb, 2_q, \si(3), \si(4), \si(5))
          + A_5^\susy(1_\qb, \si(3), 2_q, \si(4), \si(5)) \right.\cr
 &\hskip 5mm
     - \LB {\ns - \nf \over N_c}-1\RB\; \Bigl(
       A_5^{s}(1_\qb, 2_q , \sigma(3), \sigma(4), \sigma(5))
      + A_5^{s}(1_\qb, \sigma(3), 2_q, \sigma(4), \sigma(5)) \Bigr) \cr
 &\hskip 5mm
  + \LB {\nf\over N_c}+1\RB\;  \left.        \Bigl(
       A_5^{\! f}(1_\qb, 2_q , \sigma(3), \sigma(4), \sigma(5))
      + A_5^{\! f}(1_\qb, \sigma(3), 2_q, \sigma(4), \sigma(5)) \Bigr)
    \vphantom{\sum_X} \RB\ . \cr
}\eqn\AfiveB$$
In the next section we give explicit formulae for the primitive
amplitudes $A^\susy$, $A^L$, $A^s$ and $A^f$ with orderings
$(1_\qb,2_q,3,4,5)$ and $(1_\qb,2,3_q,4,5)$.
The advantage of the form~(\use\AfiveB) is that
all terms in the permutation sum may be obtained by a direct relabeling
of these primitive amplitudes.

%%%%%%%%%%%%%%%%%%%%%%%%%%%%%%%%%%%%%%%%%%%%%%%%%%

\section{Two-Quark Three-Gluon Primitive Amplitudes}
\tagsection\qqgggSection

In this section, we give explicit formul\ae\ for all the primitive
helicity amplitudes for the two-massless-quark three-gluon process
$\qb qggg$.
Using eqns.~(\use\susyAnoneformula) and (\use\AfiveB),
one can form the partial color-ordered amplitudes.
With these in hand, one can construct
either the full amplitude using equation~(\use\genqqdecomp), or the
virtual correction to the parton-level cross-section,
arising from the color-summed interference of the one-loop amplitude
with the tree amplitude, using equation~(\use\NLOCrossSection).

In calculating the primitive amplitudes for
$A_{5;1}(1_\qb^-,2_q^+,3^-,4^+,5^+)$ we used a modification of
string-based methods for gluons,
while for the other helicity configurations we found it convenient to
calculate in field theory, drawing on lessons from string theory.
In particular, we used field-theory background-field
methods~[\use\Background]
(which are an important ingredient in understanding string-based methods
in a conventional language~[\use\Mapping]).
The color-ordered Feynman rules of background-field Feynman gauge
lead to a number of calculational improvements besides the obvious
reduction in the number of terms in the three-vertex.
In particular, supersymmetric decompositions of the amplitude are more
evident~[\use\FiveGluon,\use\Tasi,\use\WeakInt], and a unitarity-based
method~[\use\Cutting] that relies on power-counting criteria
[\use\SusyFour,\use\SusyOne] is easier to apply.
We used unitarity to calculate certain ``cut-constructible''
contributions to the primitive amplitudes $A_5^L(1_\qb,2,3_q,4,5)$
which enter the subleading-color partial amplitudes
(see section~\use\IntroSection).
To perform the required loop integrations
we used the integration methods described in
refs.~[\use\IntegralsShort,\use\IntegralsLong].

We present our results in a convention where all momenta are taken to
be outgoing, that is for the process $0\rightarrow \bar{q}qggg$; helicity
conservation along the fermion line thus implies that the two fermion
legs must have opposite helicity.  Our sign conventions for the
primitive amplitudes respect the antisymmetry of the color-ordered
rules in fig.~\use\RulesFigure\ as well as the supersymmetry identities
in ref.~[\ManganoReview].
The overall sign convention for the explicit helicity amplitudes
presented here is actually opposite to that usually chosen
for fundamental representation quarks;
however, the overall sign of the loop helicity amplitude is irrelevant
as long as the tree and loop amplitudes use the same convention.
An advantage of this choice of signs is that the signs of the tree-level
gluino and quark partial amplitudes agree, so that the supersymmetry Ward
identities hold without any sign adjustments.

We shall express the primitive amplitudes in terms of the Lorentz
inner-products $s_{ij} \equiv (k_i+k_j)^2 = 2\,k_i\cdot k_j$,
and the spinor inner-products
$\spa{j}.{l} = \langle j^- | l^+ \rangle = \bar{u}_-(k_j) u_+(k_l)$
and
$\spb{j}.{l} = \langle j^+ | l^- \rangle = \bar{u}_+(k_j) u_-(k_l)$,
where $u_\pm(k)$ is a massless Weyl spinor with momentum $k$ and
chirality $\pm$~[\use\SpinorHelicity,\use\ManganoReview].

Discrete symmetries reduce the number of independent primitive
amplitudes.
Parity reverses all external helicities in a
partial amplitude; it is implemented by the ``complex
conjugation'' operation ``$\dagger$'', which
is the spinor inner-product substitution
$\spa{j}.{l} \to \spb{l}.{j}$,\ \ $\spb{j}.{l} \to \spa{l}.{j}$,
with {\it no} substitution of $i \to -i$.
For $n$-gluon amplitudes, parity takes $A \to A^\dagger$;
for two-quark $(n-2)$-gluon amplitudes, with the above sign
conventions, there is an extra minus sign, $A \to -A^\dagger$.
Using parity, we may restrict our attention to
two-quark three-gluon amplitudes having either zero or one
negative-helicity gluon.

Charge conjugation changes the identity of a quark to an
antiquark and vice-versa.  Its effect on
the primitive amplitudes is to exchange the role of `left'
and `right' contributions, $A^L \leftrightarrow A^R$,
because the direction of the fermion arrow is reversed.
Because the quark and antiquark have
opposite helicity, charge conjugation allows us to fix the
helicity of the antiquark to be negative.

Finally, equation~(\use\susyAnoneformula) for the leading-color
partial amplitude $A_{5;1}$ requires color-orderings
where the antiquark and quark are adjacent,
while equation~(\use\AfiveB) for the subleading-color partial
amplitudes $A_{5;3}$ and $A_{5;4}$ requires also
color-orderings with the antiquark and quark separated by one
gluon.  We thus need to present primitive amplitudes for eight
distinct helicity/color configurations:
two are infrared-finite, $A(-_\qb{}+_q{}+{}+{}+)$ and
$A(-_\qb{}+{}+_q{}+{}+)$;
and six are infrared-divergent,
$A(-_\qb{}+_q{}-{}+{}+)$, $A(-_\qb{}+_q{}+{}-{}+)$,
$A(-_\qb{}+_q{}+{}+{}-)$,
$A(-_\qb{}-{}+_q{}+{}+)$, $A(-_\qb{}+{}+_q{}-{}+)$ and
$A(-_\qb{}+{}+_q{}+{}-)$.
Only the latter sextet enter into the next-to-leading order corrections
to the two-quark three-gluon process.

%%%%%%%%%%%%%
% ffppp     %
%%%%%%%%%%%%%

We list the required pieces in turn, beginning with the infrared-finite
helicity configuration.  To construct $A_{5;1}$ we need
$$
\eqalign{
 A_5^{\susy}(1_\qb^-,2_q^+,3^+,4^+,5^+) &= 0\ ,  \cr
 A_5^L(1_\qb^-,2_q^+,3^+,4^+,5^+) &=
  {i\over32\pi^2} {\spa1.2 \spb2.3 \spa3.1+\spa1.4 \spb4.5 \spa5.1
      \over \spa2.3\spa3.4\spa4.5\spa5.1}
  + A_5^{s}(1_\qb^-,2_q^+,3^+,4^+,5^+)\ , \cr
 A_5^{s}(1_\qb^-,2_q^+,3^+,4^+,5^+) &=
  - {i\over 48\pi^2} \L
  {\spa1.3 \spb3.4 {\spa4.1}^2
   \over \spa1.2 \spa3.4^2 \spa4.5 \spa5.1}
  + {\spa1.4 \spa2.4 \spb4.5 \spa5.1
     \over \spa1.2 \spa2.3 \spa3.4 \spa4.5^2}
  + {\spb2.3 \spb2.5\over \spb1.2 \spa3.4 \spa4.5}\R \ ,\cr
 A_5^{f}(1_\qb^-,2_q^+,3^+,4^+,5^+) &= 0\ .\cr
}\eqn\ffppp
$$
For the construction of the subleading-color partial amplitudes
$A_{5;3}$ and $A_{5;4}$ we also need
$$
\eqalign{
  A_5^{\susy}(1_\qb^-,2^+,3_q^+,4^+,5^+) &= 0\ , \cr
  A_5^L(1_\qb^-,2^+,3_q^+,4^+,5^+) &=
     {i\over 32 \pi^2} \LB
    {\spa1.3\spa1.4\spb4.5 \over \spa1.2\spa2.3\spa3.4\spa4.5} -
    {\spa1.3^2\spb2.3\over \spa2.3\spa3.4\spa4.5\spa5.1} \RB \cr
  &\qquad + A_5^{s}(1_\qb^-,2^+,3_q^+,4^+,5^+) \ ,\cr
  A_5^{s}(1_\qb^-,2^+,3_q^+,4^+,5^+) &=
   {i \over 48 \pi^2} {\spa1.5\spa1.4\spb4.5\over\spa1.2\spa2.3\spa4.5^2}
   \ ,\cr
  A_5^{f}(1_\qb^-,2^+,3_q^+,4^+,5^+) &= 0\ .\cr
}\eqn\fpfpp$$
The tree amplitude vanishes for this helicity configuration.

The remaining helicity amplitudes are infrared-divergent, and also
require an ultraviolet subtraction.
We will present the formul\ae\ for unsubtracted amplitudes;
to carry out the \MSbar\ subtraction scheme, one should subtract
from the leading-color partial amplitudes $A_{5;1}$ the quantity
$$
  c_\Gamma \left[ {3\over2}{1\over\e}\left( {11\over3}
  - {2\over3}{\nf\over N_c} - {1\over3}{\ns\over N_c} \right) \right]
  A_5^\tree\;,
\eqn\mssubtraction
$$
where
$$
  c_\Gamma = {1\over(4\pi)^{2-\eps}}
  {\Gamma(1+\eps)\Gamma^2(1-\eps)\over\Gamma(1-2\eps)}\; ,
\eqn\cgdefn
$$
and $D= 4 - 2 \eps$.

We present our results using the dimensional-reduction
variant of dimensional regularization, with the external gluons treated
in four dimensions.
This scheme is equivalent at one-loop to the string-based
`four-dimensional helicity' scheme of ref.~[\use\StringBased].
To convert these results to the 't Hooft-Veltman scheme,
one must add to $A_{5;1}$ the quantity
$$
\delta_5\ =\ - \cg {1\over 2}\L {1 - {1\over N_c^2}} \R A_5^\tree \; ,
\eqn\schemeshift
$$
and modify the coupling constant
appropriately~[\use\KunsztFourPoint,\use\StringBased].  We obtained
the quantity~(\use\schemeshift) by direct calculation in the different
schemes, noting that only the singular terms in the integrals
contribute to this quantity.  This shift is the same as the one found
by KST [\use\KunsztFourPoint] for four-point $\bar qqgg$ amplitudes.
The universality of the shift $\delta_n = -\cg \, {1\over2}(1 -
{1\over N_c^2}) \, A_n^\tree$ for the two-quark $(n-2)$-gluon
amplitudes $A_{n;1}$ can be inferred from the invariance of
physical cross-sections under scheme shifts~[\use\KunsztFourPoint].
It can also be inferred from the universality of
collinear limits (see eqn.~(\use\loopsplit)).  The pure gluon loop
splitting amplitudes do not depend on which scheme is
used~[\use\SusyFour]; thus in equation~(\use\loopsplit) a shift
$\delta_{n-1}$ for $A_{n-1;1}$ of the form~(\use\schemeshift) implies
a shift $\delta_n$ of exactly the same form for $A_{n;1}$.  One may
also convert the expressions to conventional dimensional
regularization.  To do so one must account for the difference between
conventional and 't Hooft-Veltman schemes by having $[\eps]$-helicities
(gluon polarizations pointing into the $-2\eps$ dimensions) in
observable legs~[\use\EpsHel].
Since the amplitudes with $[\eps]$-helicities contain
an explicit overall $\eps$ only the universal poles in $\eps$ enter
and the scheme differences may be expected to affect only terms
proportional to the tree-level matrix elements.  The conversion
between the various schemes is discussed in ref.~[\use\KunsztFourPoint].

The cancellation of infrared (soft and collinear) divergences only occurs
after combining the virtual corrections presented here with
tree-level six-parton contributions to the full next-to-leading order
process.  Various general formalisms exist for constructing
infrared-finite distributions numerically~[\use\KunsztSoper,\use\GG].

For the infrared-divergent amplitudes, it is convenient to decompose
the primitive amplitudes further in a manner analogous to the
decomposition of the five-gluon amplitudes~[\use\FiveGluon],
$$
A^x = \cg\L V^x A^\tree_5 + i F^x\R,\hskip 15mm x\ =\ \susy,L,s,f\;.
\eqn\VFsplit$$
The $V$ factors are purely functions of the momentum invariants
$s_{i,i+1} = (k_i+k_{i+1})^2$, and do not contain other spinor products.
All the poles in $\eps$ are contained in the $V$ factors.
There is of course some freedom in shifting finite terms between the
$V$ and $F$ terms.
For the supersymmetric component, the $V$ factor is given by
a linear combination of $V$ functions for the all-gluon
amplitude~[\use\FiveGluon] (after adjusting for the \MSbar\ subtraction),
$$\eqalign{
V^{\rm SUSY}\ &=\ V^g_\gluon\ +\ 3 V^f_\gluon\ , \cr
}\anoneqn$$
while $F^\susy$ is related to the all-gluon $F$ terms by a supersymmetry
Ward identity. The function $V^g_\gluon$ is independent of
helicities~[\use\FiveGluon],
$$\eqalign{
V^g_\gluon &= -{1\over\e^2}\sum_{j=1}^5 \L {\mu^2\over -s_{j,j+1}}\R^\e
          +\sum_{j=1}^5 \ln\L{-s_{j,j+1}\over -s_{j+1,j+2}}\R\,
                        \ln\L{-s_{j+2,j-2}\over -s_{j-2,j-1}}\R
          +{5\over6}\pi^2\,.
  \cr
}\anoneqn$$
It will be convenient for us to define a related helicity-independent
function with the clockwise
set of double poles from the $\qb$ to the $q$ omitted,
$$
V^g_{\qb q} = V^g_\gluon
   + \sum_{j=\qb}^{q-1} {1\over\e^2} \L{\mu^2\over-s_{j,j+1}}\R^\e\ .
\anoneqn$$
All the scalar- and fermion-loop primitive
amplitudes turn out to be free of poles, so that we may take
$$
V^s = V^f = 0\;.
\anoneqn$$

We denote the dilogarithm [\use\Lewin] by $\Li_2$,
$$
  \Li_2(x)\ =\ -\int_0^x dy {\ln(1-y)\over y}\ .
\eqn\lidef
$$
In order to present the results for the remaining functions,
we also define the following functions~[\use\FiveGluon],
$$
\eqalign{
  \Ll_0(r) &= {\ln(r)\over 1-r}\,,\hskip 10mm
  \Ll_1(r) = {\Ll_0(r)+1\over 1-r}\,,\hskip 10mm
  \Ll_2(r) = {\ln(r)-\hf(r-1/r)\over (1-r)^3}\,,\cr
  \Ls_{-1}(r_1,r_2) &=
      \Li_2(1-r_1) + \Li_2(1-r_2) + \ln r_1\,\ln r_2 - {\pi^2\over6}\;,\cr
  \Ls_0(r_1,r_2) &=  {1\over (1-r_1-r_2)}\LB \Ls_{-1}(r_1,r_2)\RB\;,\cr
  \Ls_1(r_1,r_2) &= {1\over (1-r_1-r_2)}
  \LB \Ls_0(r_1,r_2) + \Ll_0(r_1)+\Ll_0(r_2)\RB\;,\cr
  \Ls_2(r_1,r_2) &= {1\over (1-r_1-r_2)}\LB \Ls_1(r_1,r_2)
                      + {1\over2} \L\Ll_1(r_1)+\Ll_1(r_2)\R \RB\;,\cr
  \Ls_3(r_1,r_2) &= {1\over (1-r_1-r_2)}\LB \Ls_2(r_1,r_2)
                      + {1\over3} \L\Ll_2(r_1)+\Ll_2(r_2)\R
                       -{1\over6r_1}-{1\over6 r_2}\RB\;.\cr
}\eqn\Lsdef
$$
The $\Ll_i$ functions are nonsingular as $r\rightarrow 1$, and
the $\Ls_i$ functions are nonsingular as $1-r_1-r_2\rightarrow 0$.

%%%%%%%%%%%%%
% ffmpp     %
%%%%%%%%%%%%%
% Reconverted back to maple and checked against
% ~bern/oneloop/five/result/ffggg.ma 07/25/94.
% SUSY WIs also checked 07/26/94.
% checked against ~bern/oneloop/five/result/ffmpptotal.ma
% and ~bern/oneloop/five/check/checkffmpp_old.ma
For $A_{5}(1_\qb^-,2_q^+,3^-,4^+,5^+)$ the functions needed in
order to construct the amplitude via equations~(\use\susyAnoneformula),
(\use\VFsplit) are:
$$
\eqalign{
A_5^\tree\ &=
i \, {{\spa1.3}^3\spa2.3 \over \spa1.2\spa2.3\spa3.4\spa4.5\spa5.1}\; ,\cr
%%%
V^\susy &= V^g_\gluon
  - {3\over2\e} \L \L{\mu^2\over-s_{34}}\R^\e
                +  \L{\mu^2\over-s_{51}}\R^\e \R\ -\ 6\ , \cr
V^L &= V^g_{12}
  -{3\over2\e}\L{\mu^2\over-s_{34}}\R^\e
  + \ln\L {-s_{51}\over -s_{12}}\R-3 \ ,\cr
}\eqn\ffmppV
$$

$$\eqalign{
F^\susy &= 3 {\spa3.2\over\spa3.1} F^f_\gluon \cr
   &= -3{\spa1.3\spa2.3\spa4.1{\spb2.4}^2 \over \spa4.5\spa5.1}
    {\Ls_1\L {-s_{23}\over -s_{51}},\,{-s_{34}\over -s_{51}}\R
         \over s_{51}^2}
   +3{\spa1.3\spa2.3\spa5.3{\spb2.5}^2 \over \spa3.4\spa4.5}
    {\Ls_1\L {-s_{12}\over -s_{34}},\,{-s_{51}\over -s_{34}}\R
         \over s_{34}^2} \cr
&\quad
 -{3\over2} {{\spa1.3}^2 \over \spa1.2\spa3.4\spa4.5\spa5.1}
    \Bigl( \spa1.5\spb5.2\spa2.3 + \spa1.2\spb2.4\spa4.3 \Bigr)
    {\Ll_0\L {-s_{34}\over -s_{51}}\R \over s_{51}}\ , \cr
}
$$
%%%%
$$
\eqalign{
F^L &= F^s -{\spa1.2\spa2.3\spa3.4\spa4.1{\spb2.4}^3
        \over\spa4.5\spa5.1}
    {\Ls_2\L {-s_{23}\over -s_{51}},\,{-s_{34}\over -s_{51}}\R
         \over s_{51}^3}
    - {\spa1.2\spa2.3{\spa3.5}^2{\spb2.5}^3
       \over\spa3.4\spa4.5}
    {\Ls_2\L {-s_{12}\over -s_{34}},\,{-s_{51}\over -s_{34}}\R
         \over s_{34}^3} \cr
  &\hskip 5mm
    - 2 {\spa1.3\spa2.3\spa4.1{\spb2.4}^2\over\spa4.5\spa5.1}
    {\Ls_1\L {-s_{23}\over -s_{51}},\,{-s_{34}\over -s_{51}}\R
         \over s_{51}^2}
    - 2 {\spa1.3\spa2.3\spa3.5{\spb2.5}^2\over\spa3.4\spa4.5}
    {\Ls_1\L {-s_{12}\over -s_{34}},\,{-s_{51}\over -s_{34}}\R
         \over s_{34}^2} \cr
  &\hskip 5mm
   - { {\spa1.3}^2\spb2.4\over\spa4.5\spa5.1}
    {\Ls_0\L {-s_{23}\over -s_{51}},\,{-s_{34}\over -s_{51}}\R
         \over s_{51}}
    - { {\spa1.3}^2\spa3.5\spb2.5\over\spa3.4\spa4.5\spa5.1}
    {\Ls_0\L {-s_{12}\over -s_{34}},\,{-s_{51}\over -s_{34}}\R
         \over s_{34}} \cr
   & \hskip 5mm
   - \L  {\spa1.3{\spa2.3}^2{\spb2.5}^2\spa1.5 \over \spa1.2\spa3.4\spa4.5}
      +{1\over2}{{\spa1.3}^2\spb1.2\spa2.3\spb2.5 \over \spa3.4\spa4.5}\R\;
      {\Ll_1\L {-s_{34}\over -s_{51}}\R \over s_{51}^2}\cr
  &\hskip 5mm
   + {1\over2} {\spa1.3\spa1.4\spa2.3{\spb2.4}^2\over\spa4.5\spa5.1}
       {\Ll_1\L {-s_{23}\over -s_{51}}\R \over s_{51}^2}
    - {1\over2} {\spa1.3\spa1.5\spa3.4{\spb4.5}^2\over\spa1.2\spa4.5}
    {\Ll_1\L {-s_{12}\over -s_{34}}\R \over s_{34}^2} \cr
  &\hskip 5mm
   + {1\over2} { {\spa1.3}^2\spb2.4
                \over\spa4.5\spa5.1}
      {\Ll_0\L {-s_{34}\over -s_{51}}\R \over s_{51}}
   -\LB 2{ {\spa1.3}^2\spb4.5\over\spa1.2\spa4.5}
        +{ {\spa1.3}^2\spa3.5\spb2.5\over\spa3.4\spa4.5\spa5.1}\RB
   { \Ll_0\L{-s_{12}\over -s_{34}}\R \over s_{34} } \cr
  &\hskip 5mm
    +{1\over2} {\spa1.4\spb2.4^2\spb4.5\over
      \spa4.5 \spb2.3\spb3.4 s_{51}}
    -  {\spa1.3\spa2.3\spb2.5\spb4.5 \over
        \spa1.2\spa3.4\spb3.4\spa4.5\spb1.5}
    -  {1\over2} {{\spa1.3}^2\spb1.2\spa2.3\spb2.5 \over
      {\spa3.4}^2\spb3.4\spa4.5\spa1.5\spb1.5} \ , \cr
}
$$
%%%%
$$
\hskip -.18 truein
\eqalign{
F^s\ &=\ {1\over 3} \Biggl[
   { \spa1.5 \spb2.5 \spa3.4 \spa3.5 {\spb4.5}^2 \over \spa4.5 }
    { 2\, \Ll_2\L{-s_{12}\over -s_{34}}\R \over s_{34}^3 }
 - { \spa1.3 \spa1.5 \spa3.4 {\spb4.5}^2
    \over \spa1.2 \spa4.5 }
    { \Ll_1\L{-s_{12}\over -s_{34}}\R \over s_{34}^2 } \cr
&\quad
- { \spa1.3 \spb2.4 \spb4.5 \over \spa1.2 \spb1.2 \spb3.4 \spa4.5 }
 + { {\spb2.4}^2 \spb2.5 \over \spb1.2 \spb2.3 \spb3.4 \spa4.5 }
                     \Biggr]\ ,\cr
%%%%
F^f\ &=\ - {{\spa1.3}^2\spb4.5 \over \spa1.2\spa4.5}
   { \Ll_0\L{-s_{12}\over -s_{34}}\R \over s_{34} } \ .  \cr
}\eqn\ffmppF$$

%%%%%%%%%%%%
%  ffpmp   %
%%%%%%%%%%%%
% Reconverted back to maple and checked against
% ~bern/oneloop/five/result/ffpmptotal.ma 07/28/94.
% SUSY WIs also checked 07/28/94.

For $A_{5}(1_\qb^-,2_q^+,3^+,4^-,5^+)$ the various functions are
$$
\eqalign{
A_5^\tree &=
i \, {{\spa1.4}^3\spa2.4 \over \spa1.2\spa2.3\spa3.4\spa4.5\spa5.1} \ , \cr
%%%
V^\susy &= V^g_\gluon
  - {3\over2\e} \L \L{\mu^2\over-s_{12}}\R^\e
                +  \L{\mu^2\over-s_{34}}\R^\e \R\ -\ 6 \ , \cr
V^L &= V^g_{12}
  - {3\over2\e}  \L{\mu^2\over-s_{34}}\R^\e -2 \ , \cr
}\eqn\ffpmpV$$

$$\eqalign{%
F^\susy &= 3 {\spa4.2\over\spa4.1} F^f_\gluon\cr
  &= - 3 {\spa1.3\spa1.4\spa2.4{\spb3.5}^2 \over \spa1.2\spa2.3}
    {\Ls_1\L {-s_{34}\over -s_{12}},\,{-s_{45}\over -s_{12}}\R
         \over s_{12}^2}
  -3  {\spa1.4{\spa2.4}^2{\spb2.5}^2 \over \spa2.3\spa3.4}
    {\Ls_1\L {-s_{51}\over -s_{34}},\,{-s_{12}\over -s_{34}}\R
         \over s_{34}^2} \cr
&\quad
  - {3\over2} {{\spa1.4}^2\spa2.4
       \over \spa1.2\spa2.3\spa3.4\spa4.5\spa5.1}
      \Bigl( \spa1.2\spb2.5\spa5.4 + \spa1.5\spb5.3\spa3.4 \Bigr)
    {\Ll_0\L {-s_{12}\over -s_{34}}\R \over s_{34}}\ , \cr
}$$
%%%%
$$
\eqalign{
F^L &= F^s - {\spa1.3\spa3.4\spa4.5{\spb3.5}^3 \over\spa2.3}
   {\Ls_2\L {-s_{34}\over -s_{12}},\,{-s_{45}\over -s_{12}}\R
         \over s_{12}^3}
   -{\spa1.2{\spa2.4}^2\spa4.5{\spb2.5}^3 \over \spa2.3\spa3.4}
   {\Ls_2\L {-s_{51}\over -s_{34}},\,{-s_{12}\over -s_{34}}\R
         \over s_{34}^3}  \cr
  &\hskip 5mm
   +{\spa1.4{\spb3.5}^2\L \spa1.4\spa2.3\spa1.5
                 +2\spa1.3\spa1.5\spa2.4 +\spa1.3\spa1.4\spa2.5\R
           \over\spa1.2\spa2.3\spa5.1}
    {\Ls_1\L {-s_{34}\over -s_{12}},\,{-s_{45}\over -s_{12}}\R
         \over s_{12}^2}\cr
  &\hskip 5mm
   +{\spa1.4\spa2.4{\spb2.5}^2\L2\spa1.5\spa2.4+\spa1.4\spa2.5\R
     \over\spa2.3\spa3.4\spa5.1}
    {\Ls_1\L {-s_{51}\over -s_{34}},\,{-s_{12}\over -s_{34}}\R
         \over s_{34}^2}
  + {{\spa1.4}^3\spa1.3{\spb1.3}^2 \over \spa2.3\spa4.5\spa5.1}
    {\Ls_1\L {-s_{12}\over -s_{45}},\,{-s_{23}\over -s_{45}}\R
         \over s_{45}^2}\cr
&\hskip 5mm
  - {3\over2} {{\spa1.4}^2\spa2.4
       \over \spa1.2\spa2.3\spa3.4\spa4.5\spa5.1}
      \Bigl( \spa1.2\spb2.5\spa5.4 + \spa1.5\spb5.3\spa3.4 \Bigr)
    {\Ll_0\L {-s_{12}\over -s_{34}}\R \over s_{34}}
  - {{\spa1.4}^2\spb3.5 \over \spa2.3\spa1.5}
    {\Ll_0\L {-s_{12}\over -s_{34}}\R \over s_{34}} \cr
&\hskip 5mm
  + {{\spa1.4}^2\spa2.4\spb2.3 \over \spa4.5\spa1.5\spa2.3}
    {\Ll_0\L {-s_{12}\over -s_{45}}\R \over s_{45}}
  - {{\spa1.4}^3\spb1.3 \over \spa2.3\spa4.5\spa1.5}
    {\Ll_0\L {-s_{23}\over -s_{45}}\R \over s_{45}}
  - {{\spa1.4}^2\spa2.4\spb2.5 \over \spa2.3\spa3.4\spa1.5}
    {\Ll_0\L {-s_{34}\over -s_{51}}\R \over s_{51}} \cr
% from finite part of "scalarnons":
&\hskip 5mm
 + {1\over 2} \Biggl[
   {\spa1.4\spa3.4{\spb3.5}^2 \over \spa2.3}
    {\Ll_1\L {-s_{12}\over -s_{34}}\R \over s_{34}^2}
  + {\spa1.2\spa1.4\spa2.4\spa4.5{\spb2.5}^2
        \over \spa2.3\spa3.4\spa1.5}
    {\Ll_1\L {-s_{34}\over -s_{12}}\R \over s_{12}^2}\cr
&\hskip 10mm
  + {\spa1.4{\spa2.4}^2{\spb2.5}^2 \over \spa2.3\spa3.4}
    {\Ll_1\L {-s_{34}\over -s_{51}}\R \over s_{51}^2}
  + {\spa2.4{\spb2.5}^2\spb3.5 \over \spb1.2\spa2.3 s_{34} \spb4.5}
  + {\spa1.3\spa1.4\spb1.5{\spb3.5}^2
        \over s_{12} \spa2.3 s_{34} \spb4.5}
  + {\spa1.4{\spa2.4}^2{\spb2.5}^2 \over \spa2.3\spa3.4 s_{34} s_{51}}
  \cr
&\hskip 10mm
  - {{\spa1.4}^2\spa2.4\spb2.3 \over s_{12} \spa2.3\spa4.5\spa1.5}
           \Biggr]\ , \cr
}
$$
%%%%
$$
\eqalign{
F^s\ &=
    2{\spa1.3\spa3.4\spa4.5\spa1.5{\spb3.5}^4
             \over \spa1.2}
   {\Ls_3\L {-s_{34}\over -s_{12}},\,{-s_{45}\over -s_{12}}\R
          \over s_{12}^4}  \cr
&\quad
  +{2\over3}\, {\spa1.4 \spa1.5 \spa3.4 {\spb3.5}^3 \over \spa1.2}
    {\Ll_2\L {-s_{12}\over -s_{34}}\R \over s_{34}^3}
  +{2\over3} \, {\spa1.3 \spa1.4\spa4.5 {\spb3.5}^3 \over \spa1.2}
    {\Ll_2\L {-s_{12}\over -s_{45}}\R \over s_{45}^3}
  -{1\over3} {\spa1.3\spa1.5{\spb3.5}^4
        \over \spa1.2\spb3.4\spb4.5 s_{12}^2}
           \,, \cr
%%%%
F^f\ &=\
   -{{\spa1.4}^2{\spb3.5}^2 \over \spa1.2}
    {\Ls_1\L {-s_{34}\over -s_{12}},\,{-s_{45}\over -s_{12}}\R
         \over s_{12}^2} \ . \cr }
\eqn\ffpmpF
$$

%%%%%%%%%%
% ffppm  %
%%%%%%%%%%
% Reconverted back to maple and checked against
% ~bern/oneloop/five/result/ffggg.ma 07/27/94.
% SUSY WIs also checked 07/27/94.

For $A_{5}(1_\qb^-,2_q^+,3^+,4^+,5^-)$ the various functions are
$$
\eqalign{
A_5^\tree &=
i \, {{\spa1.5}^3\spa2.5 \over \spa1.2\spa2.3\spa3.4\spa4.5\spa5.1}\ , \cr
%%%
V^\susy &= V^g_\gluon
  - {3\over2\e} \L \L{\mu^2\over-s_{12}}\R^\e
                +  \L{\mu^2\over-s_{45}}\R^\e \R\ -\ 6, \cr
V^L &= V^g_{12}
  - {3\over2\e}  \L{\mu^2\over-s_{45}}\R^\e -{5\over2} \ , \cr
}\eqn\ffppmV$$

$$\eqalign{%
F^\susy &= 3 {\spa5.2\over\spa5.1} F^f_\gluon\cr
  &=   {3\over2} {\spa1.5\spa2.5 \over \spa1.2\spa2.3\spa3.4\spa4.5}
     \Bigl( \spa5.4\spb4.3\spa3.1 + \spa5.3\spb3.2\spa2.1 \Bigr)
    {\Ll_0\L {-s_{12}\over -s_{45}}\R \over s_{45}}\ , \cr
}
$$
%%%%
$$
\eqalign{
F^L &= F^s -
  {{\spa1.5}^2\spa1.4{\spb1.4}^2 \over \spa2.3\spa3.4}
    {\Ls_1\L {-s_{45}\over -s_{23}},\,{-s_{51}\over -s_{23}}\R
         \over s_{23}^2}
 - {{\spa1.5}^2\spa3.5\spa1.3{\spb1.3}^2 \over \spa2.3\spa3.4\spa4.5}
    {\Ls_1\L {-s_{12}\over -s_{45}},\,{-s_{23}\over -s_{45}}\R
         \over s_{45}^2}  \cr
&\hskip 5mm
 - {{\spa1.5}^2\spb1.4 \over \spa2.3\spa3.4}
    {\Ll_0\L {-s_{23}\over -s_{51}}\R \over s_{51}}
 + {{\spa1.5}^2\spb1.2\spa2.5 \over \spa2.3\spa3.4\spa4.5}
    {\Ll_0\L {-s_{23}\over -s_{45}}\R \over s_{45}}\cr
&\hskip 5mm
  - \LB{{\spa1.5}^2\spb3.4 \over \spa1.2\spa3.4}
  -{1\over2} {\spa1.5\spa2.5\L3\spa5.4\spb4.3\spa3.1 +\spa5.3\spb3.2\spa2.1\R
            \over \spa1.2\spa2.3\spa3.4\spa4.5}\RB\;
    {\Ll_0\L {-s_{12}\over -s_{45}}\R \over s_{45}} \cr
&\hskip 5mm
 + {1\over 2} \LB
 - {\spa1.3\spa3.5\spa4.5{\spb3.4}^2 \over \spa2.3\spa3.4}
    {\Ll_1\L {-s_{12}\over -s_{45}}\R \over s_{45}^2}
 + {\spb2.4\spb3.4 \over \spa3.4\spb4.5\spb1.5} \RB\ ,   \cr
}
$$
%%%%
$$
\eqalign{
F^s\ &=\ {1\over 3} \Biggl[
 - {{\spa1.3}^2 {\spa4.5}^2 {\spb3.4}^3 \over \spa1.2 \spa3.4}
    { 2\, \Ll_2\L{-s_{45}\over -s_{12}}\R \over s_{12}^3 }
 + {\spa1.3 \spa4.5 \spa1.5 {\spb3.4}^2 \over \spa1.2 \spa3.4}
    { 3\, \Ll_1\L{-s_{45}\over -s_{12}}\R \over s_{12}^2 }   \cr
&\quad
 + 2 { {\spa1.5}^2 \spb3.4 \over {\spa1.2}^2 \spb1.2 \spa3.4}
 + {\spa1.5\spb1.3\spb2.4 \over\spa1.2\spb1.2\spa3.4\spb1.5}
 + {\spa1.4\spb1.4\spb2.4\spb3.4
     \over \spa1.2\spb1.2\spa3.4\spb4.5\spb1.5}  \Biggr]\ ,\cr
%%%%
F^f\ &=\ -{{\spa1.5}^2\spb3.4 \over \spa1.2\spa3.4}
   { \Ll_0\L{-s_{12}\over -s_{45}}\R \over s_{45} } \ .  \cr
}\eqn\ffppmF$$

%%%%%%%%%%
% fmfpp  %
%%%%%%%%%%

\def\text{\textstyle}
For use in constructing the subleading-color partial amplitudes,
the $V$ and $F$ functions for the components of
$A_{5}(1_\qb^-,2^-, 3_q^+,4^+,5^+)$ are
$$\eqalign{
A_5^\tree &=
  i {\spa1.2^3\spa3.2\over\spa1.2\spa2.3\spa3.4\spa4.5\spa5.1} \ , \cr
V^\susy &= V^g_\gluon
  - {3\over2\e} \L \L{\mu^2\over-s_{23}}\R^\e
                +  \L{\mu^2\over-s_{51}}\R^\e \R - 6 \, , \cr
V^L &= V^g_{13}
  - {3\over2\e} \L{\mu^2\over-s_{23}}\R^\e
  +\Ls_{-1}\L\text {-s_{34} \over -s_{51}}, {-s_{23}\over -s_{51}} \R
  +\Ls_{-1} \L \text {-s_{12} \over -s_{34}}, {-s_{51}\over -s_{34}} \R
  -3 \ , \cr
\cr
}\eqn\fmfppV$$

and
$$\eqalign{
F^\susy &= 3 {\spa3.2\over\spa1.2} F^f_\gluon\cr
 &= {3\over 2}
   {{\spa1.2} \L\spa2.3\spb3.4\spa4.1+\spa2.4\spb4.5\spa5.1\R\over
       \spa3.4\spa4.5\spa5.1}
     {\Ll_0\L {-s_{23}\over -s_{51}}\R\over s_{51}} \ , \cr
%%%%
F^L &=
  F^s + {1\over 2} {\spa1.5 \spb5.4^2 \spa4.2^2 \over \spa3.4 \spa4.5 }
       {\Ll_1 \L {- s_{23}\over -s_{51}} \R \over s_{51}^2 }
  +2 {\spa1.2 \spa2.4\spb4.5\spa5.1\over\spa3.4\spa4.5\spa5.1}
          {\Ll_0\L{-s_{23} \over -s_{51}} \R \over s_{51} } \cr
 &\hskip 5mm
 + {1\over 2} {\spb4.5 \spb3.5 \spb1.3 \over
    \spb2.3 \spb1.2 \spb1.5 \spa4.5} \ , \cr
%%%%
F^s &=
  {1\over 3} {\spb3.4 \spb3.5 \over \spb1.2 \spb2.3 \spa4.5} \ , \cr
%%%%
F^f &= 0 \ . \cr
}\eqn\fmfppF$$

%%%%%%%%%%
% fpfmp  %
%%%%%%%%%%

For $A_{5}(1_\qb^-,2^+, 3_q^+,4^-,5^+)$, the functions are
$$\eqalign{
A_5^\tree &=
   i {\spa1.4^3\spa3.4 \over \spa1.2\spa2.3\spa3.4\spa4.5\spa5.1}\ , \cr
V^\susy &= V^g_\gluon
  - {3\over 2 \e} \LB \L{\mu^2 \over -s_{12}}\R^\e +
        \L{\mu^2 \over -s_{34}}\R^\e \RB - 6 \; ,  \cr
V^L &= V^g_{13}
  - {3\over 2 \e} \L{\mu^2 \over -s_{34}}\R^\e
  -{1\over2}\ln\L{-s_{12}\over -s_{34}}\R
  +\Ls_{-1}\L \text {-s_{23} \over -s_{45}}, {-s_{12}\over -s_{45}} \R
  - 3 \; ,  \cr
}\eqn\fpfmpV$$

and
$$\eqalign{
F^\susy &= 3 {\spa3.4\over\spa1.4} F^f_\gluon\cr
        &= -3{\spa1.3\spa1.4\spa3.4\spb3.5^2 \over \spa1.2\spa2.3}
  {\Ls_1 \L{-s_{45}\over -s_{12}}, {-s_{34} \over -s_{12}} \R \over s_{12}^2}
  - 3{\spa1.4\spa2.4\spb2.5^2 \over \spa2.3}
   {\Ls_1 \L {-s_{51}\over -s_{34}}, {-s_{12}\over -s_{34}} \R
                 \over s_{34}^2} \cr
 & \hskip 5mm
  + {3\over 2} {\spa1.4^2\spa3.4
         (\spa1.2\spb2.5\spa5.4 + \spa1.5\spb5.3\spa3.4) \over
          \spa1.2\spa2.3\spa3.4\spa4.5\spa1.5}
     {\Ll_0\L{-s_{12} \over -s_{34} }\R \over s_{34}} \ , \cr
}
$$
%%%%
$$
\eqalign{
F^L &=
 - {\spa1.3^2\spa3.4\spa4.5\spb3.5^3\over\spa1.2\spa2.3}
   {\Ls_2\L{-s_{34} \over -s_{12}}, { -s_{45}\over -s_{12}} \R\over s_{12}^3}
 - {\spa1.2\spa4.5\spa2.4\spb2.5^3\over\spa2.3}
   {\Ls_2\L {-s_{12} \over -s_{34}}, {-s_{51}\over -s_{34}} \R\over s_{34}^3}
   \cr
 &\hskip 5mm
 - 2 {\spa1.4 \spa3.4\spa1.3 \spb3.5^2\over \spa1.2\spa2.3}
   {\Ls_1 \L {-s_{34}\over -s_{12}}, {-s_{45}\over -s_{12}} \R\over s_{12}^2}
 -2{\spa1.4\spa2.4\spb2.5^2\over\spa2.3}
 {\Ls_1 \L {-s_{12}\over -s_{34}}, {-s_{51}\over -s_{34}} \R\over s_{34}^2}\cr
 &\hskip 5mm
 - {\spa1.4^2 \spa1.3\spb3.5  \over \spa1.2\spa2.3\spa5.1}
   {\Ls_0 \L {-s_{34}\over -s_{12}}, {-s_{45}\over -s_{12}} \R\over s_{12}}
 -  {\spa1.4^2 \spb2.5\over \spa2.3\spa5.1}
   {\Ls_0 \L {-s_{12}\over -s_{34}}, {-s_{51}\over -s_{34}} \R\over s_{34}}\cr
 &\hskip 5mm
 - {1\over2} {\spa1.4\spa2.4\spb2.5^2\over\spa2.3}
   {\Ll_1\L{-s_{51}\over-s_{34}}\R\over s_{34}^2}
 + {1\over2} {\spa1.2\spa1.4\spa4.5\spb1.5\spb2.5\over\spa2.3\spa2.5}
   {\Ll_1\L{-s_{12}\over-s_{34}}\R\over s_{34}^2}\cr
 &\hskip 5mm
 + {1\over2} {\spa1.3\spa1.4\spa3.4\spb3.5^2\over\spa1.2\spa2.3}
   {\Ll_1\L{-s_{12}\over-s_{34}}\R\over s_{34}^2}
 -\LB 2{\spa1.4^2\spa3.4\spb3.5\over\spa1.2\spa2.3\spa4.5}\,
    -{1\over2} {\spa1.2 \spa1.4\spa4.5\spb2.5\over
                \spa1.5\spa2.3\spa2.5}\RB\;
     {\Ll_0\L{-s_{12}\over-s_{34}}\R\over s_{34}} \cr
 &\hskip 5mm
  +{1\over2}{\spa1.4 \spa2.4 \spb2.5\over
             \spa2.3 \spa2.5 \spa3.4 \spb3.4}
  +{1\over2}{\spa1.3^2 \spb3.5^3\over
             s_{12} \spa1.2 \spa2.3 \spb3.4 \spb4.5} \ , \cr
%%%%
F^s &= 0\ , \cr
%%%%
F^f &= 0\ . \cr
}\eqn\fpfmpF$$

%%%%%%%%%%
% fpfpm  %
%%%%%%%%%%
% Reconverted back to maple and checked against
% ~bern/oneloop/five/resultsublead/ffggg_lance.ma 07/29/94.
% SUSY WIs also checked 07/29/94.

For $A_{5}(1_\qb^-,2^+, 3_q^+,4^+,5^-)$, the functions are
$$\eqalign{
A_5^\tree &=
   i {\spa1.5^3\spa3.5 \over \spa1.2\spa2.3\spa3.4\spa4.5\spa5.1}\ , \cr
V^\susy &= V^g_\gluon
  - {3\over2\e} \L \L{\mu^2\over-s_{12}}\R^\e
                +  \L{\mu^2\over-s_{45}}\R^\e \R - 6 \, , \cr
V^L &= V^g_{13}
       -{3\over2\e} \L{\mu^2\over-s_{12}}\R^\e
       + \Ls_{-1} \L \text {-s_{23}\over -s_{45}}, {-s_{12}\over -s_{45}} \R
       -3 \;,\cr
}\eqn\fpfpmV$$

and
$$\eqalign{
F^\susy &= 3 {\spa3.5\over\spa1.5} F^\susy_\gluon\cr
   &=  {3\over 2} {\spa1.5\spa3.5
     (\spa5.4\spb4.3\spa3.1 + \spa5.3\spb3.2\spa2.1) \over
       \spa1.2 \spa2.3\spa3.4\spa4.5}
        {\Ll_0 \L {-s_{12} \over -s_{45}}\R \over s_{45} } \ , \cr
F^L &=
   {{\spa1.5}^2 \spb2.4\over \spa1.2\spa3.4}
    {\Ls_0\L {-s_{34} \over -s_{51}}, {-s_{23}\over -s_{51}} \R
     \over s_{51}}
   +{\spa1.5^2 \spa1.3\spb1.4\over \spa1.2\spa2.3\spa3.4}
    {\Ls_0\L {-s_{51} \over -s_{23}}, {-s_{45}\over -s_{23}} \R
     \over s_{23}}\cr
 &\hskip 5mm
  -{1\over 2} {\spa3.5\spa4.5\spb3.4^2\spa1.3^2\over \spa1.2\spa2.3\spa3.4}
       {\Ll_1\L{-s_{12}\over -s_{45}} \R \over s_{45}^2}
  -2{\spa1.5\spa3.5\spa1.3 \spb3.4\over \spa1.2\spa2.3\spa3.4}
           {\Ll_0\L{-s_{12}\over -s_{45}}\R \over s_{45}}
  - {1\over 2} {\spb2.4^2 \over \spa3.4\spb1.5\spb4.5} \ , \cr
%%%%
F^s &= 0 \ , \cr
%%%%
F^f &= 0 \ .\cr
}\eqn\fpfpmF$$

The remaining amplitudes are related by discrete symmetries to those
presented above; consider, for example,\hfil\break
 $A_5^L(1_\qb^\pm,2_q^\mp,3^-,4^-,5^-)$:
$$\eqalign{
A_5^L(1_\qb^+,2_q^-,3^-,4^-,5^-) &=
          - \LB A_5^L(1_\qb^-,2_q^+,3^+,4^+,5^+)\RB^\dagger\cr
A_5^L(1_\qb^-,2_q^+,3^-,4^-,5^-) &=
           A_5^R(2_\qb^+,3^-,4^-,5^-,1_q^-)
      = -A_5^L(2_\qb^+,1_q^-,5^-,4^-,3^-)\cr
   &= +\LB A_5^L(2_\qb^-,1_q^+,5^+,4^+,3^+)\RB^\dagger\ . \cr
}\anoneqn$$
The overall signs in these relations drop out in the
cross-section, because the signs are the same for loop
amplitudes and tree amplitudes.

We have performed a variety of checks on the amplitudes:

\item{1)} a check of collinear factorization for all primitive
amplitudes in all channels, illustrated in appendix~\use\CollinearAppendix,
providing an extremely stringent check of the primitive amplitudes;

\item{2)} a verification of the
supersymmetry identities [\use\Susy] for the primitive
amplitudes $A_{5}(1_\qb^-,2_q^+,3^+,4^+,5^-)$ and
$A_{5}(1_\qb^-,2_q^+,3^+,4^-,5^+)$ by explicitly calculating all terms
in the supersymmetry relation (\use\susysum);

\item{3)} a verification of some of the cuts in amplitudes
that were not calculated via cutting methods;

\item{4)} a check on eqn.~(\use\AfiveB) giving $A_{5;3}$ and $A_{5;4}$,
by comparing the poles in $\eps$ in these quantities
against explicit formulae for such singular terms in
ref.~[\use\KunsztSingular];

\item{5)} a check of formula~(\use\NLOCrossSection)
for the virtual part of the color-summed cross-section,
by showing that the infrared poles in $\eps$
properly cancel against singular terms in the $2 \rightarrow 4$ matrix
elements arising from the integration over soft and collinear phase
space [\use\KunsztSoper,\use\GG,\use\KunsztSingular];

\item{6)} a check of the permutation formula for $A_{5;4}$ in
eqn.~(\use\AfiveB), by comparing the fermion loop ($n_{\! f}$)
contributions to the corresponding contribution of the previously
calculated [\use\Morgan] process $Z \rightarrow 3
\gamma$, but with the $Z$ polarization vector replaced with a fermion
bilinear (and gluon propagator connecting it to the loop); it is not
difficult to see that the contributing diagrams are identical for
these two cases.  (The axial coupling does not contribute to $Z
\rightarrow 3 \gamma$ so it does not affect the comparison.)  We have
explicitly verified that the fermion loop contributions on the
right-hand side of eqn.~(\use\AfiveB) agree with the appropriately
modified expressions contained in ref.~[\use\Morgan] for vanishing
fermion masses.

%%%%%%%%%%%%%%%%%%%%%%%%%%%%%%%%%%%%%%%%%%%%%%

\section{Conclusions}
\tagsection\ConclusionSection

In this paper, we presented all one-loop QCD amplitudes for two
external quarks and three external gluons.  Combining these results
with the ones for five gluons [\use\FiveGluon] and four quarks and
one gluon [\use\Kunsztqqqqg],
this completes the set of one-loop amplitudes required for
calculating next-to-leading order corrections to three-jet
production at hadron colliders.  The computation made use of a number
of techniques, including spinor helicity [\use\SpinorHelicity], color
decompositions~[\use\TreeColor,\use\TreeColorB,\use\Color], string-based
methods~[\use\StringBased,\use\Subsequent], supersymmetry
methods~[\use\Susy,\use\FiveGluon,\use\Tasi,\use\WeakInt], improved
gauge choices [\use\Background,\use\GN,\use\Mapping], perturbative
unitarity [\use\SusyFour,\use\SusyOne], and collinear limits
[\use\TreeCollinear,\use\ManganoReview,\use\AllPlus,\use\GordonConf].

We also introduced {\it primitive amplitudes} as gauge-invariant
building blocks from which amplitudes containing fundamental
representation external legs may be constructed.  The usefulness of
primitive amplitudes follows from their relatively simple analytic
structure. In a previous paper we obtained a formula [\use\SusyFour]
valid for adjoint representation states which allows one to obtain all
subleading-color partial amplitudes from the leading-color partial
amplitudes.  Using primitive amplitudes we generalized this formula to
the case of two external fundamental representation quarks and
$(n-2)$-gluons.  Further generalizations to larger numbers of external
quarks are straightforward.

In calculating the amplitudes we made extensive use of supersymmetry
identities [\use\Susy], both as a check and to reduce the number of
independent amplitudes to be calculated.  We verified that
as the momenta of two adjacent external legs become collinear, the
amplitudes presented here properly reduce to sums of lower-point
amplitudes multiplied by universal splitting amplitudes.  This
provides a stringent consistency check on the amplitudes.

The one-loop two-quark three-gluon amplitudes presented in this paper
constitute, along with the five-gluon~[\use\FiveGluon] and four-quark
one-gluon~[\use\Kunsztqqqqg] ones,
one of the major ingredients required for the construction of
a next-to-leading order program for the prediction of three-jet physics
at hadron colliders.
The infrared singularities in these amplitudes must be cancelled by
adding the singular contributions from real emission,
for example using the formalisms of refs.~[\use\GG,\use\KunsztSoper].
The program also requires the full form of the real emission
contributions in non-singular regions, given here by the known
six-point tree amplitudes~[\use\TreeColor,\use\ManganoReview].
Such a program would allow the study of three-jet distributions to
next-to-leading order;
as with the two-jet case studied extensively by various collider
detector collaborations, large-statistics data are available.
In the three-jet case, one may study a richer variety
of distributions.  The comparison of three-jet rates to two-jet rates,
in conjunction with such an NLO program, offers the first possibility
of measuring the strong coupling constant $\alpha_s$ in a purely
hadronic process deep in the perturbative regime.  Beyond three-jet
studies, such a program also incorporates the elements required to
study jet structure beyond the leading non-vanishing order available
in two-jet NLO programs.

\vskip .2 cm

\noindent{\bf Acknowledgements.}

\vskip .1 cm

We thank D.C. Dunbar for useful discussions, and A. Signer for pointing
out an error in an earlier version of equation~(\use\loopphotonbb);
one of us (L.D.) also thanks Z. Kunszt for useful discussions.
We are grateful for the support of NATO Collaborative Research Grant
CRG--921322 (L.D. and D.A.K.).

%%%%%%%%%%%%%%%%%%%%%%%%%%%%%%%%%%%%%%%%%%%%%%%
%%%%%%%%%%%%%%%%%%%%%%%%%%%%%%%%%%%%%%%%%%%%%%%

\appendix{Subleading-Color Partial Amplitudes
from Primitive Amplitudes}
\tagappendix\SubleadAppendix

In this appendix we prove that the subleading partial amplitudes
$A_{n;j>1}$ are given by sums over permutations of primitive
amplitudes.  We will use a similar result for the case of $n$ external
adjoint particles~[\use\SusyFour] as an intermediate step in deriving
the result for the case that the two external fermions are in the
fundamental representation.

First we discuss  the  distinction  between  the gauge  groups
$SU(N_c)$ and $U(N_c)=SU(N_c)\times  U(1)$.   If  all particles in  an
amplitude  transform as the adjoint   representation of $SU(N_c)$, and
all vertices are  given  by (combinations of)  the structure constants
$f^{abc}$, then  there  is   essentially  no    distinction,   because
$f^{abc}=0$  whenever  $a$  corresponds   to  the  $U(1)$   generator,
$T^{a_{U(1)}}={\bf 1}/\sqrt{N_c}$.
In other words,  the  $U(1)$ `photon' is
automatically    projected   out  by   vertices  such   as  nonabelian
vector-boson self-interactions.  The   $-1/N_c$ term  in the $SU(N_c)$
Fierz  identity~(\use\SUNFierz) removes the  `photon'   explicitly  by
projecting onto  traceless  hermitian matrices.  It  can be ignored if
only adjoint-representation particles are present.

More generally, the $-1/N_c$ Fierz term only affects those diagrams
for $\bar{q}qg\ldots g$
where a gauge boson propagator is attached at both ends to a line in the
fundamental representation --- a fermion or scalar line.
In those diagrams where both of these ends are in the loop, that is
with exactly one gauge boson propagator in the loop
itself (the diagrams contributing to $A_n^R(1_{\bar{q}},2_q,3,\ldots,n)$),
the $-1/N_c$ term leads to $A_{n;1}$ contributions only, as discussed above.
The only other diagrams affected are fermion or scalar loop contributions,
where a gauge boson attaches the pinched-off external fermion line
to the loop.
In summary, the $n_{\!f}$- and $n_s$-independent parts of the subleading
$\bar{q}qg\ldots g$ partial amplitudes $A_{n;j>1}$, and all
of the $\gluino\gluino g\ldots g$ partial amplitudes, can be analyzed
as if the gauge group were $U(N_c)$, neglecting the $-1/N_c$ Fierz term
when working out the color flow in the double-line formalism.
This result in turn implies that such $\bar{q}qg\ldots g$ double-line
diagrams can be obtained from a subset of $\gluino\gluino g\ldots g$
diagrams by `color-stripping' --- removing a color line that flows
directly from one gluino to the other, thereby
converting the adjoint representation gluino
into a fundamental representation quark.
Consequently, conversion of a subleading-color formula for
$\gluino\gluino g\ldots g$ to the $\bar{q}qg\ldots g$ case is quite
straightforward.

The formula we want to re-derive, and then modify,
expresses the subleading-color partial amplitudes $A_{n;j>2}^{\rm adj}$
for $\gluino\gluino g\ldots g$ (or $ggg\ldots g$)
as a sum over permutations of leading-color partial amplitudes
$A_{n;1}$,
$$
 A_{n;j}^{\rm adj}(1,2,\ldots,j-1;j,j+1,\ldots,n)\ =\
 (-1)^{j-1} \sum_{\sigma\in COP\{\alpha\}\{\beta\}}
 A_{n;1}^{\rm adj} \L \sigma(1,2,\ldots,n) \R,
\eqn\sublanswer
$$
where $\alpha_i \in \{\alpha\} \equiv \{j-1,j-2,\ldots,2,1\}$,
$\beta_i \in \{\beta\} \equiv \{j,j+1,\ldots,n-1,n\}$,
and the set of permutations $COP\{\alpha\}\{\beta\}$ is defined
below eqn.~(\use\subltotal).
We have added the superscript ``adj'' to ~(\use\sublanswer) to avoid
confusion with the $\bar{q}qg\ldots g$ amplitudes denoted by $A_{n;j}$.

In ref.~[\use\SusyFour] a string-theory based proof of
eqn.~(\use\sublanswer) was presented.  We now review this proof, but in
terms of the color-ordered Feynman rules in fig.~\use\RulesFigure.
The leading-color partial amplitude $A_{n;1}^{\rm adj}(1,2,\ldots n)$
associated with $\Tr(T^{a_1} T^{a_2} \cdots T^{a_n})$ is given by the
sum of all color-ordered planar diagrams whose external legs follow
the cyclic ordering of the color trace, as depicted in
fig.~\use\ColorOrderFigure.
Examples of five-point diagrams, dressed with color
flow lines, are shown in \fig\ColorFlowFigure.
Our convention for the direction of the color flow is to follow
the reverse ordering of the color trace; this will lead to the standard
convention for the color arrow following the fermion arrow after
conversion to fundamental-representation fermions.
To compute the subleading-color amplitudes $A_{n;j}^{\rm adj}$
associated with the color structure $\Tr(T^{a_1} T^{a_2} \cdots
T^{a_{j-1}})$$\times$$\Tr(T^{a_j} T^{a_{j+1}} \cdots T^{a_n})$ directly,
one must sum over all planar color-ordered diagrams whose corresponding
Feynman diagrams can give rise to this color structure.
These are the diagrams where the cyclic ordering of legs that
belong to each trace follows the ordering of that trace,
but where the ordering in one trace is reversed because the two color
lines associated with a adjoint particle flow in opposite directions
around the loop.  For example, in \fig\TraceReverseFigure\
the legs follow ordering 123456, but the color structure is
$\Tr(T^{a_4} T^{a_3} T^{a_2})$$\times$$\Tr(T^{a_5} T^{a_6} T^{a_1})$.

\LoadFigure\ColorFlowFigure{\baselineskip 13 pt
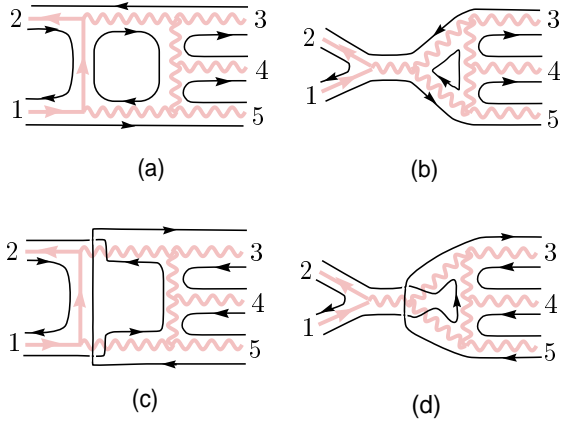
\noindent\narrower\ninerm Examples of the color flow for adjoint
representation fermions, using the double-line formalism.
The Feynman diagrams are gray, and the oriented color lines dressing
them are black.}
{\epsfysize 2.2truein}
{ColorFlow.psd}{}

Thus we sum over color-ordered diagrams with the legs permuted
over $COP\{\alpha\}\{\beta\}$, where $\alpha =\{j-1, j-2, \ldots, 1\},
\;\beta = \{j,j+1, \ldots, n\}$.  We must, however, explicitly exclude
one class of diagrams whose color flow is incorrect.  This is the
class of diagrams where indices from {\it both} sets $\{\alpha\}$ and
$\{\beta\}$ label leaves of the same tree attached to the loop.
The color flow in these diagrams cannot produce the desired trace
structure, because the line attaching the tree to the loop can carry
only a single pair of color indices, whereas two pairs would be
required to join both $\alpha$ and $\beta$ indices to those elsewhere
on the loop.
Examples of diagrams contributing to the color
structure $\Tr(T^{a_1} T^{a_2})$$\times$$\Tr(T^{a_5} T^{a_4} T^{a_3})$
are shown in figs.~\ColorFlowFigure{c}\ and~\ColorFlowFigure{d},
and one should exclude diagrams such as the one depicted in
\fig\ExcludeFigure .
(In the string-based derivation~[\use\SusyFour],
another class of diagrams --- those where a single tree
contains {\it all} elements of either $\{\alpha\}$ or $\{\beta\}$ ---
is also excluded explicitly.  This class automatically cancels out of
the field-theory calculation, so long as the set $\{\alpha\}$ or
$\{\beta\}$ in question contains only legs transforming under the
adjoint representation.)

\LoadFigure\TraceReverseFigure{\baselineskip 13 pt
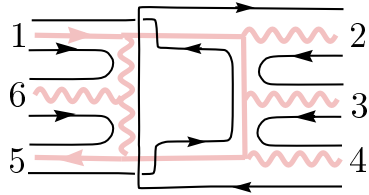
\noindent\narrower\ninerm The ordering of one trace is reversed
as compared to the other.}
{\epsfysize 1.0truein}{TraceReverse.psd}{}

\LoadFigure\ExcludeFigure{\baselineskip 13 pt
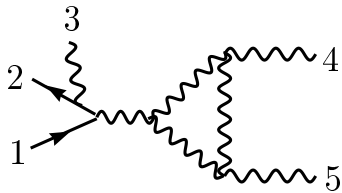
\noindent\narrower\ninerm An example of a diagram that does not
contribute to the coefficient of the color structure
$\Tr(T^{a_1} T^{a_2})$$\times$$\Tr(T^{a_5} T^{a_4} T^{a_3})$.}
{\epsfysize 1.1truein}{Exclude.psd}{}

One can divide the set of all diagrams into a
`parent' subset, which have only three-point vertices and
no non-trivial trees (depicted in fig.~\NoTreeFigure),
and all the remaining diagrams.
We refer to the latter as `daughter' diagrams, because
each can be derived from some `parent' diagram via a continuous
`pinching' process, in which two lines attached to the loop are
brought together to a four-point interaction, or further pulled out
from the loop, and left as the branches of a tree attached to the
loop.  Repeating this process in all inequivalent ways yields all
graphs contributing to the same color ordering as the parent diagram,
i.e. all of its daughter diagrams.  For instance, the color-ordered
diagram in fig.~\use\ExcludeFigure\ is a daughter of the parent with
ordering 12345 depicted in fig.~\use\ColorFlowFigure{a}.  Daughter
diagrams of different parents can be essentially the same diagram.
For example, if one swaps legs 2 and 3 in fig.~\use\ExcludeFigure, one
obtains a daughter of the parent with ordering 13245, but the two
daughter diagrams differ only by an overall minus sign coming from the
antisymmetric color-ordered three-vertex.
Such relations are important for proving eqns.~(\use\sublanswer) and
(\use\subltotal).

Let us first focus on the `parent' subset of diagrams.
As mentioned in section~\use\PartialAmplSection,
by performing a color decomposition of ordinary Feynman diagrams
and using eqns.~(\use\structure) and (\use\SUNFierz) one can show that all
`parent' diagrams feed into both $A_{n;1}^{\rm adj}$ and
$A_{n;j>1}^{\rm adj}$ in the correct way so that
eqn.~(\use\sublanswer) is satisfied for this class of diagrams.
The same arguments apply to those `daughter'
diagrams where each pinched-off tree contains only members of the
$\{\alpha\}$ set, or only members of the $\{\beta\}$ set.

The only thing left to prove is that the class of daughter diagrams
specifically excluded from $A_{n;j}$ --- where individual trees have
both $\{\alpha\}$ and $\{\beta\}$ members --- does yield a vanishing
contribution when summed over the permutations in
$COP\{\alpha\}\{\beta\}$.  Eqn.~(\use\sublanswer) then follows.  When
such diagrams have only three-point vertices on the trees, the
diagrams can be arranged so they cancel in pairs.  The pairs are
related by the exchange of an $\{\alpha\}$ leg with a $\{\beta\}$ leg
on a tree; the cancellation follows from the anti-symmetry of the
three-point color-ordered Feynman vertices in fig.~\use\RulesFigure\
under the interchange of the ordering of the two outer legs.  For
example, the pairs of diagrams in \fig\SubleadFigureA\ cancel in the sum.
For diagrams with trees containing four-point vertices,
using the color-ordered rules in fig.~\use\RulesFigure,
the cancellations in the sum over $COP\{\alpha\}\{\beta\}$ occur in
triplets, such as those shown in \fig\SubleadFigureB.  Diagrams with
four-point vertices attached to the loop can be decomposed into the
same color structures encountered above.  This provides a purely
field-theoretic proof of eqn.~(\use\sublanswer), where all external
and internal states are in the adjoint representation, independent of
whether they are are gluons or adjoint fermions; in particular it
applies to the subleading-color $\gluino\gluino g\ldots g$
super-Yang-Mills partial amplitudes.

\LoadFigure\SubleadFigureA{\baselineskip 13 pt
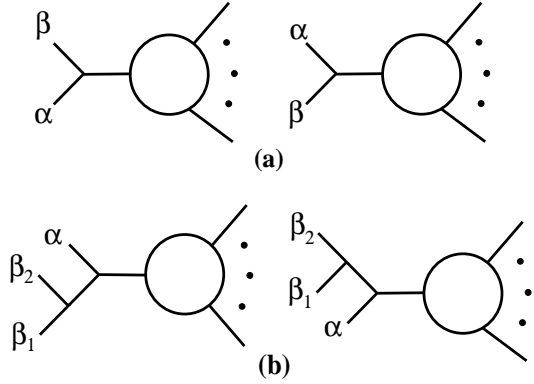
\noindent\narrower\ninerm Examples of pairs of diagrams that cancel in the
permutation sum.}
{\epsfysize 2.0truein}
{SubleadCancelA.psd}{}

\LoadFigure\SubleadFigureB{\baselineskip 13 pt
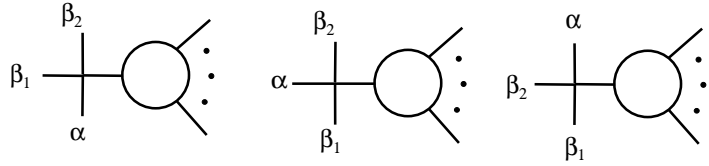
\noindent\narrower\ninerm Diagrams with four-point vertices cancel
in the permutation sum in triplets.}
{\epsfysize 0.9truein}
{SubleadCancelB.psd}{}

Now consider the modifications necessary for $\bar{q}qg\ldots g$
QCD amplitudes, where the two fermions are fundamental representation
quarks instead of adjoint representation gluinos.
We will show that the subleading-color partial
amplitudes $A_{n;j>2}(1_\qb, 2_q; 3, \ldots, n)$ in
the full amplitude~(\use\genqqdecomp), omitting for now the
$n_{f,s}$-dependent terms from closed fermion or scalar loops,
are given by exactly the same type of formula as eqn. (\use\sublanswer),
$$
 A_{n;j}(1_\qb,2_q; 3,\ldots,j+1;j+2,j+3,\ldots,n)
   |_{{\rm no}\ n_{f,s}} \ =\
 (-1)^{j-1} \sum_{\sigma\in COP\{\alpha\}\{\beta\}}
    A_n^{L,[1]} \L \sigma(1_{\bar{q}},2_q,3,\ldots,n)\R,
\eqn\sublanswerfund
$$
where $\alpha_i \in \{\alpha\} \equiv \{j+1,j,\ldots,4,3\}$,
$\beta_i \in \{\beta\} \equiv \{1,2,j+2,j+3,\ldots,n-1,n\}$,
$COP\{\alpha\}\{\beta\}$ is defined in exactly the same
way as for the adjoint representation formula,
and $n_{f,s}$ means either $n_s$ or $n_{\! f}$.

The main difference between the adjoint formula~(\use\sublanswer) and
the fundamental formula~(\use\sublanswerfund) is that the adjoint
formula lumps `left' diagrams, where the fermion goes around the loop
on the left side, together with otherwise identical `right'
diagrams, whereas the fundamental formula keeps the two separate,
as required by their different color flows.
To derive~(\use\sublanswerfund) from~(\use\sublanswer) we remove a
single color line from a special subset of the
$\gluino\gluino g\ldots g$ partial amplitudes $A_{n;j}^{\rm adj}$
on the left-hand side of~(\use\sublanswer), those where the two gluino
charge matrices are in the same trace and are adjacent to each other;
this subset is in one-to-one correspondence with the
$\bar{q}qg\ldots g$ partial amplitudes $A_{n;j}$.
We show that this color-line removal corresponds to dropping the
unwanted `right' set of diagrams on the right-hand side
of~(\use\sublanswer), thus converting $A_{n;1}^{\rm adj}$ to $A_n^L$.
As discussed above, the $-1/N_c$ Fierz corrections can be ignored
here.

Let us focus on the coefficient of the same-trace color structure
$\Gr_{n;j}^s$ (for some fixed $j$) in the modified color
decomposition~(\use\gengluinodecompII),
$$
\eqalign{
 &\Tr(T^{a_3}\ldots T^{a_{j+1}})\ \
  \Tr(T^{f_1} T^{f_2} T^{a_{j+2}}\ldots T^{a_n}),
  \qquad\quad j=3,\ldots,n-2, \cr
 &\Tr(T^{a_3}\ldots T^{a_n})\ \ \Tr(T^{f_1} T^{f_2}),
  \qquad\quad j=n-1, \cr
  }
\eqn\TraceA
$$
whose corresponding partial amplitude is
$A_{n;j}^{\rm adj}(3,\ldots,j+1;1_\gluino,2_\gluino,j+2,j+3,\ldots,n)$.
Our convention for drawing the
color-dressed diagrams is to impose a clockwise ordering on the legs
associated with the color trace containing the fermion color matrices
and a counterclockwise ordering on the legs associated with the other
trace. In every double-line diagram contributing to this
partial amplitude, there is a color line that runs directly from
gluino 2 to gluino 1 along the gluino line, as examples in
fig.~\use\ColorFlowFigure\ illustrate.
If we remove this color line (as shown in
\fig\ColorStripFigure), then~(\TraceA) is converted to
$$
\eqalign{
 \Gr_{n;j}^{(\bar{q}q)}(3,\ldots,j+1;j+2,\ldots,n)
 \ &=\ \Tr(T^{a_3}\ldots T^{a_{j+1}})\ \
   (T^{a_{j+2}}\ldots T^{a_n})_{i_2}^{~\ib_1}\ ,
  \quad j=3,\ldots,n-2, \cr
 \Gr_{n;n-1}^{(\bar{q}q)}(3,\ldots,n)\ &=\ \Tr(T^{a_3}\ldots T^{a_n})\ \
     \delta_{i_2}^{~\ib_1} \ ,\cr}
\eqn\TraceB
$$
that is to the fundamental representation
color structures given in eqn.~(\use\grqq), with coefficients
$A_{n;j}$.

\LoadFigure\ColorStripFigure{\baselineskip 13 pt
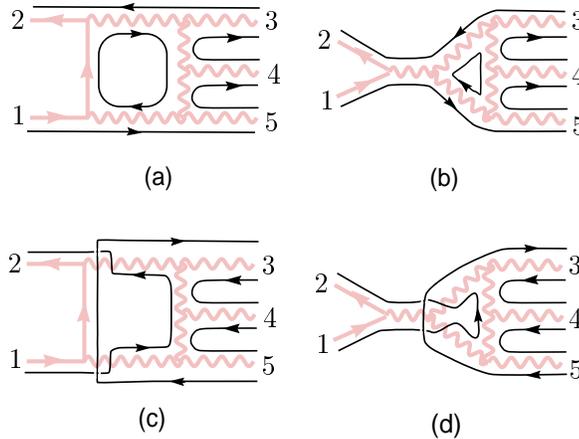
\noindent\narrower\ninerm Examples of color stripping; the color lines
running directly from gluino 2 to gluino 1 in
fig.\ \use\ColorFlowFigure\ have been removed.}
{\epsfysize 2.3truein}{ColorStrip.psd}{}

A given $\gluino\gluino g\ldots g$ diagram, with one color line
running between the two gluinos removed, can be interpreted as a
$\bar{q}qg\ldots g$ diagram, but only if the color line to be removed
runs along the fermion side of the loop.  This requirement eliminates
half the diagrams, namely the `right' diagrams contributing to
$A_n^R$, leaving precisely the desired `left' diagrams contributing to
$A_n^L$.
As a particular example, the two
routings of the gluino through the diagrams depicted in
\fig\GluinoFigure{a,b} contribute to the color structure~(\TraceA).
But in fig.~\use\GluinoFigure{b} the color line
running directly between the two gluinos must run along the gluon side
of the loop in order to generate~(\TraceA); therefore this diagram
should be dropped in converting to $\bar{q}qg\ldots g$, while
fig.~\use\GluinoFigure{a} should be kept.  This shows that replacing
$A_{n;1}$ by $A_n^L$ on the right-hand side of eqn.~(\use\sublanswer)
is the correct prescription for the `parent' diagrams.  However, we
must again show that all unwanted diagrams with attached trees cancel
in the permutation sum.  As was the case for the adjoint
representation case, in each term on the right-hand side of
eqn.~(\use\sublanswerfund) we are including diagrams which do not
belong on the left-hand side, because the color flow they represent
does not allow them to contribute.  The argument is similar to the one
for the adjoint case.  If diagrams contain trees with leaves labeled
by indices from both sets $\{\alpha \}$ and $\{
\beta \}$, then such diagrams cancel in the permutation sum exactly as
for the adjoint case.  These cancellations are due to the antisymmetry
of the color-stripped vertices, and are thus independent of the color
representation of the fermions.
Note that `left' diagrams cancel against `left', and `right'
against `right'.
This completes the conversion of the adjoint
formula~(\sublanswer) to the fundamental formula~(\sublanswerfund),
excluding the contributions of closed fermion or scalar loops.

\LoadFigure\GluinoFigure{\baselineskip 13 pt
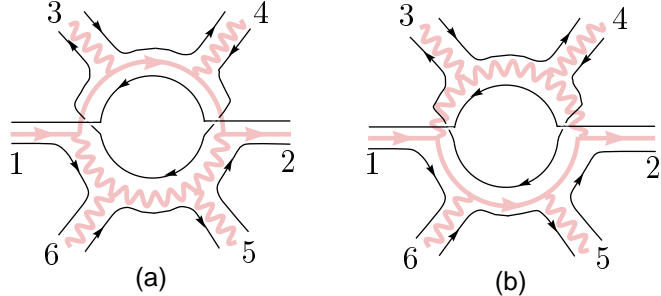
\noindent\narrower\ninerm Both `left' (a) and `right' (b) diagrams
contribute to adjoint representation gluino partial amplitudes, but
only the `left' diagram contributes for fundamental representation quarks.}
{\epsfysize 1.6truein}
{Gluino.psd}{}

We turn next to the contributions of closed fermion or scalar loops in
the fundamental representation.  In this case, neglecting the $-1/N_c$
term in the $SU(N_c)$ Fierz identity~(\SUNFierz) leads only to the
$n_{f,s}$-dependent terms in $A_{n;1}$ in eqn.~(\Anoneformula), and
does not yield a contribution to $A_{n;j>1}$.  Including the $-1/N_c$,
or $U(1)$ subtraction, term in the gluon propagator connecting the
external $\bar{q}q$ line to the fermion/scalar loop decouples the
color flow for the tree containing $\bar{q}q$ from the loop color
flow.  However, the color-stripping argument can still be applied to
these terms, if we allow the stripped color line to propagate down the
`photon', around the loop, and back again along the `photon'.  We
start with the $\gluino\gluino g\ldots g$ double-line configuration
where a color line starts at gluino 2, flows down the gauge boson line
connecting the external fermion legs to an adjoint fermion (or scalar)
loop, flows around the loop and then returns through the gauge boson
line to wind up at gluino 1, as shown in \fig\PhotonFigure{a}.  We
remove this color line, as shown in fig.~\use\PhotonFigure{b}, to
obtain the $U(1)$ subtraction contribution for $\bar{q}qg\ldots g$.
In this case it is the `right' type diagrams that have the correct
color flow to be strippable, so the appropriate formula (including the
Fierz factor of $-1/N_c$) is
$$
\eqalign{
 &A_{n;j}(1_\qb,2_q; 3,\ldots,j+1;j+2,j+3,\ldots,n)
   |_{n_{f,s}\ {\rm terms}} \cr
 &\qquad =\
 (-1)^{j-1} \sum_{\sigma\in COP\{\alpha\}\{\beta\}}
    \Biggl[ - {\nf\over N_c}
    A_n^{R,[1/2]} \L \sigma(1_{\bar{q}},2_q,3,\ldots,n) \R
   - {\ns\over N_c}
    A_n^{R,[0]} \L \sigma(1_{\bar{q}},2_q,3,\ldots,n) \R \Biggr]\ . \cr}
\eqn\sublnfns
$$
The cancellation of unwanted diagrams with attached trees from
the right-hand side of eqn.~(\sublnfns)
works as in the no-$n_{f,s}$ case.

\LoadFigure\PhotonFigure{\baselineskip 13 pt
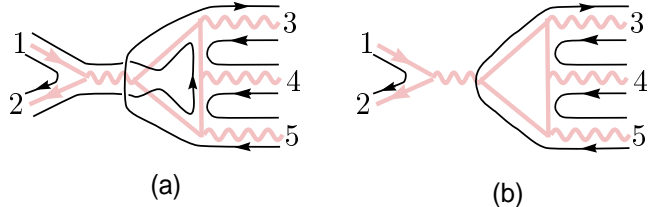
\noindent\narrower\ninerm Color stripping the gluon line connecting the
external fermion line to the fermion loop in (a)
produces the `photon' subtraction diagram in (b).}
{\epsfysize 1.1truein}
{Photon.psd}{}

Thus the final formula for subleading-color $\bar{q}qg\ldots g$ partial
amplitudes in terms of primitive amplitudes is the one given in
eqn.~(\use\subltotal).

%%%%%%%%%%%%%%%%%%%%%%%%%%%%%%%%%%%%%%%%%%%%%%%

\appendix{Four-Point Amplitudes}
\tagappendix\FourPointAppendix

In this appendix, we collect the four-gluon and two-quark two-gluon
amplitudes, needed for checking the collinear limits of the two-quark
three-gluon amplitudes.  These amplitudes agree with the results of
KST~[\use\KunsztFourPoint].  Note, however, that these authors used a
different color decomposition and overall sign convention
than used in this paper.

We begin by listing all tree amplitudes that
appear in the collinear limits, eqn.~(\use\loopsplit), of the five-point
amplitudes presented in this paper,
%%%%%%%%%%%%%%%%%%%%%%
% gggg and ffgg tree %
%%%%%%%%%%%%%%%%%%%%%%
$$
\eqalign{
A_4^{\rm tree}(1^-,2^-,3^+,4^+) &=
i {\spa1.2^4 \over \spa1.2\spa2.3\spa3.4\spa4.1}  \, , \cr
A_4^{\rm tree}(1^-,2^+,3^-,4^+) &=
i{\spa1.3^4\over \spa1.2\spa2.3\spa3.4\spa4.1} \, , \cr
%
%treeqqggmpmp :=
A_4^{\rm tree} (1_{\bar{q}}^-,2_q^+,3^-,4^+) &=
i{\spa1.3^3\spa2.3 \over \spa1.2\spa2.3\spa3.4\spa4.1} \, , \cr
%
%treeqqggmppm :=
A_4^{\rm tree} (1_{\bar{q}}^-,2_q^+,3^+,4^-) &=
i{\spa1.4^3\spa2.4\over \spa1.2\spa2.3\spa3.4\spa4.1} \, , \cr
%treeqgqgmmpp
A_4^{\rm tree} (1_\qb^-, 2^-, 3^+_q, 4^+) &=
 i {\spa1.2^3 \spa3.2 \over \spa1.2 \spa2.3\spa3.4\spa4.1} \, . \cr
}\eqn\TreeAmpl
$$

Now consider the one-loop four-point amplitudes, beginning with the
four-gluon amplitudes.  Amplitudes with only external gluons may
be decomposed in terms of contributions to supersymmetric multiplets
$$
A_{n;1} (1,2,\ldots,n) =
  A_n^g + \Bigl(4-{n_{\! f}\over N_c} \Bigr) A_n^f
   + \Bigl(1+ {n_s \over N_c} - {n_{\! f}\over N_c} \Bigr) A_n^s\ ,
\eqn\ngluonsusydecomp
$$
where $A^g$ is the contribution of an $N=4$ multiplet, $-A^f$ the
contribution of an $N=1$ chiral multiplet, and $A^s$ the contribution
of a complex scalar.

%%%%%%%%
% pppp %
% mppp %
%%%%%%%%

The finite four-gluon amplitudes are
$$
\eqalign{
%ggggpppp :=
A_4^g (1^+, 2^+, 3^+, 4^+) &= 0 \, , \cr
A_4^f (1^+, 2^+, 3^+, 4^+) &= 0 \, , \cr
A_4^s (1^+, 2^+, 3^+, 4^+) &=
 {i\over 48 \pi^2} {s_{12}s_{23}\over\spa1.2\spa2.3\spa3.4\spa4.1} \, , \cr
%
%ggggmppp :=
A_4^g (1^-, 2^+, 3^+, 4^+) &= 0 \, , \cr
A_4^f (1^-, 2^+, 3^+, 4^+) &= 0 \, , \cr
A_4^s (1^-, 2^+, 3^+, 4^+) &=
{i \over 48 \pi^2}
           {\spa2.4\spb2.4^3\over \spb1.2\spa2.3\spa3.4\spb4.1} \, .\cr
}
\eqn\ggggfinite
$$
The tree amplitudes vanish for these helicity configurations.

For the amplitudes containing infrared and ultraviolet
singularities, we further decompose the amplitude into $V$ and $F$
pieces of the sort given in eqn.~(\use\VFsplit)
$$
A^x_{4} = \cg (V^x A^{\rm tree}_4 + i F^x) \, , \hskip 2 cm
x = g, f, s \, .
\anoneqn
$$
The universal functions
$$
\eqalign{
V^g_{4, \rm gluon} &= - {2\over \e^2}  \LB \L {\mu^2 \over -s_{12}}\R^\e
                +\L {\mu^2 \over -s_{23}}\R^\e \RB
                 + \ln^2\Bigl({-s_{12}\over -s_{23}}\Bigr) + \pi^2 \, , \cr
}
\eqn\vfourgluon
$$
appear in the four-point amplitudes with infrared divergences.

For the four-gluon amplitude $A_{4;1} (1^-, 2^-, 3^+, 4^+)$ we have
$$
\eqalign{
V^g &= V^g_{4, \gluon} \, ,
\hskip 3 truecm F^g = 0  \, ,  \cr
V^f &= -{1\over \eps} \L \mu^2 \over -s_{23}\R^\e - 2 \, ,
\hskip 1.4 truecm F^f  = 0 \, , \cr
V^s &= - {V^f\over 3} + {2\over 9} \, ,
\hskip 2.68 truecm F^s = 0 \, .\cr
}\eqn\ggggmmpp
$$

For $A_{4;1} (1^-, 2^+, 3^-, 4^+)$ we have
$$
\eqalign{
V^g &= V^g_{4, \gluon} \, ,
\hskip 3 truecm F^g = 0  \, ,  \cr
V^f &= -{1\over 2 \e} \L \L \mu^2 \over -s_{12}\R^\e +
\L\mu^2 \over -s_{23}\R^\e \R - 2 \, , \cr
V^s &= - {V^f\over 3} + {2\over 9} \, , \cr
F^f &= {\spa1.3^4\over \spa1.2\spa2.3\spa3.4\spa4.1}
\biggl[ - {s_{12}s_{23} \over 2 s_{13}^2}
          \Bigl(\ln^2\Bigl({-s_{12} \over -s_{23}}\Bigr) + \pi^2\Bigr)
          + {1\over 2}{s_{12}-s_{23}\over s_{13}}
                  \ln\Bigl({-s_{12}\over -s_{23}}\Bigr)\biggr]\, , \cr
F^s &= {\spa1.3^4\over \spa1.2\spa2.3\spa3.4\spa4.1}
\biggl( -{s_{12}s_{23} \over s_{13}^2} \biggr)
   \biggl[ - {s_{12}s_{23} \over s_{13}^2 }
     \Bigl(\ln^2\Bigl({-s_{12}\over -s_{23}}\Bigr) +\pi^2\Bigr) \cr
\null & \hskip 4.3 cm
             + {s_{12}-s_{23}\over s_{13}}
              \L 1 + {s_{13}^2 \over 6 s_{12} s_{23} } \R
                    \ln\Bigl({-s_{12}\over -s_{23}} \Bigr)
             + 1  \biggr] \, . \cr
}
\eqn\ggggmpmp
$$

%\bar{q}qgg:     (A_{4;1} terms)
%%%%%%%%
% ffpp %
%%%%%%%%

For the finite amplitudes with two quarks and two gluons we have
$$
\eqalign{
A_4^\susy(1^-_\qb, 2^+_q, 3^+, 4^+) &= 0 \, ,\cr
A_4^L(1^-_\qb, 2^+_q, 3^+, 4^+) &=
-{i\over 32 \pi^2} {\spa1.2\spb2.4 \over \spa2.3\spa3.4}
+ A_4^s (1^-_\qb, 2^+_q, 3^+, 4^+) \, ,\cr
A_4^s(1^-_\qb, 2^+_q, 3^+, 4^+) &=
-{i\over 48 \pi^2} {s_{23}\over s_{12}}
            {\spa1.2\spb2.4 \over \spa2.3\spa3.4} \, ,\cr
A_4^f(1^-_\qb, 2^+_q, 3^+, 4^+) &= 0 \, .\cr
}
\eqn\qqggffpp
$$

%%%%%%%%
% fpfp %
%%%%%%%%

$$
\eqalign{
A_4^\susy(1_\qb^-,2^+,3_q^+,4^+) & = 0 \, , \cr
A_4^L(1_\qb^-,2^+,3_q^+,4^+) & =
-{i\over 32 \pi^2} {\spa1.3\spb2.4 \over \spa2.3\spa3.4} \, , \cr
A_4^s(1_\qb^-,2^+,3_q^+,4^+) & = 0 \, , \cr
A_4^f(1_\qb^-,2^+,3_q^+,4^+) & = 0 \, . \cr
}\eqn\qgqgfpfp
$$

%%%%%%%%
% ffmp %
%%%%%%%%

For the helicities containing infrared and ultraviolet singularities
we decompose the amplitude as in the five-point case
(eqn.~(\use\VFsplit)), and again we have $V^s = V^f = 0$.
For $A_{4} (1^-_\qb, 2^+_q, 3^-, 4^+)$ we have
$$
\eqalign{
V^\susy  &=  V^g_{4, \gluon}
 - {3\over 2\e} \L  \L{\mu^2 \over -s_{12} }\R^\e
              +     \L{\mu^2 \over -s_{23} }\R^\e \R  -  6 \, , \cr
V^L &= \LB V^g_{4, \gluon} + {1\over \e^2} \L{\mu^2 \over - s_{12}} \R^\e
       \RB
 - {3\over 2 \e} \L {\mu^2\over -s_{12}} \R^\e - {5\over 2} \, ,  \cr
}\eqn\qqggffmpV
$$
$$
\eqalign{
F^\susy &=
 3 \, {\spa1.3^3\spa2.3 \over \spa1.2\spa2.3\spa3.4\spa4.1}
\biggl[ - {s_{12}s_{23} \over 2 s_{13}^2}
               \Bigl(\ln^2\Bigl({-s_{12} \over -s_{23}}\Bigr) + \pi^2\Bigr)
          + {1\over 2}{s_{12}-s_{23}\over s_{13}}
                  \ln\Bigl({-s_{12}\over -s_{23}}\Bigr)\biggr] \, , \cr
F^L  &= {\spa1.3^3\spa2.3 \over \spa1.2\spa2.3\spa3.4\spa4.1}
        {s_{12} \over 2 s_{13}}
 \biggl[ \L 3 - {2 s_{23} \over s_{13}} \R
         \ln\Bigl({-s_{12}\over -s_{23}} \Bigr)
  \ +\ \L {s_{12}^2 \over s_{13}^2}  - {3 s_{23} \over s_{13} } \R
           \Bigl(\ln^2 \Bigl({-s_{12}\over -s_{23}} \Bigr) + \pi^2 \Bigr)
           + 1 \biggr] \, ,\cr
F^s &= 0 \, ,\cr
F^f &= 0 \, .\cr
}\eqn\qqggffmpF
$$

%%%%%%%%
% ffpm %
%%%%%%%%

For $A_{4} (1^-_\qb, 2^+_q, 3^+, 4^-)$ we have
$$
\eqalign{
V^\susy &= V^g_{4, \gluon}
- {3\over\e} \L {\mu^2 \over -s_{12} }\R^\e - 6 \, , \cr
V^L &= \LB V^g_{4, \gluon} + {1\over \e^2} \L{\mu^2 \over -s_{12}} \R^\e\RB
 - {3\over 2 \e} \L {\mu^2\over -s_{12}} \R^\e - {5\over 2} \, ,\cr
}\eqn\qqggffpmV
$$
$$
\eqalign{
F^\susy &= 0 \, , \cr
F^L &=  {\spa1.4^3\spa2.4\over \spa1.2\spa2.3\spa3.4\spa4.1}
{s_{12} \over 2 s_{13}} \LB\ln^2\Bigl({-s_{12} \over -s_{23}} \Bigr)
+ \pi^2 \RB \, ,\cr
F^s &= 0 \, ,  \cr
F^f &= 0 \, . \cr
}
\eqn\qqggffpmF
$$

%%%%%%%%
% fmfp %
%%%%%%%%

For $A_4 (1^-_\qb, 2^-, 3_q^+, 4^+)$ we have
$$
\eqalign{
V^\susy &= V^g_{4, \gluon}
        - {3\over\e} \L {\mu^2\over -s_{23}} \R^\e - 6 \,, \cr
V^L &= {1\over 2} V^\susy \,,\cr
}\eqn\qgqgfmfpV
$$
$$
\eqalign{
F^\susy &= 0 \, ,\hskip 2 cm
F^L =  0 \, ,\cr
F^s &= 0 \, ,   \hskip 2 cm
F^f = 0 \, . \cr
}
\eqn\qgqgfmfpF
$$
The remaining amplitude, $A_4 (1^-_\qb, 2^+, 3_q^+, 4^-)$,
may be obtained via a reflection,
$$
A_4 (1^-_\qb, 2^+, 3_q^+, 4^-) =
A_4 (1^-_\qb, 4^-, 3_q^+, 2^+) \, .
\eqn\qgqgreflection
$$

%%%%%%%%%%%%%%%%%%%%%%%%%

\appendix{Collinear Limit Checks}
\tagappendix\CollinearAppendix

In this appendix, we illustrate the use of collinear limits in
verifying the correctness of explicitly calculated amplitudes,
including signs and normalizations.  As the momenta of two
color-adjacent external
legs become collinear, the amplitudes factorize into sums of lower
point amplitudes multiplied by `splitting amplitudes'.  The
constraints provided by the collinear limits are sufficiently
restrictive that they can be used to construct ans\"atze for higher-point
amplitudes based on known lower point amplitudes
[\use\AllPlus,\use\SusyFour].  The collinear-limit checks may also be
used in numerical programs, when converting the amplitudes presented
in this paper to physical cross-sections.

At one loop, the collinear limits of color-ordered one-loop QCD
amplitudes have the form [\use\AllPlus,\use\SusyFour]
$$
\eqalign{
A_{n}^{\rm loop} \mathop{\longrightarrow}^{a \parallel b}
\sum_{\lambda=\pm}  \biggl(
  \Split^{\rm tree}_{-\lambda}(a^{\lambda_a},b^{\lambda_b})\,
&
      A_{n-1}^{\rm loop}(\ldots(a+b)^\lambda\ldots)
\cr
&  +\Split^{\rm loop}_{-\lambda}(a^{\lambda_a},b^{\lambda_b})\,
      A_{n-1}^{\rm tree}(\ldots(a+b)^\lambda\ldots) \biggr) \;,
\cr}
\eqn\loopsplit
$$
where $k_a \rightarrow z P$ and $k_b \rightarrow (1-z) P$,
with $P=k_a+k_b$, $P^2=s_{ab} \to 0$.
The tree and loop splitting amplitudes, $\Split^{\rm tree}_{-\lambda}$
and $\Split^{\rm loop}_{-\lambda}$, behave as $1/\sqrt{s_{ab}}$
in this limit.
This formula holds for any of the primitive amplitudes
--- which at loop level may carry the additional labels
$J=1,1/2,0$ (for $n$-gluon amplitudes), or $x=\susy,L,s,f$
(for two-quark $(n-2)$-gluon amplitudes) ---
as well as for the leading-color amplitudes $A_{n;1}$,
provided that one assigns the correct labels to all `loop' quantities
in~(\use\loopsplit).
A tabulation of the splitting amplitudes appearing in
massless QCD computations was given in appendix~B of
ref.~[\use\SusyFour].
The results were given in terms of leading-color
partial amplitudes $A_{n;1}$ rather than primitive amplitudes;
below we shall convert the results to the latter form.
We describe the collinear behavior of amplitudes before subtraction
of the ultraviolet pole.
A proof of the universality of the splitting amplitudes,
limited to external gluons and scalar loops, was outlined in
refs.~[\use\AllPlus,\use\GordonConf]; a more complete treatment will
be given elsewhere.
The power of the collinear-limit constraint arises from the fact
that relationship~(\use\loopsplit) must hold in every channel.

In order to use the splitting amplitudes in appendix~B of
ref.~[\use\SusyFour] for the primitive amplitudes presented here,
we must take into account
that the overall sign convention used in ref.~[\use\SusyFour]
for the helicity amplitudes $A(1_\qb^-,2_q^+,3,\ldots,n)$ is opposite
to that used for primitive amplitudes in the present paper.
(The overall sign convention for $A(1_\qb^+,2_q^-,3,\ldots,n)$ is
the same.)
With the former sign convention, one cannot relabel $\qb \to q$,
$q \to \qb$ in amplitudes without introducing extra signs.
Equivalently, one cannot directly interpret the tree-level quark
amplitudes as gluino amplitudes, because the extra constraints on
gluino amplitudes imposed by the Majorana nature of the gluino are
not satisfied with the former choice of sign.
The change in amplitude conventions implies a sign change in the second
and fourth tree splitting amplitudes in eqn.~(B.5) of ref.~[\use\SusyFour].
With these sign changes the tree splitting amplitudes in that appendix
may be used for the primitive amplitudes presented in this paper;
furthermore, one may relabel $\qb \to q$, $q \to \qb$ in the splitting
amplitudes to get the additional orderings
$\Split_{-\lambda}(\qb,g)$ and $\Split_{-\lambda}(g,q)$
which appear in limits of the primitive amplitudes.

For all splitting amplitudes which do not vanish at tree level,
the proportionality constant $r_S$ is defined by
$$
 \Split^{\rm loop}_{-\lambda}(a^{\lambda_a},b^{\lambda_b})
 \ =\ \cg
 \times \Split^{\rm \tree}_{-\lambda}(a^{\lambda_a},b^{\lambda_b})
 \times r_S(-\lambda,a^{\lambda_a},b^{\lambda_b})\;.
\eqn\genrsdef
$$
For any collinear limit of a supersymmetric primitive amplitude
$A_n^\susy$, $r_S$ is given by the function $r_S^\susy$
given in eqn.~(B.13) of ref.~[\use\SusyFour].

For the non-supersymmetric primitive amplitudes ($x=L,s,f$)
we first discuss the case $g\rightarrow gg$.
Because the fermion line is routed on a definite side of the loop
in these components,
the loop splitting amplitudes depend on whether the pair of adjacent
collinear gluons $a,b$ are between $\qb$ and $q$
($A_n^x(1_\qb,\ldots,a,b,\ldots,2_q,\ldots,n)$)
or between $q$ and $\qb$
($A_n^x(1_\qb,\ldots,2_q,\ldots,a,b,\ldots,n)$).
In the first case the $x=L,s,f$ loop splitting amplitudes all
vanish.
In the second case they are given by the pure-gluon ($J=1$) contribution
for $x=L$, the negative of the sum of the $J=1/2$ and $J=0$ contributions
for $x=f$, and the $J=0$ contribution for $x=s$:
$$
\Split^{x}_{+}(a^{+},b^{+})\ =\ \left\{\eqalign{0,&\qquad x=\susy,f\;;\cr
 \Split^{[1]}_{+}(a^{+},b^{+}),&\qquad x=L, s\;\cr}\RP
\anoneqn$$
(this is the one case where the tree splitting amplitude vanishes),
and
$$\eqalign{
r_S^L(\pm,a,b)\ &=\ r_S^{[1]}(\pm,a,b)\;,\cr
r_S^f(\pm,a,b)\ &=\ 0\;,\cr
r_S^s(\pm,a,b)\ &=\ r_S^{[0]}(\pm,a,b)\;.\cr
}\anoneqn$$
The quantities $\Split^{[1]}_{+}(a^+,b^+)$ and $r_S^{[J]}$ are given in
eqns.~(B.8) and (B.9) of ref.~[\use\SusyFour].

The $r_S$ functions for the non-supersymmetric $g\rightarrow \qb q$
splitting amplitudes are
$$\hskip-10pt\eqalign{
r_S^{L}(\pm,\bar{q}^{\mp},q^{\pm})\ &=\
  -{1\over\e^2} \LB \L{\mu^2\over z (1-z)(-s_{\qb q})}\R^{\e}
                  - \L{\mu^2\over (-s_{\qb q})}\R^{\e} \RB \cr
&\qquad\qquad
  + {13\over6\e} \L{\mu^2\over (-s_{\qb q})}\R^{\e}
  + 2 \ln(z)\,\ln(1-z) - {\pi^2\over6} + {83\over18}\;,\cr
r_S^{s}(\pm,\bar{q}^{\mp},q^{\pm})\ &=\
  - {1\over3\e} \L{\mu^2\over (-s_{\qb q})}\R^{\e}
  - {8\over9}\;,\cr
r_S^{f}(\pm,\bar{q}^{\mp},q^{\pm})\ &=\
    {1\over\e} \L{\mu^2\over (-s_{\qb q})}\R^{\e} + 2\;.\cr
}\anoneqn$$

For the non-supersymmetric $q\rightarrow q g$ splitting amplitudes,
one obtains different results depending on whether the gluon
(denoted by $a$)
is before or after the quark, with respect to the clockwise ordering
of the vertices.  The difference is again due to the routing of the
fermion line in the primitive amplitudes.
We have
$$\eqalign{
r_S^{L}(q^{-},a^{+})\ &=\ r_S^{L}(q^{+},a^{-})
  \ =\ f^L(1-z,s_{qa})\ ,\cr
r_S^{L}(q^{-},a^{-})\ &=\ r_S^{L}(q^{+},a^{+})
  \ =\  f^L(1-z,s_{qa}) + {1-z\over2}\ ,\cr
r_S^{L}(a^{+},\qb^{+})\ &=\ r_S^{L}(a^{-},\qb^{-})
  \ =\ f^L(z,s_{a\qb}) + {z\over2}\ ,\cr
r_S^{L}(a^{-},\qb^{+})\ &=\ r_S^{L}(a^{+},\qb^{-})
  \ =\ f^L(z,s_{a\qb})\ ,\cr
r_S^{L}(a^{+},q^{-})\ &=\ r_S^{L}(a^{-},q^{+})
  \ =\ f^R(z,s_{aq})\ ,\cr
r_S^{L}(a^{-},q^{-})\ &=\ r_S^{L}(a^{+},q^{+})
  \ =\  f^R(z,s_{aq}) - {z\over2}\ ,\cr
r_S^{L}(\qb^{+},a^{+})\ &=\ r_S^{L}(\qb^{-},a^{-})
  \ =\ f^R(1-z,s_{\qb a}) - {1-z\over2}\ ,\cr
r_S^{L}(\qb^{+},a^{-})\ &=\ r_S^{L}(\qb^{-},a^{+})
  \ =\ f^R(1-z,s_{\qb a})\ ,\cr
}\anoneqn
$$
where the functions $f^L$ and $f^R$ correspond
to the leading- and subleading-color parts of the function $f$ defined
in eqn.~(B.11) of ref.~[\use\SusyFour],
$$\eqalign{
  f^L(z,s)\ &=\ - {1\over\e^2}\L{\mu^2\over z(-s)}\R^{\e} - \Li_2(1-z)
  \;, \cr
  f^R(z,s)\ &=\ - {1\over\e^2}\L{\mu^2\over(1-z)(-s)}\R^{\e}
     + {1\over\e^2}\L{\mu^2\over(-s)}\R^{\e} - \Li_2(z)\;, \cr
  f(z,s)\ &=\ f^L(z,s)-{1\over N_c^2} f^R(z,s)\;.\cr
}\anoneqn$$
The $x=s,f$ parts of the $q\rightarrow q g$ splitting amplitudes
all vanish.

%%%%%%%%%%%%%%%%%

As an illustration of the collinear limits consider the
amplitude $A_5^f(1_{\bar q}^-, 2_q^+, 3^-, 4^+, 5^+)$ given by
$$
\eqalign{
A_5^f(1_{\bar q}^-, 2_q^+, 3^-, 4^+, 5^+)\ &\equiv\
\ i \, \cg \, F^f
\ =\ -i \, \cg \, {\spa1.3^2 \spb4.5 \over \spa1.2 \spa4.5  }
{\ln\Bigl( {-s_{12} \over -s_{34}} \Bigr) \over s_{34} - s_{12}} \; .\cr
}
\eqn\CollExampleA
$$
The collinear limits of $A_5^f$ may be considered independently of
all other primitive amplitudes because they are separately gauge
invariant; in QCD amplitudes $A_5^f$ enters with a coefficient
proportional to the number of fermions $n_{\! f}$, which distinguishes
it from the other primitive amplitudes.
(Although $A_5^s$ also enters QCD amplitudes with coefficients
containing $n_{\! f}$, it does so only in the combination
$n_s - n_{\! f}$ and it may therefore be treated independently.)

First consider the limit as gluon 3 becomes collinear with gluon 4
($3\parallel 4$), and also the $4\parallel 5$ collinear limit.
In either of these cases, eqn.~(\use\CollExampleA) does not contain
a collinear singularity.
Compare this result to the expectation from eqn.~(\use\loopsplit).
These limits are particularly simple to analyze because the
$A_4^f$ amplitudes with two quarks and two gluons vanish for all
helicities, $A_4^f(1_{\bar q}, 2_q, 3, 4) = 0$
(see eqns.~(\use\qqggffpp), (\use\qqggffmpF) and (\use\qqggffpmF)).
Additionally, the loop splitting amplitudes
vanish for all helicities, $\Split^f_{\pm} (a^\pm, b^\pm) = 0$, where
$\Split^f_{\pm}\equiv -(\Split^{[1/2]}_{\pm}+\Split^{[0]}_{\pm})$,
and $\Split^{[1/2]}_{\pm}$, $\Split^{[0]}_{\pm}$ are defined in
appendix B of ref.~[\use\SusyFour].
Thus, all contributions to the right-hand side of eqn.~(\use\loopsplit)
vanish for the $3\parallel 4$ and $4\parallel 5$ collinear limits,
in agreement with the lack of collinear singularities in these
channels in the amplitude~(\use\CollExampleA).

The $1_\qb\parallel 2_q$ collinear limit is a bit less trivial
since the expression~(\use\CollExampleA) does contain a collinear pole.
Using the explicit value of $A_5^f$ in eqn.~(\use\CollExampleA)
we have
$$
A_5^f (1_{\bar q}^-, 2_q^+, 3^-, 4^+, 5^+)
\ \mathop{\longrightarrow}^{1 \parallel 2}\
-i \, \cg \, {z \over \spa1.2} {\spa P.3^2 \spb4.5 \over \spa4.5}
 {\ln\Bigl({-P^2 \over -s_{34} }\Bigr) \over s_{34}}  \, ,
\eqn\CollLim
$$
where we have taken $k_1 = z P$ and $k_2 = (1-z) P$.
Now compare this to the result obtained from the collinear formula
(\use\loopsplit).
The necessary four-gluon amplitudes are
$$
\eqalign{
A_4^{\rm tree}(P^+, 3^-, 4^+, 5^+) & = 0 \, , \cr
A_4^{\rm tree}(P^-, 3^-, 4^+, 5^+) & =
i {\spa P.3^4 \over \spa P.3 \spa3.4 \spa4.5 \spa5.P} \, , \cr
A_4^f(P^+, 3^-, 4^+, 5^+) & = 0 \, , \cr
A_4^f(P^-, 3^-, 4^+, 5^+) & =
-\cg \L {1\over \e} \L {\mu^2 \over -s_{34}} \R^\e +2 \R
A_4^{\rm tree} (P^-, 3^-, 4^+, 5^+) \, , \cr
}
\anoneqn
$$
while the relevant splitting functions are
$$
\eqalign{
\Split_+^{\rm tree} (\bar q^-, q^+) & = {z \over \spa{\bar q}.{q}} \; ,
\cr
\Split_+^{f} (\bar q^-, q^+) & = \cg
\L {1\over \e} \L {\mu^2 \over -s_{\qb q}} \R^\e +2 \R
{z \over \spa{\bar q}.{q}} \; . \cr
}\anoneqn
$$
Plugging these results into eqn.~(\use\loopsplit) we obtain
for the non-vanishing terms
$$
\eqalign{
A_5^f(& 1_{\bar q}^-, 2_q^+, 3^-, 4^+, 5^+) \cr
&\mathop{\longrightarrow}^{1 \parallel 2}\
 \Split_+^{\rm tree} (1^-_{\bar q}, 2^+_q)\  A^f_4(P^-, 3^-, 4^+, 5^+)
+ \Split_+^f (1^-_{\bar q}, 2^+_q)\ A^{\rm tree}_4(P^-, 3^-, 4^+, 5^+) \cr
& = -i \, \cg \, {z \over \spa1.2}
\ln\Bigl( {-P^2 \over -s_{34} } \Bigr)
 {\spa{P}.3^4 \over \spa{P}.3\spa3.4\spa4.5\spa5.P} \cr
}\anoneqn
$$
which reproduces eqn.~(\use\CollLim), after spinor helicity
simplifications.

One can continue in this way, systematically verifying that
the collinear limits in all channels are correct.  We have done
so for all amplitudes presented in this paper.

In checking the collinear limits, the behavior of most of the
functions appearing in the amplitudes is straightforward to obtain.
We present here only the one function with a slightly complicated
limit.
We take the collinear limit of two color-adjacent momenta $k_c$ and
$k_{c+1}$; denoting the sum by $P$, the momentum fraction $z$
satisfies $k_c = zP$ and $k_{c+1} = (1-z)P$ in the limit.
Furthermore, we will denote the limit of $s_{c+2,c-2}$ by $s$ and the
limit of $s_{c-2,c-1}$ by $t$.
With these definitions we have
$$
\eqalign{
{\Ls_2\L{-s_{c-2,c-1}\over-s_{c,c+1}},\,{-s_{c+2,c-2}\over-s_{c,c+1}}\R
 \over s_{c,c+1}^3}
\ &\mathop{\longrightarrow}^{c \parallel c+1}\
  {1\over 2 s t} {1\over s_{c,c+1}} \cr
  &\hskip 5mm
   + {\ln^2\L{-s\over -t}\R+\pi^2 \over 2(s+t)^3}
                +{\ln\L{-s_{c,c+1}\over -t}\R}
                      \LB{1\over t (s+t)^2}+{1\over 2 t^2 (s+t)}\RB\cr
  &\hskip 5mm
                +{\ln\L{-s_{c,c+1}\over -s}\R}\;
                      \LB{1\over s (s+t)^2}+{1\over 2 s^2 (s+t)}\RB\cr
  &\hskip 5mm
   +{1\over 2 t^2 (s+t)}+{1\over 2 s^2 (s+t)}+{1\over 2 st (s+t)}\ .\cr }
\anoneqn
$$

%%%%%%%%%%%%%%%%%%%%

\appendix{Mixed Photon-Gluon Amplitudes}
\tagappendix\MixedAppendix

The same primitive amplitudes used to construct the two-quark
$(n-2)$-gluon amplitudes can also be used to construct amplitudes with
one to $(n-2)$ photons replacing gluons.
The two-quark $(n-2)$-gluon color
decompositions~(\use\gentreeqqdecomp) and (\use\genqqdecomp)
are valid for external gauge bosons in $U(N_c)$ as well as $SU(N_c)$,
because the Fierz subtraction term in the gluon propagator
(eqn.~(\use\SUNFierz)) does not contribute unless the gluon
is sandwiched between two fundamental representation lines.
Thus partial amplitudes with photons may be obtained by substituting the
photon generator matrix (which is proportional to the identity matrix)
into the appropriate color decomposition formula,
and grouping together terms with the same color structure.
We illustrate the construction explicitly for
two-quark one-photon $(n-3)$-gluon amplitudes.

At tree level, amplitudes with one photon have a color decomposition
similar to that of pure nonabelian ones,
$$
{\cal A}^{1\gamma\,\tree}_n = Q{\sqrt{2} e} g^{n-3}
   \sum_{\sigma\in S_{n-3}}
  \L T^{a_{\sigma(3)}}\cdots T^{a_{\sigma(n-1)}}\R_{i_1}{}^{\ib_2}
   A^{1\gamma\,\tree}_n(1_\qb,2_q;\sigma(3),\ldots,\sigma(n-1);n)\; ,
\eqn\treephotondecomp
$$
where $e$ is the QED coupling constant, and $Q$ is the charge of
the quark.  The photon is taken to be the last leg, $n$.
The partial amplitude $A^{1\gamma\,\tree}_n$ is related by a `decoupling'
equation~[\use\ManganoReview] to the pure nonabelian partial amplitudes
$A^\tree_n$,
$$\eqalign{
A^{1\gamma\,\tree}_n(1_\qb,2_q;3,\ldots,n-1;n)
 \ =\ A^\tree_n(1_\qb,2_q;n,3,4,\ldots,n-1)
          &+ A^\tree_n(1_\qb,2_q;3,n,4,\ldots,n-1) \cr
 + \cdots &+ A^\tree_n(1_\qb,2_q;3,4,\ldots,n-1,n) \; . \cr
}\eqn\treephoton
$$
Equation~(\use\treephoton) is obtained by substituting the photon
generator matrix $T^{a_n} \propto {\bf 1}$ into
eqn.~(\use\gentreeqqdecomp) and collecting terms.
Note that unwanted diagrams on the right-hand side of~(\use\treephoton)
--- those coupling the photon to gluon lines --- cancel out in the sum.

We can perform similar decompositions at one loop.
We keep only $\Ord(e)$ contributions; we do not include the $\Ord(e^3)$
contributions where an internal photon line is exchanged between two
charged lines.
If we set aside the $\nf$- and $\ns$-dependent pieces,
then the color factors are again simply the color factors for
the two-quark $(n-3)$-gluon amplitude~(\use\grqq),
$$
\hskip - .4 cm
\LP {\cal A}^{1\gamma}_n\RV_{\nf,\ns=0} = Q{\sqrt{2} e} g^{n-1}
   \sum_{j=1}^{n-2} \sum_{\sigma\in S_{n-3}/S_{n-1;j}}
    \Gr_{n-1;j}^{(\qb q)}(\sigma(3,\ldots,n-1))\
  \LP A_{n;j}^{1\gamma}(1_\qb,2_q;\sigma(3,\ldots,n-1);n)
   \RV_{\nf,\ns=0}\ ,
\eqn\loopphotondecompa
$$
and the partial amplitudes with one photon are again
given by sums over the two-quark rest-gluon partial amplitudes,
inserting the photon in all inequivalent gluon locations,
$$\eqalign{
& \LP A^{1\gamma}_{n;j}(1_\qb,2_q;3,\ldots,j+1;j+2,\ldots,n-1;n)
   \RV_{\nf,\ns=0} \cr
& =\
 A_{n;j+1}(1_\qb,2_q;n,3,\ldots,j+1;j+2,\ldots,n-1)
+ \cdots + A_{n;j+1}(1_\qb,2_q;3,\ldots,n,j+1;j+2,\ldots,n-1) \cr
&\qquad
+ A_{n;j}(1_\qb,2_q;3,\ldots,j+1;n,j+2,\ldots,n-1)
+ \cdots + A_{n;j}(1_\qb,2_q;3,\ldots,j+1;j+2,\ldots,n-1,n) \; ; \cr
}\eqn\loopphotona
$$
the unwanted diagrams again cancel.

For those contributions that do contain a (charged) fermion or scalar
loop, we must in effect consider separately the diagrams where the photon
couples to the external fermion line, and the diagrams where it couples
to the closed fermion (or scalar) loop, since these different fermion lines
may have different charges.
This can be accomplished (for $j>1$) by considering
separately the contributions where the photon replaces a gluon within
the first set, $3\ldots j+1$, associated with the trace of gluon matrices
in $\Gr_{n-1;j}^{(\qb q)}$, or within the second set, $j+2\ldots n$,
associated with the string of gluon matrices whose $i_2,\ib_1$ component
appears in the color factor.  In the former case, the factor of $Q$
along with $\nf$
(or $\ns$) will be replaced by the trace over the fermion charge matrix
(respectively the scalar charge matrix); in the latter case, the
amplitude will continue to appear with factors of $Q$ and $\nf$ (or
$\ns$ for scalar-loop contributions).

More concretely, define $A_{n;j}^{[1/2]}$
to be the coefficient of $\nf/N_c$ in $A_{n;j}$,
$$\eqalign{
A_{n;j}^{[1/2]}(1_\qb,2_q;3,\ldots,n-1,n) = N_c
{\partial\over\partial \nf} A_{n;j} (1_\qb,2_q;3,\ldots,n-1,n) \; .
}\anoneqn$$
Using equations~(\use\susyAnoneformula), (\use\subltotal), we may
relate $A_{n;j}^{[1/2]}$ to primitive amplitudes,
$$\eqalign{
A_{n;1}^{[1/2]} (1_\qb,2_q;3,\ldots,n-1,n)\ &=\
 - A_{n}^f (1_\qb,2_q;3,\ldots,n-1,n)
 - A_{n}^s (1_\qb,2_q;3,\ldots,n-1,n), \cr
A_{n;j>1}^{[1/2]} (1_\qb,2_q;3,\ldots,n-1,n)\ &=\
 (-1)^j \sum_{\sigma\in COP\{\alpha\}\{\beta\}}
     A_n^{R,[1/2]} \L \sigma(1_{\bar{q}},2_q,3,\ldots,n) \R. \cr
}\anoneqn$$
We can then construct the two different parts of the one-photon
partial amplitude for each $j>1$, as described above,
$$\eqalign{
& A_{n;j}^{1\gamma,[1/2]:I}(1_{\bar{q}},2_q;3,\ldots,j+1;j+2,\ldots,n-1;n)
  \cr
& =\
  A_{n;j+1}^{[1/2]}(1_{\bar{q}},2_q;n,3,\ldots,j+1;j+2,\ldots,n-1)
+ \cdots
+ A_{n;j+1}^{[1/2]}(1_{\bar{q}},2_q;3,\ldots,n,j+1;j+2,\ldots,n-1)
  \; ,\cr\cr
& A_{n;j}^{1\gamma,[1/2]:II}(1_{\bar{q}},2_q;3,\ldots,j+1;j+2,\ldots,n-1;n)
  \cr
& =\
  A_{n;j}^{[1/2]}(1_\qb,2_q;3,\ldots,j+1;n,j+2,\ldots,n-1)
+ \cdots
+ A_{n;j}^{[1/2]}(1_\qb,2_q;3,\ldots,j+1;j+2,\ldots,n-1,n)\;.\cr
}\eqn\loopphotonba
$$
For $j=1$ the color flow is different, and we get instead
$$\eqalign{
A_{n;1}^{1\gamma,[1/2]:I}(1_\qb,2_q;3,\ldots,n-1;n)
\ &=\ A_{n;1}^{[1/2]}(1_\qb,2_q;n,3,\ldots,n-1)
   + \cdots + A_{n;1}^{[1/2]}(1_\qb,2_q;3,\ldots,n-1,n)\cr
&\qquad  + A_n^{L,[1/2]}(1_\qb,n,2_q,3,\ldots,n-1), \cr
A_{n;1}^{1\gamma,[1/2]:II}(1_\qb,2_q;3,\ldots,n-1;n)
   \ &=\ - A_n^{L,[1/2]}(1_\qb,n,2_q,3,\ldots,n-1). \cr
}\eqn\loopphotonbb
$$

Using these pieces, we write out the $\nf$-dependent pieces in the
full amplitude,
$$\hskip -.5 cm
\eqalign{
\LP {\cal A}^{1\gamma}_n\RV_{\nf} &=
  {\Tr_{\rm f}(Q_f)\over N_c}\,{\sqrt{2} e} g^{n-1}
   \sum_{j=1}^{n-2} \sum_{\sigma\in S_{n-3}/S_{n-1;j}}
    \Gr_{n-1;j}^{(\qb q)}(\sigma(3,\ldots,n-1))\
 A_{n;j}^{1\gamma,[1/2]:I}(1_{\bar{q}},2_q;\sigma(3,\ldots,n-1);n) \cr
& +{Q\nf\over N_c}\,{\sqrt{2} e} g^{n-1}
   \sum_{j=1}^{n-2} \sum_{\sigma\in S_{n-3}/S_{n-1;j}}
    \Gr_{n-1;j}^{(\qb q)}(\sigma(3,\ldots,n-1))\
 A_{n;j}^{1\gamma,[1/2]:II}(1_{\bar{q}},2_q;\sigma(3,\ldots,n-1);n)
  \; ,\cr
}\eqn\loopphotondecompb
$$
where $Q_f$ is the fermion charge matrix, and $\Tr_{\rm f}$ represents
the trace over flavors.
An analogous decomposition holds for the contributions
proportional to $\ns$.

\listrefs
\listfigs
\end